\documentclass[useAMS,usenatbib,usegraphicx]{mn2e}
\usepackage{threeparttable}
\usepackage{rotating}

\title[X-ray monitoring of NGC~5408~X-1 by {\it Swift}]{A long-term X-ray monitoring of the ultraluminous X-ray source NGC~5408~X-1 with {\it Swift} reveals the presence of dips but no orbital period}
\author[F. Gris\'e, P. Kaaret, S. Corbel, D. Cseh, and H. Feng]{F. Gris\'e,$^{1,2,3}$\thanks{E-mail:
fgrise@iac.es} P. Kaaret, $^{3}$ S. Corbel, $^{4}$ D. Cseh, $^{4,5}$ and H. Feng~$^{6}$\\
$^{1}$Instituto de Astrof\'isica de Canarias, E-38200 La Laguna, Tenerife, Spain\\
$^{2}$Departamento de Astrof\'isica, Universidad de La Laguna, Avda. Astrof\`isico Francisco Sanchez s/n, E-38271 La Laguna, Tenerife, Spain\\
$^{3}$Department of Physics and Astronomy, Iowa City 52240, USA\\
$^{4}$Laboratoire Astrophysique des Interactions Multi-\'echelles (UMR 7158), CEA/DSM-CNRS-Universit\'e Paris Diderot, CEA Saclay,\\
F-91191 Gif sur Yvette, France\\
$^{5}$ Department of Astrophysics/IMAPP, Radboud University Nijmegen, P.O. Box 9010, 6500 GL Nijmegen, The Netherlands\\
$^{6}$Department of Engineering Physics and Center for Astrophysics, Tsinghua University, Beijing 100084, China}

\begin{document}

\date{Accepted . Received ; in original form }

\pagerange{\pageref{firstpage}--\pageref{lastpage}} \pubyear{2012}

\maketitle

\label{firstpage}

\begin{abstract}
NGC~5408~X-1 is a well-studied ultraluminous X-ray source (ULX) that has been seen to emit in X-rays persistently above the Eddington limit of a stellar-mass black hole for years. In this paper we report on the most extensive X-ray monitoring of a ULX, using more than four years of observations from the {\it Swift} satellite. We find that the 115~day periodicity reported by \citet{Strohmayer09} disappeared after only a few cycles, confirming the suspicion of \citet{Foster10} that the periodicity is most likely super-orbital and not the orbital period of the system. 
We also report on a clear dipping behaviour of the source that may be related to a (super)-orbital phenomenon. All these features are reminiscent of Galactic X-ray binaries and strengthen their link with ULXs. Deeper observations of a dip could help resolve the ambiguity about the interpretation of the spectral components of ULXs.

\end{abstract}

\begin{keywords}
accretion, accretion discs -- X-rays: binaries -- X-rays: individual (NGC 5408 X-1)
\end{keywords}

\section{Introduction}

In the last few years the number of known extragalactic, non-nuclear X-ray sources emitting well above the Eddington limit of a 20 $M_{\odot}$ black hole (L$_X \sim 3 \times 10^{39}\ \mathrm{erg \; s^{-1}}$) has been dramatically increasing with now hundreds of candidates \citep{Liu11, Swartz11, Walton11} thanks to the wealth of data acquired by {\it Chandra}, {\it XMM-Newton}, {\it Suzaku} and {\it Swift}. However, the nature of these ultraluminous X-ray sources (ULXs) is still a matter of great debate. Most are probably accreting black holes, but it is still unclear if they represent a new class of intermediate mass black holes (IMBHs ; $M \ga 100\ M_{\odot}$) or if they are stellar mass black holes (StMBHs ; $M < 100\ M_{\odot}$) with super-Eddington and/or mildly beamed emission (see \citealt{Feng11} for a recent review on ULXs).

The compact nature of ULXs was first confirmed by detection of X-ray variability on short timescales, down to minutes (e.g.\ \citealt{Okada98}). Evidence for long-term X-ray variability came with the repetition of pointed observations, which showed that they are also variable on timescales of months and years (e.g.\ \citealt{LaParola01}). The first long-term X-ray monitoring of a ULX was done with the Rossi X-ray Timing Explorer (RXTE) on M82~X-1 \citep{Kaaret06a,Kaaret06b}. M82~X-1 displays a 62-day period over 10 cycles that is likely the orbital period of the system \citep{Kaaret07} because the high coherence of the signal over a 3-year monitoring is consistent with a strictly periodic signal. However, other ULXs are usually fainter (in terms of observed flux) or more distant and could not be monitored by RXTE. The arrival of the {\it Swift} X-ray telescope \citep{Gehrels04} finally opened the way to follow the brightest of these sources with coverage with sufficient cadence and effective area to set statistically meaningful limits on underlying behaviour. Although based on a small sample, it revealed that ULXs were in general highly variable on timescales of days, by factors of a few to 15 \citep{Kaaret09,Grise10}. However, their variability does not usually show any clear periodic behaviour, with some exceptions. In addition, the peculiar HLX-1, the most luminous confirmed ULX, shows strong outbursts that appear to repeat every year \citep{Farrell09,Servillat11} with the X-ray count rate varying by a factor of 40.

Apart from M82~X-1 and HLX-1, strictly periodic X-ray variability has been reported for only a handful of ULXs. For example, M~51~X-7 displays a 2.1~hour periodicity \citep{Liu02, Yoshida10} and a ULX in NGC~3379 shows a 12.6~hour periodicity \citep{Fabbiano06}. However, these periods have usually been observed in single, short observations and have not been confirmed. In addition, these few objects are usually observed with an X-ray luminosity that is only slightly above $10^{39}\ \mathrm{erg \; s^{-1}}$ and are probably low-mass, or intermediate-mass X-ray binaries seen to radiate at or slightly above their Eddington luminosity. Their short periods 
favour this explanation \citep{Liu02}. Brighter ULXs that show evidence for periodicities are quite rare. NGC~5408~X-1 (hereafter N5408X1) displayed a long, 115~day periodicity in monitoring done with {\it Swift} for $\sim 500\ \mathrm{days}$ (\citealt{Strohmayer09}, hereafter S09). The latter object is the subject of this paper.

A very luminous and unresolved X-ray source in NGC~5408 ($d=4.8$~Mpc, \citealt{Karachentsev02}) was discovered with the Einstein observatory \citep{Stewart82} in the early 1980s that revealed itself later to be part of a population of extragalactic accreting black hole candidates \citep{Colbert99}.
NGC~5408~X-1 has since been observed multiple times at different wavelengths (see \citealt{Kaaret03,Soria06,Lang07,Cseh12} for radio studies, \citealt{Pakull03,Kaaret09a, Cseh11} for optical studies, \citealt{Grise12} for the spectral energy distribution of the source from ultraviolet to near-infrared). It has been seen to display an average X-ray luminosity (0.3-10 keV) of $1\times 10^{40}\ \mathrm{erg \; s^{-1}}$ (S09) with clear variability on timescales of minutes, days, months, and years within a factor of $\sim 2$--$3$ \citep{Soria04,Kaaret09,Strohmayer09,Kong10}. The source is also one of the few ULXs where quasi-periodic oscillations (QPOs) have been discovered \citep{Strohmayer07,Strohmayer09,Heil09}, although a clear result on their meaning is still a matter of debate \citep{Middleton11}. This is mainly due to the constancy of the X-ray spectral properties of N5408X1 \citep{Kaaret09} which does not allow a study of the timing properties over a wide range of spectral parameters \citep{Dheeraj12}, hampering a clear observational result on the behaviour of the QPOs with regard to that in GBHBs.

Intensive {\it Swift} monitoring of N5408X1 has shown a $\sim 115\ \mathrm{day}$ periodicity that has been interpreted as the orbital period of the ULX system (S09, \citealt{Han12}), although \citet{Foster10} suggests that this modulation may instead be super-orbital. A study based on high-resolution optical spectra of this ULX \citep{Cseh11} also shows some inconsistency between the available constraints on the mass function and an orbital period of $\sim 115\ \mathrm{days}$.
In this paper, we examine again the long term X-ray behaviour of N5408X1 using the large number of {\it Swift} observations performed since that report up until the end of June 2012\footnote{While this paper was reviewed, a paper \citep{Pasham13} was published reporting part of the same {\it Swift} data set including our daily monitoring from 2011. Here, we analysed in addition all data available up until the end of June 2012}.  This represents a factor of 3 increase in time coverage. We also discuss peculiarities that are visible in the light curves and suggest new observations that would lead to a better understanding of this ULX.

\section[]{Observations and data analysis}
All observations from the {\it Swift} data archive for NGC~5408~X-1, up until the end of June 2012, have been retrieved, which amount to 354 observations that span 1532~days (April 9, 2008 - June 19, 2012, see Table~\ref{tab_xrtdata} for more details). This corresponds to the most detailed X-ray monitoring of an ultraluminous X-ray source ever, including two periods of quasi-daily observations\footnote{One observation per day was performed in these periods, except for 9 days in the first period, and 5 days in the second} spanning respectively 116 days (May 1, 2011 - August 24, 2011) and 29 days (January 29, 2012 - February 26, 2012). This is the best sampling used to date to monitor a ULX on long periods, which compares favorably to the $3$--$7$~days cadence of previous observations (see Table~\ref{tab_xrtdata}) and allows us to look for short duration phenomenons in more details.

\begin{table*}
\centering
\caption[]{The {\it Swift}/XRT Observations for NGC~5408~X-1}
\label{tab_xrtdata}
\begin{threeparttable}[t]
\centering
\small
\begin{tabular}{lllcl}
\hline
\hline
Year & Number of Obs.\tnote{a} & Typical cadence & Typical exposure time & Typical fractional coverage \\
     &		      &       (days)	&	(s)	        &				\\
\hline
2008      &  55          & 3.4	        &	2115		& 0.23				\\
2009      &  77          & 3.6		&	1990		& 0.24				\\
2010      &  52          & 6.4		&	1920		& 0.28				\\
2011      &  80 (117)\tnote{b}    & 1.7 (1.0)\tnote{b}	&	1568		& 0.38 (0.83)\tnote{b}			\\
2012      &  47 (53)\tnote{b}     & 2.9		&	1047		& 0.61				\\
\hline
\end{tabular}
\begin{tablenotes}
\footnotesize
\item[a] Number of observations, based on the 1-day binning.
\item[b] In brackets, number of observations (or cadence, or fractional coverage) per civil day if significantly different from the 1-day binning.
\end{tablenotes}
\normalsize
\end{threeparttable}
\end{table*}

Given the (low) count rate of the source, {\it Swift}/XRT \citep{Burrows05} was used in its photon-counting (PC) mode.
The reduction process we used is the same as in \citet{Grise10}. Specifically, we retrieved level 2 event files from the archive, and then analysed each snapshot separately, rejecting the ones with an exposure time below 100 s.
Source counts were extracted from a circular region with a radius of 20 pixels (90 per cent of the point-spread function at 1.5 keV). The background was extracted using a circular region with a radius of 60 pixels located away from the ULX and also avoiding other, faint X-ray sources visible on a deep, combined image made from all observations. The spectra were corrected for the loss of flux due to bad pixels and bad columns. For this, an exposure map was generated for each snapshot and used to create an auxiliary response file (ARF). The hardness ratio (HR) used throughout this paper was defined as the ratio between the net count rate in the 1.5--10 keV band versus the 0.3--1.5 keV band.

We extensively used periodograms to look for periodicities in the data set. We mainly used the Lomb-Scargle (LS) implementation \citep{Lomb76,Scargle82} available in the {\it aitlib} set of IDL subroutines\footnote{http://astro.uni-tuebingen.de/software/idl/aitlib/timing/scargle.html} as suitable for unevenly sampled time series, and we used the method discussed in \citet{Horne86} for normalization of the periodogram. The shortest period that we can look for is given by the Nyquist frequency which is actually not so well defined for unevenly spaced time series. It has been shown that frequencies up to several times the Nyquist limit can be recovered from such data sets, without much frequency aliasing \citep{Mignard05}. Here, we will use a rather conservative lower limit of 1 day when considering this data set. Usually, a set of independent frequencies is calculated using, for instance, equation 13 of \citet{Horne86}. However, a set of independent frequencies only exists when dealing with an evenly spaced time series. Since the false alarm probability function is directly linked to this set of frequencies, it appears that false alarms derived in this manner could be inadequate. Therefore, we decided to follow the method outlined by \citet{Frescura08} which consists of using Monte-Carlo simulations to derive an empirical cumulative distribution function (CDF), that will be used to derive the number of frequencies to look at, as well as to directly determine the false alarm probabilities.
White noise simulations were produced keeping the time-sampling of the real data, and using the variance of the original light curve.
Practically, for each data set, we compute the CDF of $10^4$ white noise simulations, starting with the Scargle sampling rate (i.e number of frequencies $N_0/2$, where $N_0$ is the number of data points). At each iteration, we increase the sampling rate by an integer factor and calculate the corresponding CDF, stopping the iterative process when the limiting CDF is attained, i.e when the sum of square deviations (from the limiting CDF) of the CDF for $\nu$ times over-sampling is below 1 (see Figures 5 \& 6 in \citealt{Frescura08}). This amounts to applying an oversampling factor $\nu$ that ranges from factors 5 to 10 in our simulations. For simplicity, and due to the large number of Monte-Carlo simulations used in this study, we decided to use a constant oversampling factor of $10$. As noted in \citet{Frescura08}, we also observed that ``over-sampling the
periodogram does not dramatically increase the number of large peaks expected''.
From the limiting CDF, we can directly get the power thresholds corresponding to false alarm probabilities (fap), that we choose here as 0.01, 0.001, and 0.000064 (or 2.6, 3.3, and $4 \sigma$).

The periodograms do not show evidence of increasing slope at long periods and therefore the effect of red noise on the significance of the periodicities is at most limited. We investigated the effect of binning on the light curves and decided to mainly use the unbinned light curve (using all 791 separate snapshots). We also compare with the case of 1-day binning (314 data points). The largest effect of binning is to average the snapshots separated by several hours from each other, but belonging to the same observation performed during a given day. Only at times of very good sampling (the daily observations) do snapshots from two consecutive days get averaged, if they were performed within 24 hours of each other. In any case, given the long periodicities expected this should not change dramatically the significance of periodicities expected at some hundred days. Regardless, binned light curves are shown here for comparison, and a full characterization of the time series was done on the unbinned data set. The Lomb-Scargle periodogram does not take into account uncertainties, so we decided to discard data points with an error greater than $0.02\ \mathrm{count \; s^{-1}}$ which reduces the number of individual data points to 723 and 311 for the two cases considered. This removes points throughout the light curve, but preferentially around MJDs 55000--55100 and 55700--55800 where the exposure time of several snapshots was quite short, $\la 400$~s. However, data points removed this way are fairly well distributed in count rate ranges, i.e. between 0.04 and 0.15 $\mathrm{count\ s^{-1}}$, therefore not introducing any significant bias.

We also examined the spectral behaviour as a function of source brightness. We co-added spectra using different count rate ranges (see Table~\ref{tab_fp} for details) using the {\small ADDSPEC FTOOL}. We used the response matrix files (RMFs) from the calibration database according to the date of the observations. All {\it Swift} observations of N5408X1 were carried out after 2008, so the same RMF file was used for all observations (\verb=swxpc0to12s6_20010101v013.rmf=). The ARFs were then co-added separately using ADDARF and weighted accordingly to their counts. {\small GRPPHA} was used to bin the spectra with at least 20 counts in each bin. Finally, we fitted the spectra using XSPEC 12.7.1 \citep{Arnaud96}. Errors on the spectral parameters were estimated using the {\it error} command with 90 per cent confidence intervals for one interesting parameter. Fluxes and their associated errors (also at the 90 per cent confidence level) for different model components were calculated using the {\it cflux} command (see the XSPEC user's guide for more information).

\section[]{Results}

\subsection{Periodicities}

\begin{figure*}
\centering
   \includegraphics[width=9.5cm,angle=270]{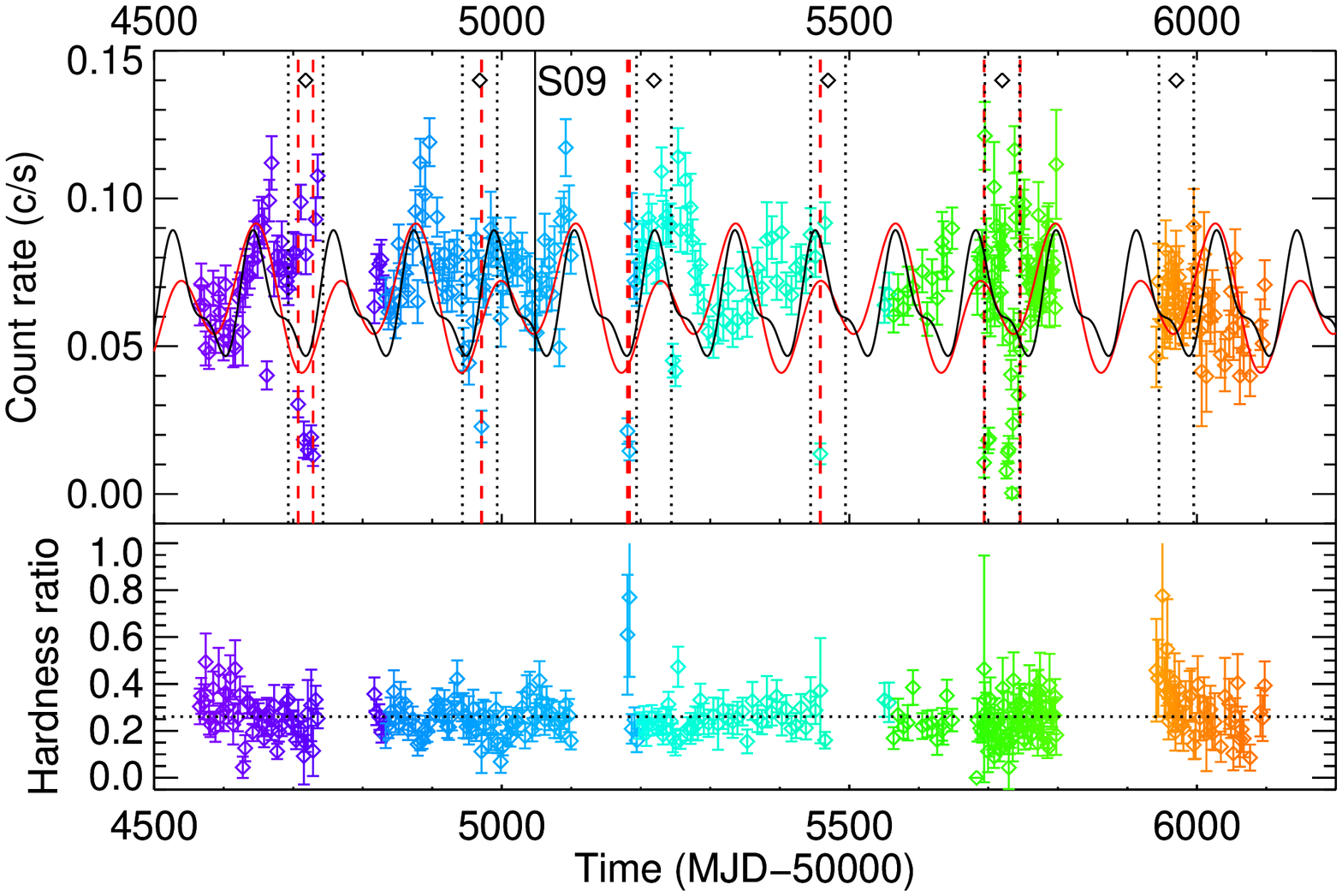}
   \begin{tabular}{cc}
   \includegraphics[width=5cm,angle=270]{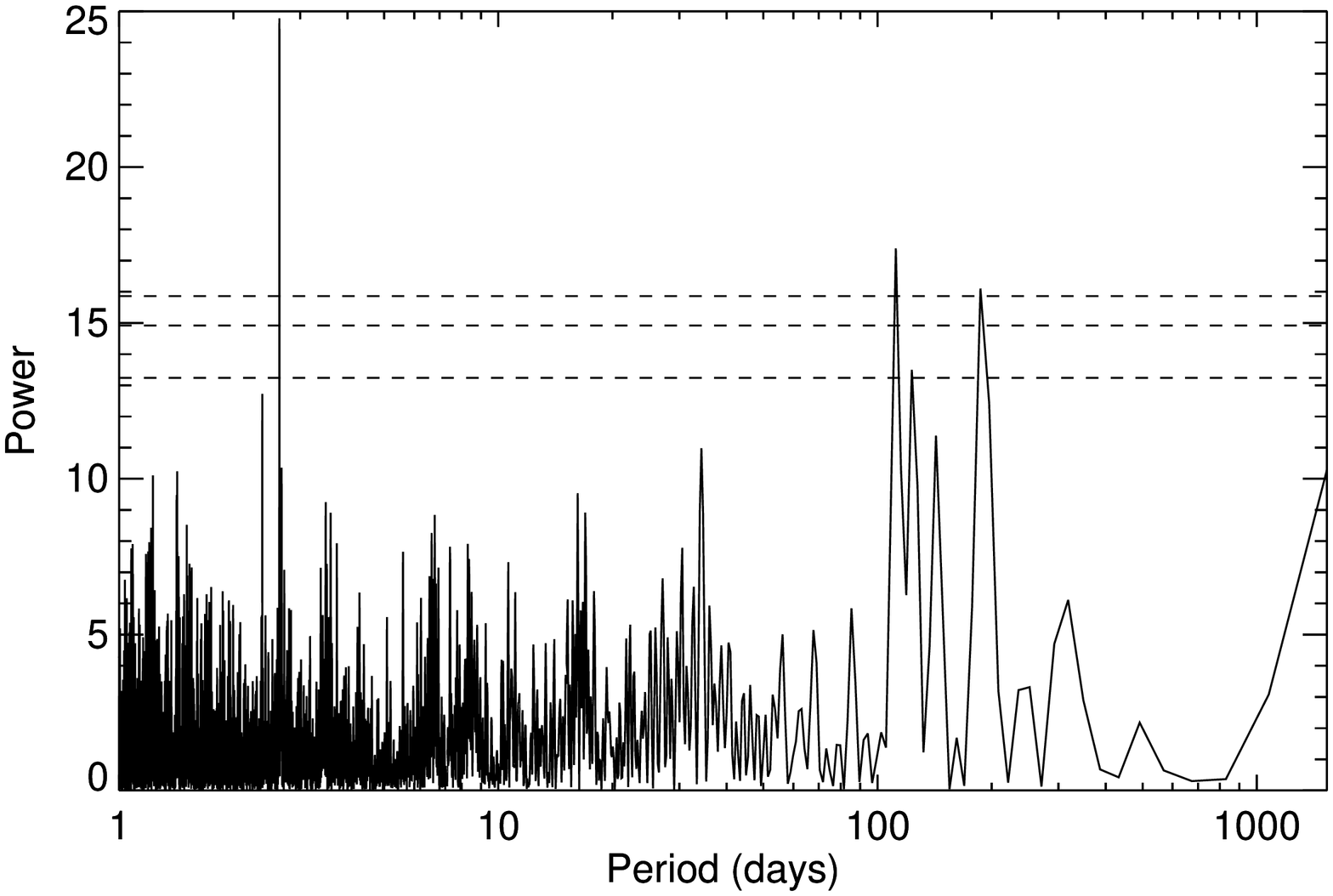}					 
   &\includegraphics[width=5cm,angle=270]{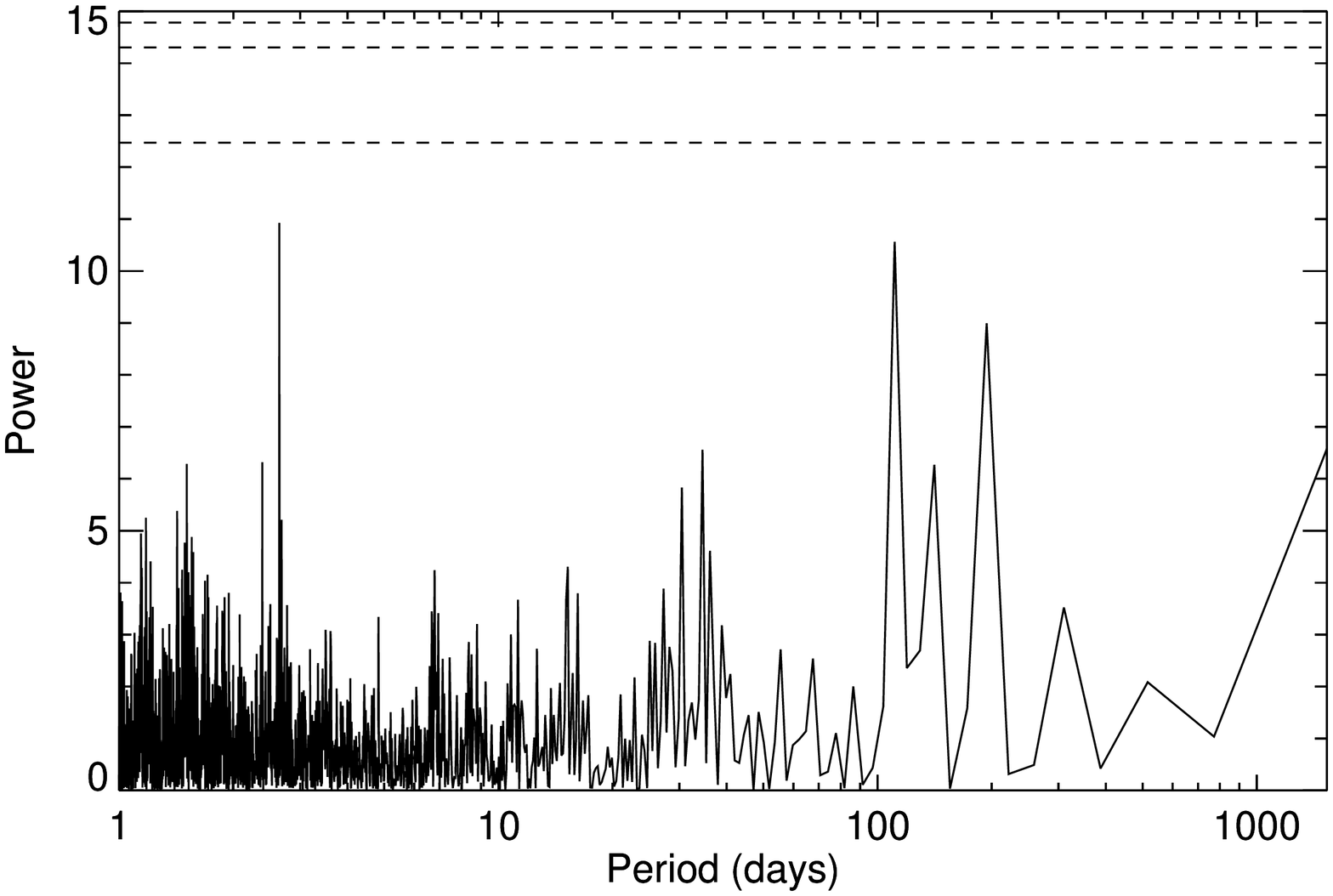}\\
   \includegraphics[width=5cm,angle=270]{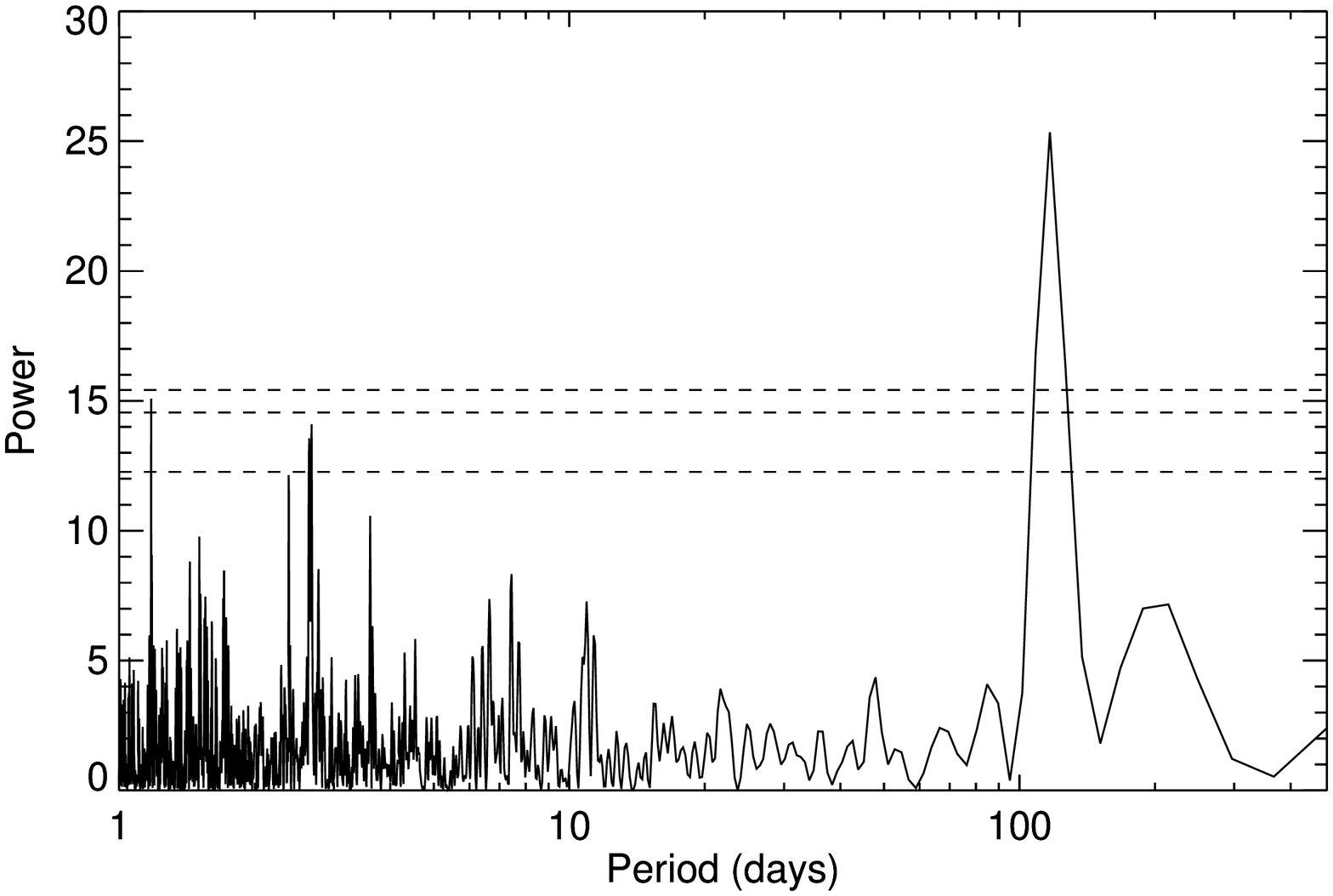}					 
   &\includegraphics[width=5cm,angle=270]{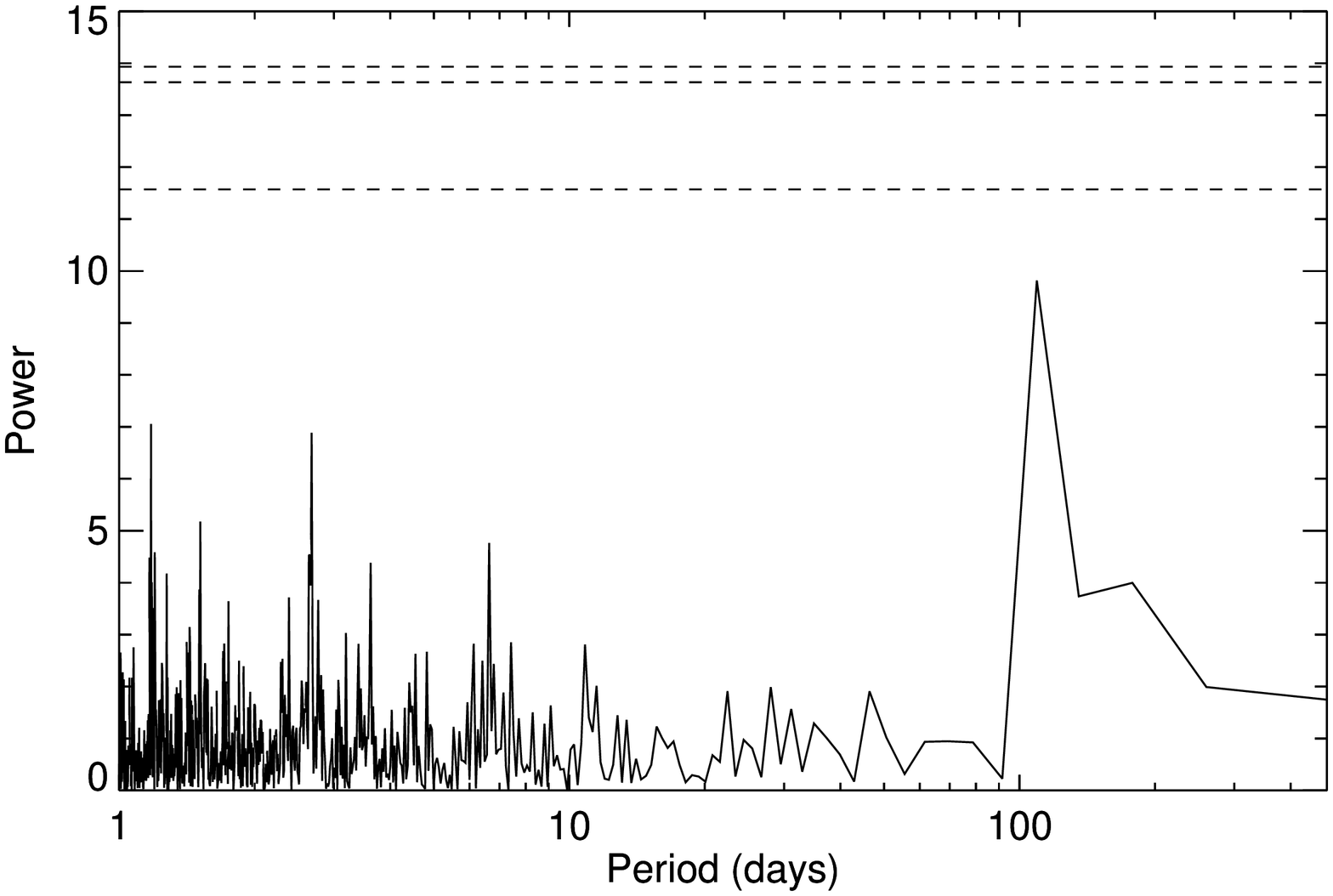}\\
    \end{tabular}
   \caption[]{Top: X-ray light curve in the 0.3--10 keV band for NGC~5408~X-1, and associated hardness ratios. Only the light curve re-binned with one observation per day is shown here, for clarity. The color scheme represents the civil year the data were taken (from 2008 to 2012, see Table~\ref{tab_xrtdata}). The black curve is the expected light curve under the S09 orbital period (115.5~days) and the red curve is the expected light curve under the \citet{Pasham13} orbital period (230.0~days), showing their inadequacy to fit the light curve past $\sim \mathrm{MJD}\ 55300$. The vertical red dashed lines show the extent of observed dipping intervals while the black dashed lines represents the expected dipping intervals based on a recurrent periodicity of 250.5~days (and the black diamonds the expected midpoints of the dipping intervals), using the most detailed dipping interval seen in 2011 as the origin. Middle: Lomb-Scargle periodograms of the unbinned light curve (left) and of the rebinned light curve (right). Dashed, horizontal lines denote the false alarm probability (fap) of 1-0.99, 1-0.999, and 1-0.999936 (or significance of 2.6, 3.3, and $4 \sigma$), based on white noise simulations. The two main peaks are at $\sim 2.6$ and $\sim 112$~days. Bottom: Lomb-Scargle periodograms, but only considering the S09 data set.}
   \label{lc_periodogram_main}
\end{figure*}

\begin{figure*}
\centering
   \begin{tabular}{cc}
   \includegraphics[width=5cm,angle=270]{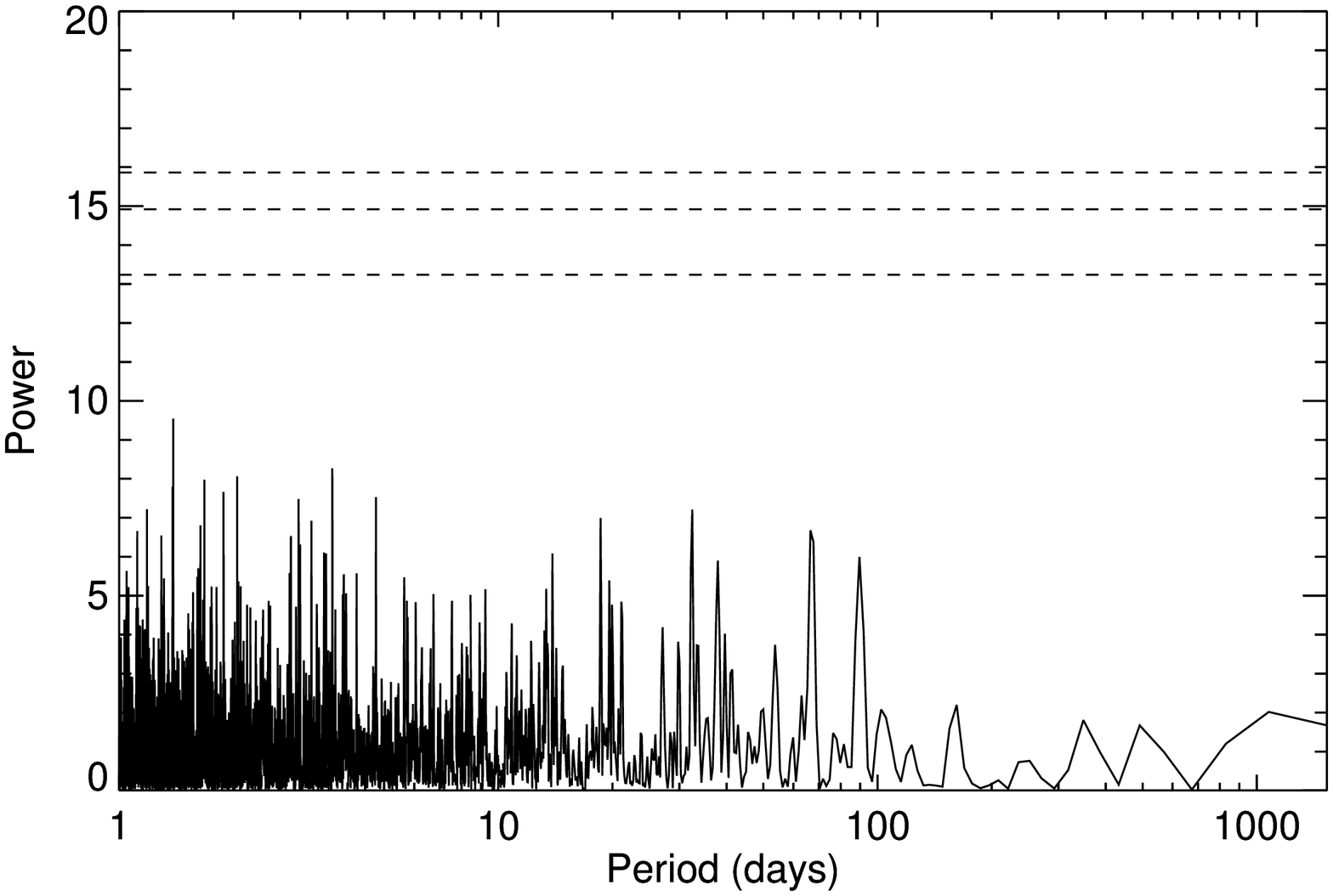}					 
   &\includegraphics[width=5cm,angle=270]{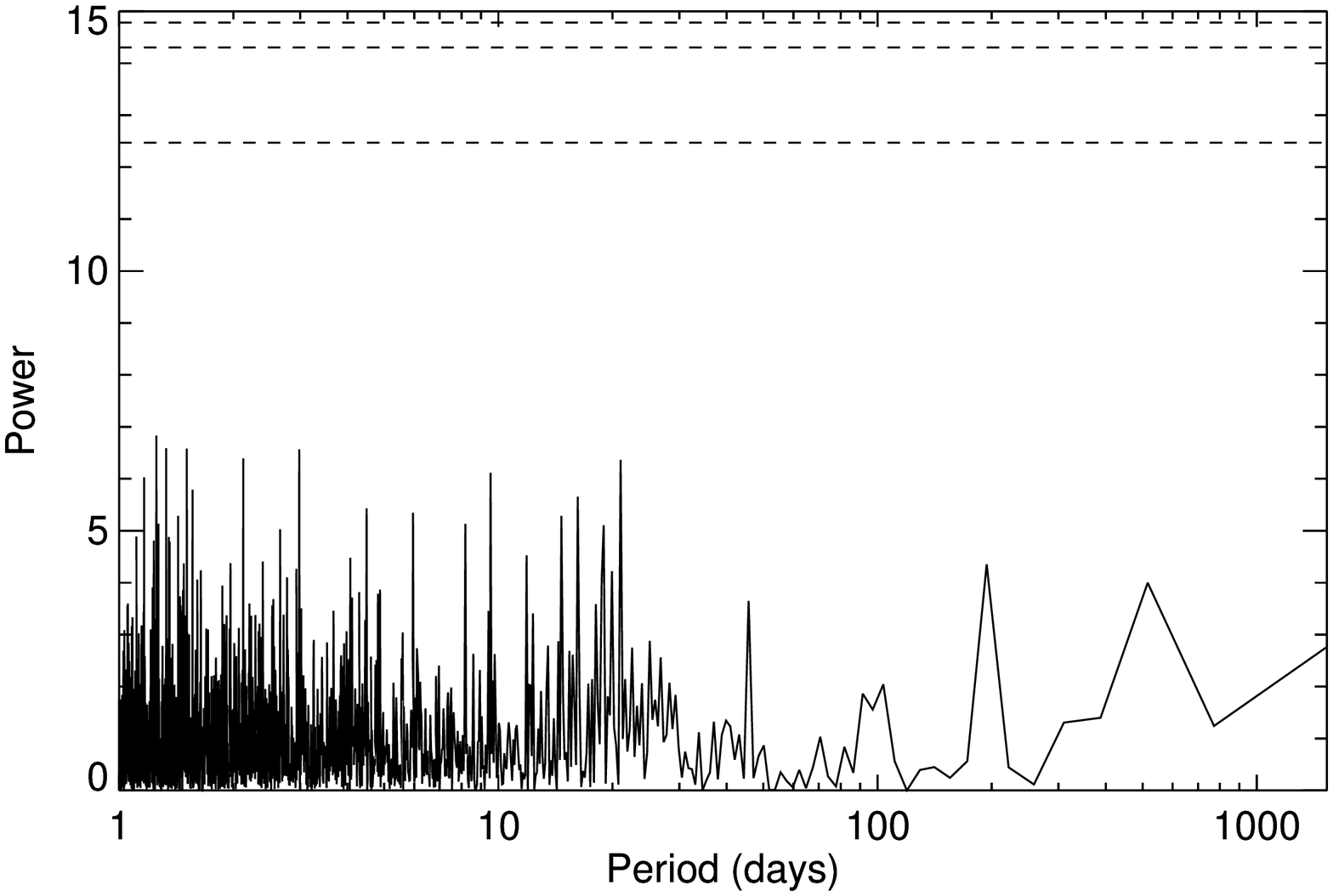}\\
    \end{tabular}
   \caption[]{Lomb-Scargle periodograms of a randomly selected white-noise light curve using the unbinned time-sampling (left), and using the rebinned time-sampling (right). Dashed, horizontal lines denote the false alarm probability (fap) of 1-0.99, 1-0.999, and 1-0.999936 (or significance of 2.6, 3.3, and $4 \sigma$), based on $10^4$ white noise simulations.}
   \label{wn_periodograms}
\end{figure*}

The 4.2 yr light curve of the ULX is shown in Figure~\ref{lc_periodogram_main}. There are obvious short and long-term variations of the count rate that are mostly restricted to the range [0.04-0.12]~$\mathrm{count \; s^{-1}}$, with a few observations showing lower count rates -- the X-ray dips described below.

To confirm and refine the possible periodicity of 115~days interpreted as the orbital period of the system (S09) or as a super-orbital period \citep{Foster10}, we calculated periodograms from the 0.3--10 keV light curve from the whole, unbinned data set. The resulting plot (Figure~\ref{lc_periodogram_main}, left central panel) shows several peaks, with the strongest one being at $P \sim 2.65$~days for a power of 24.8. A second series of (less) significant peaks is present between periods of 100 and 200~days. More specifically, two other peaks above the $4 \sigma$ significance level of the white noise can be seen with powers of 17.4 and 16.1, and periods of $P \sim 112$ and 187~days, respectively. The peak at $\sim 112$~days is within the errors of the periodicity published in S09 ($115.5 \pm 4$~days). However, our periodogram, with a temporal baseline three times as long, shows that this peak became less significant over time, when compared with the S09 data set only (Figure~\ref{lc_periodogram_main}, central and bottom left panels). Also, we find more power at higher frequencies than S09.
Rebinning the light curve with one point per day (Figure~\ref{lc_periodogram_main}, right central panel) shows that the three strongest peaks (with powers of 10.9, 10.6, and 9.0 at respectively periods of 2.64, 111, and 194~days) are not even significant at the 0.99 confidence level which is a strong drop in significance considering the long periodicities of two of the peaks. This behaviour is well reproduced in our simulations using a fake sinusoidal signal, and therefore we will only use the unbinned data set for searching periodicities in the rest of this paper. We thereby caution other users that binning may significantly alter the significance of periodicities, even for long ones. There are also significant gaps in the {\it Swift} light curve. To look for a possible effect on the periodograms, we show (Figure~\ref{wn_periodograms}) a randomly selected white noise light curve from the Monte-Carlo simulations. From these, we can see that the gaps do not lead to the presence of significant peaks.

\begin{figure*}
\centering
   \begin{tabular}{cc}
    \includegraphics[width=5cm,angle=270]{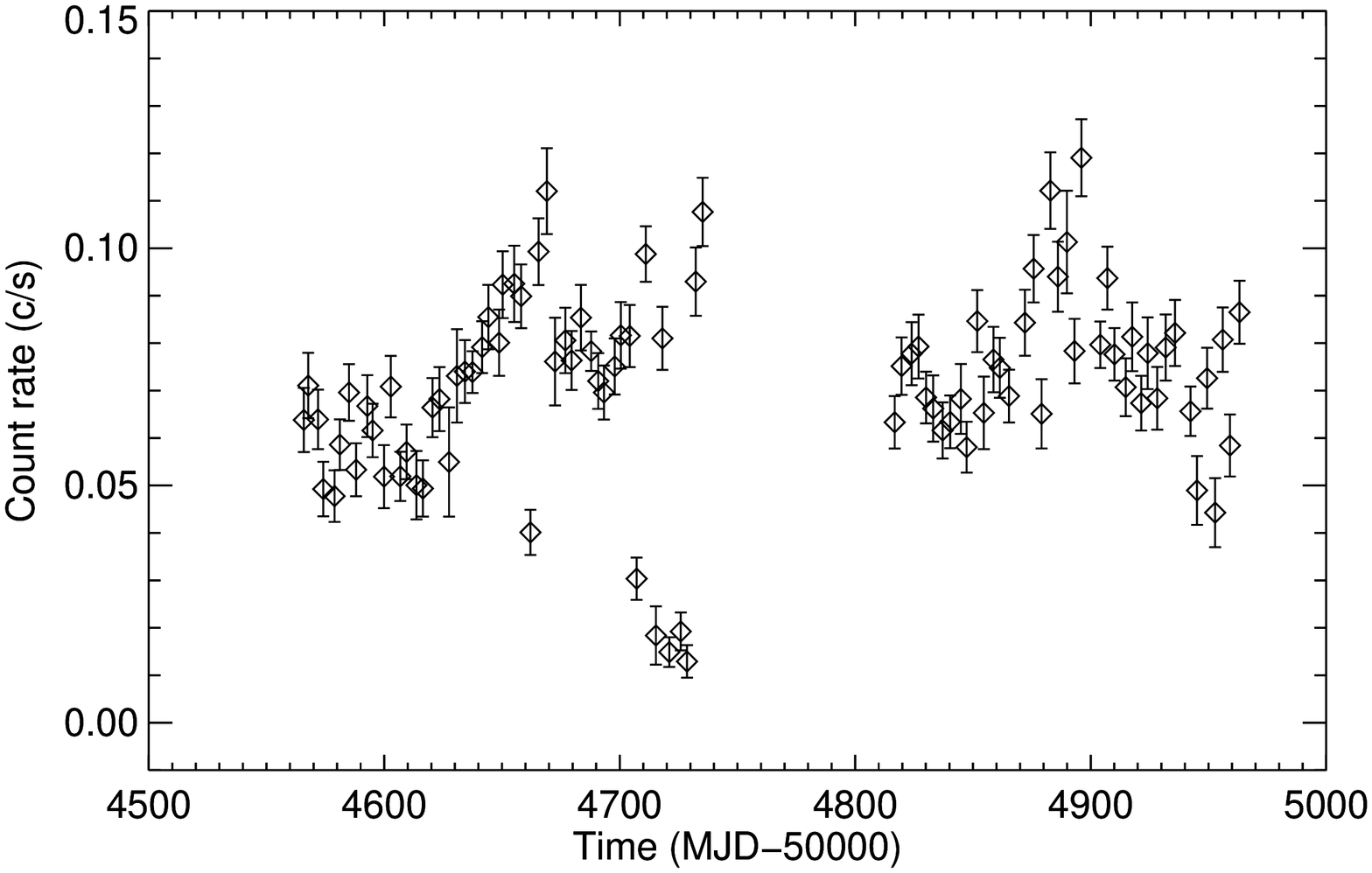}
   &\includegraphics[width=5cm,angle=270]{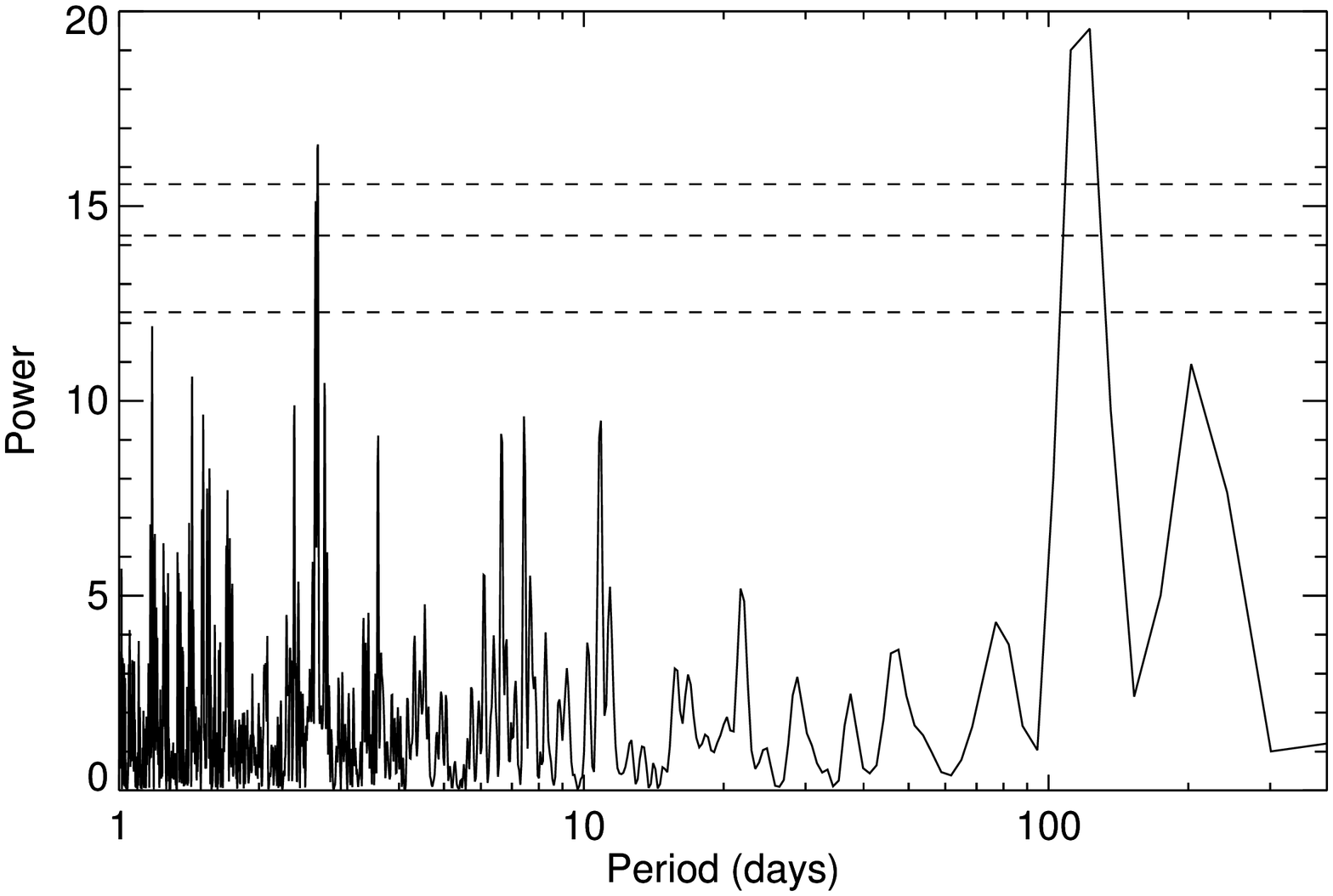}\\
    \includegraphics[width=5cm,angle=270]{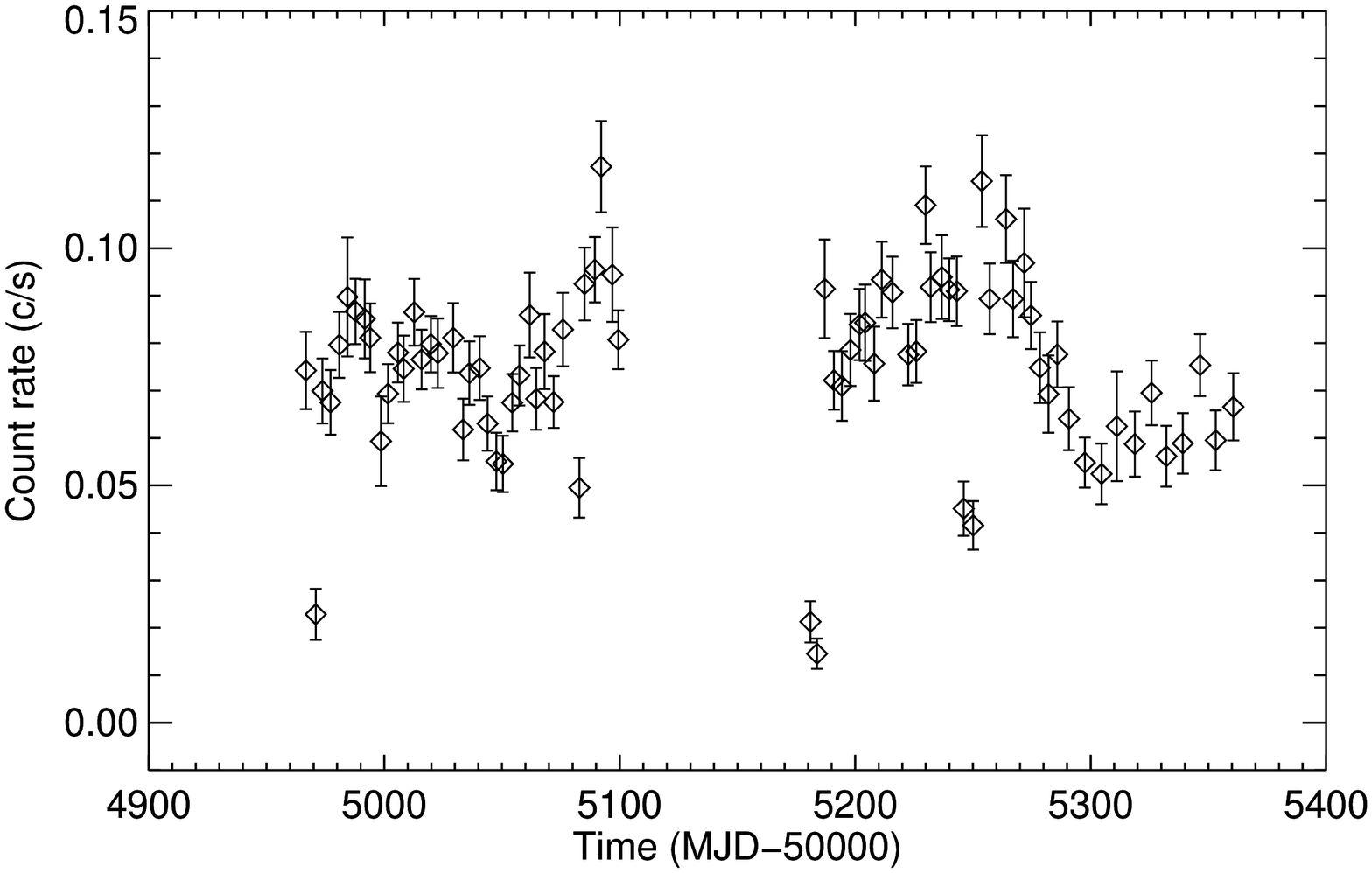}
   &\includegraphics[width=5cm,angle=270]{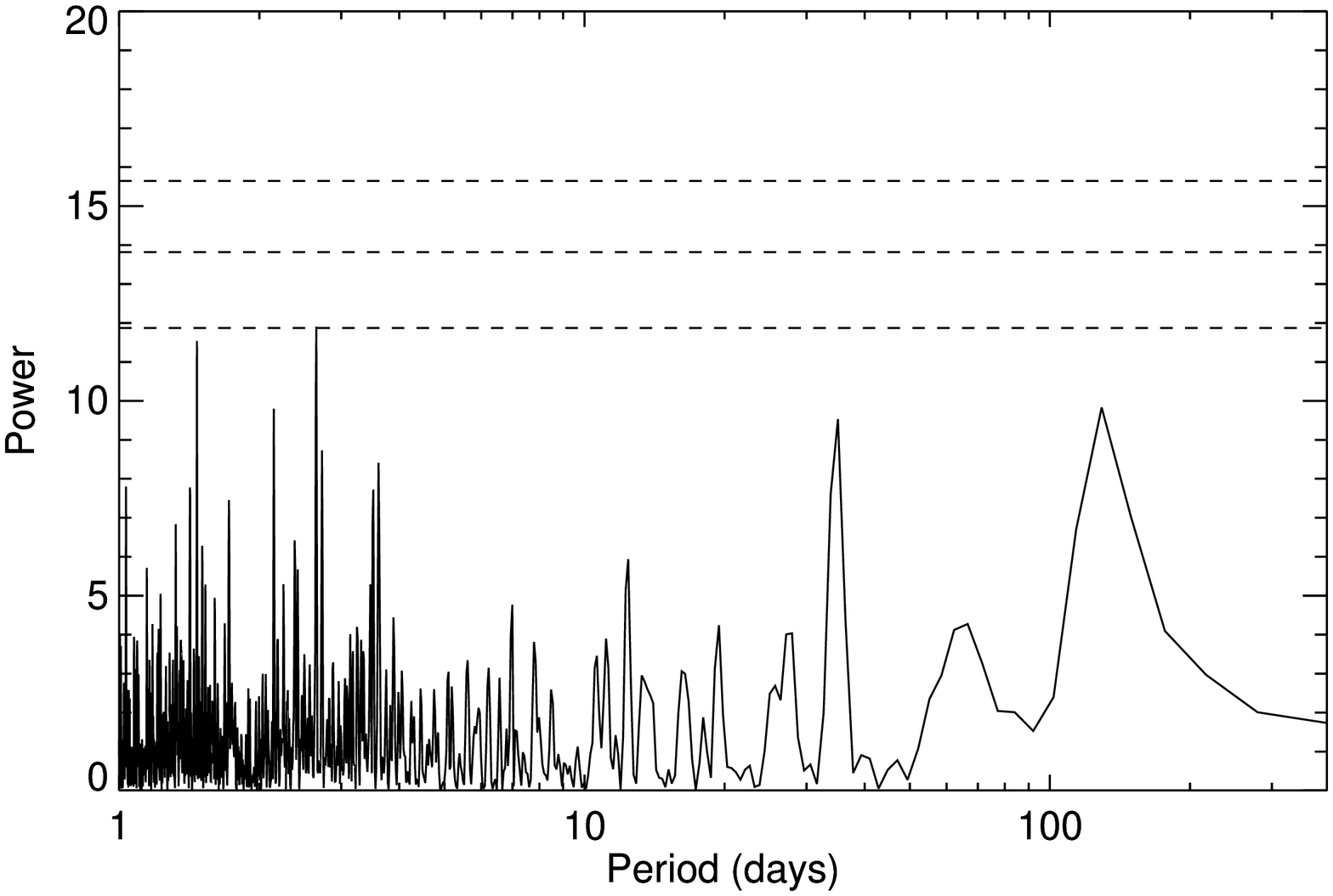}\\
    \includegraphics[width=5cm,angle=270]{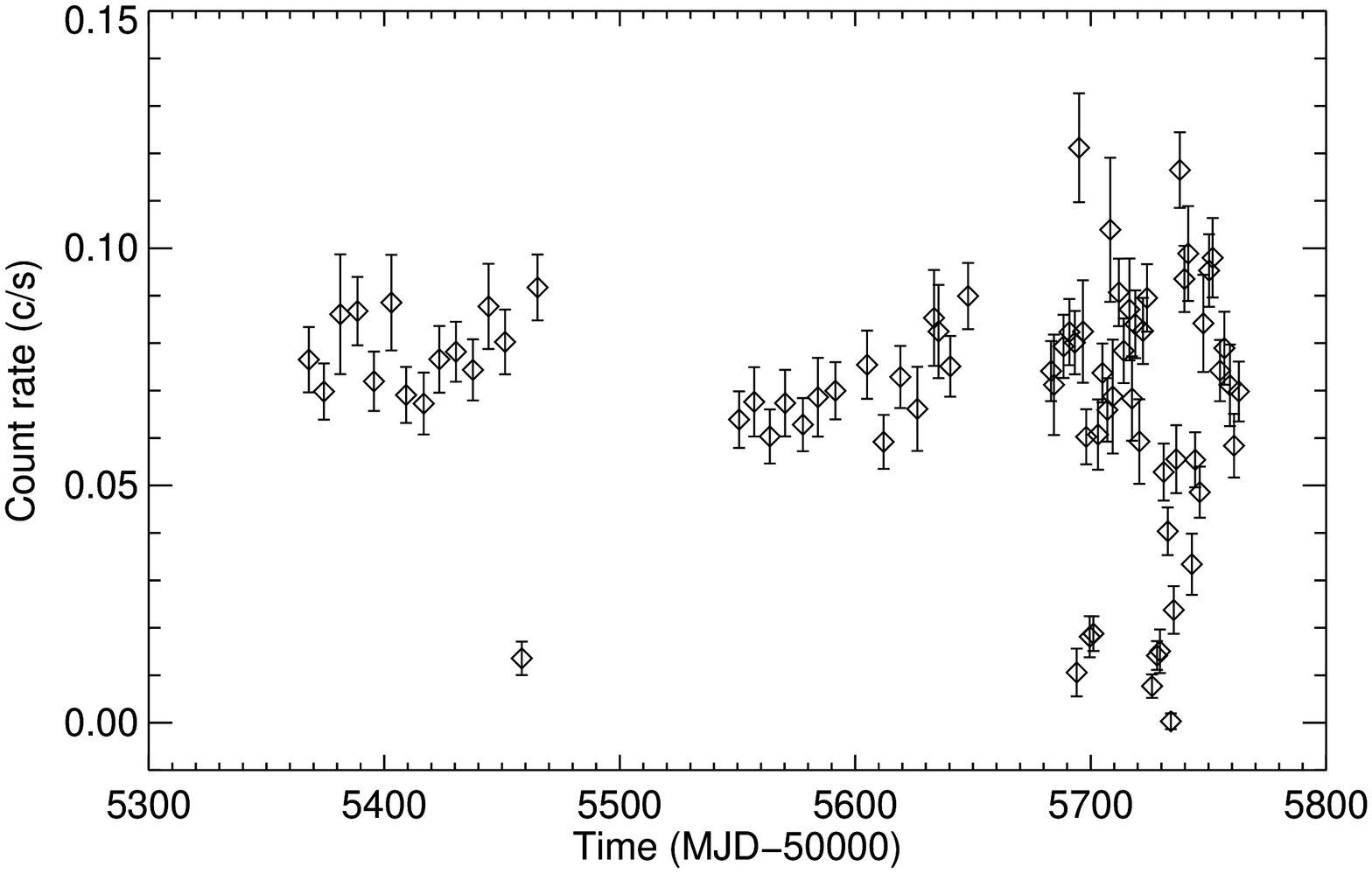}
   &\includegraphics[width=5cm,angle=270]{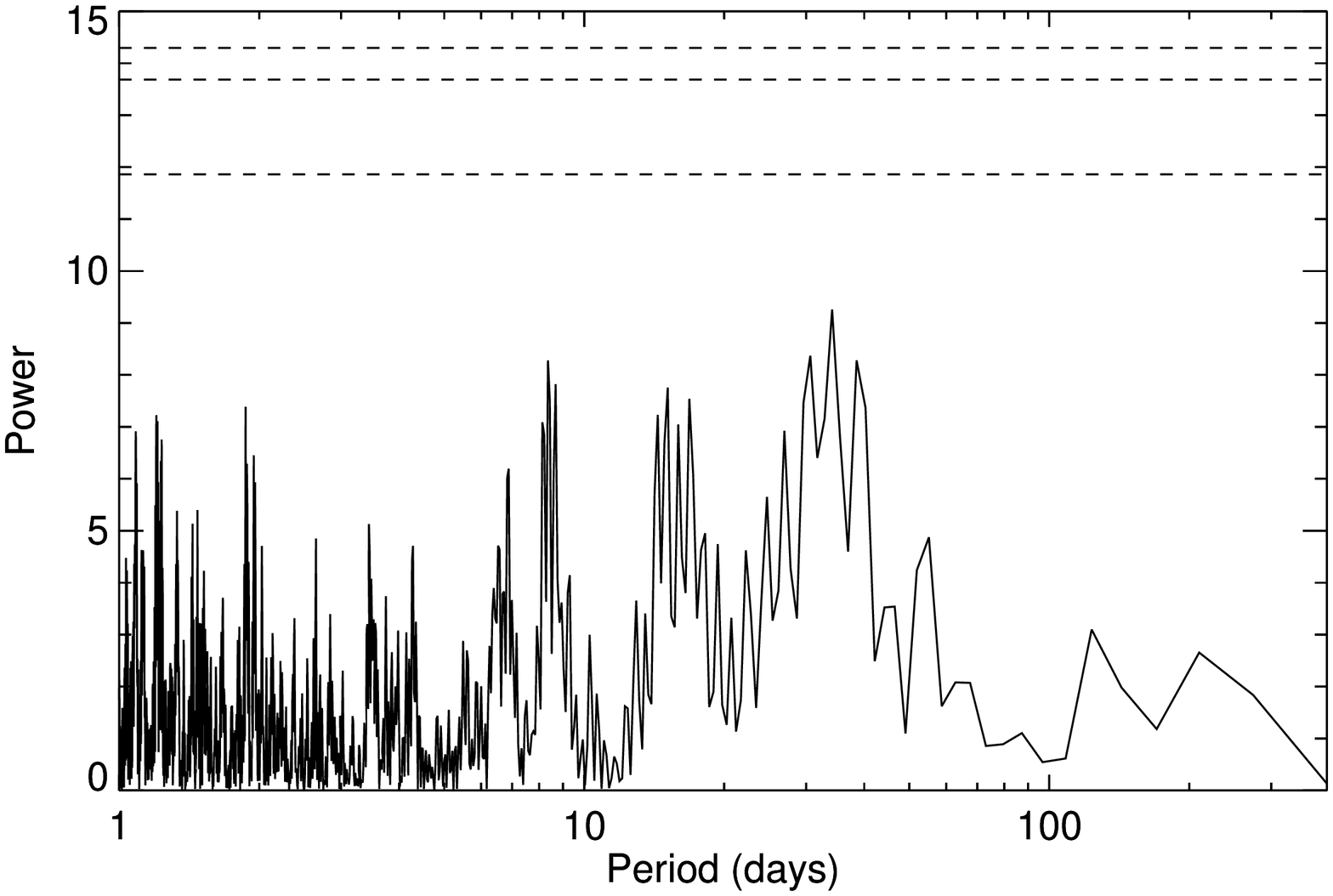}\\
    \includegraphics[width=5cm,angle=270]{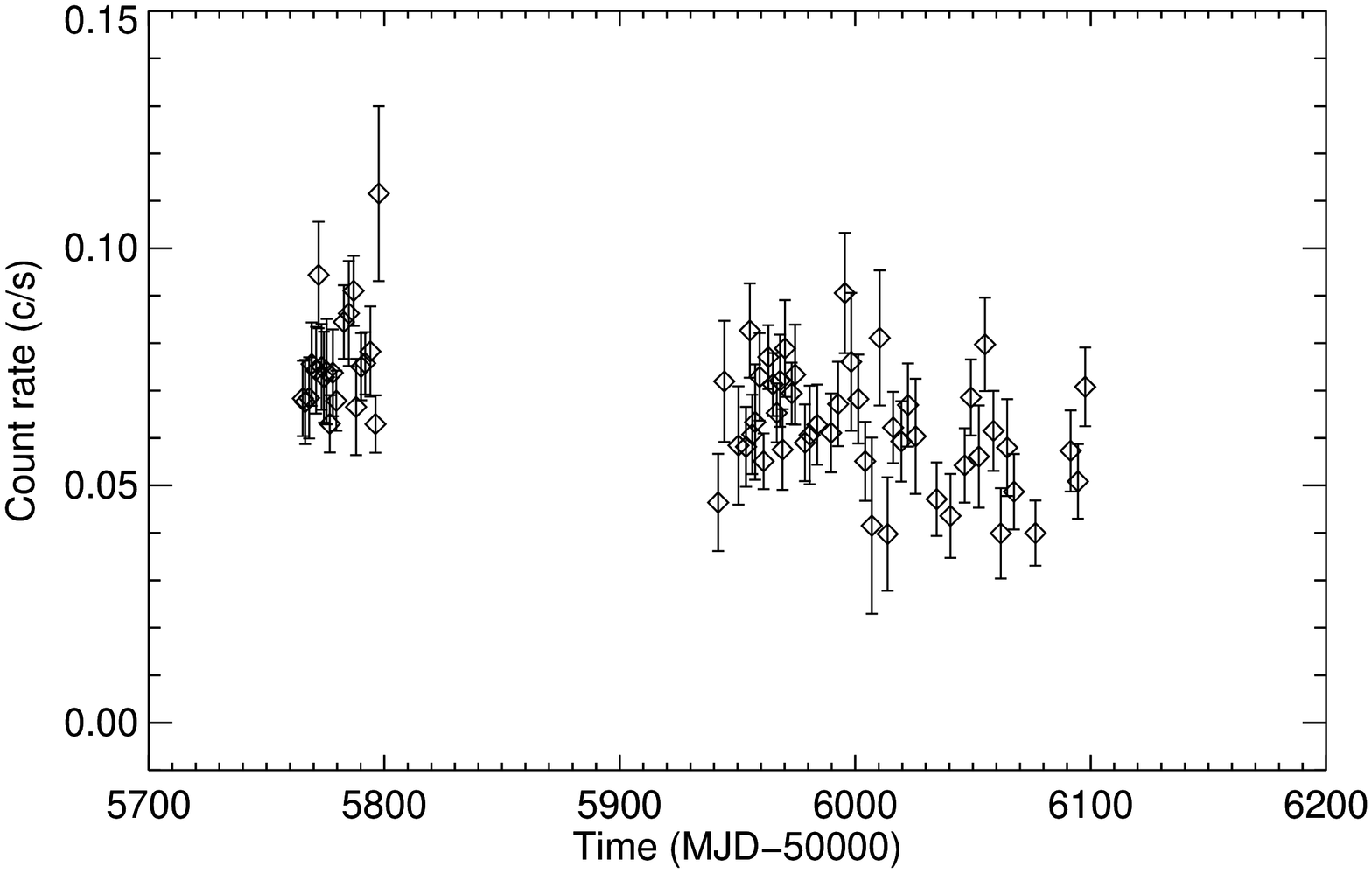}
   &\includegraphics[width=5cm,angle=270]{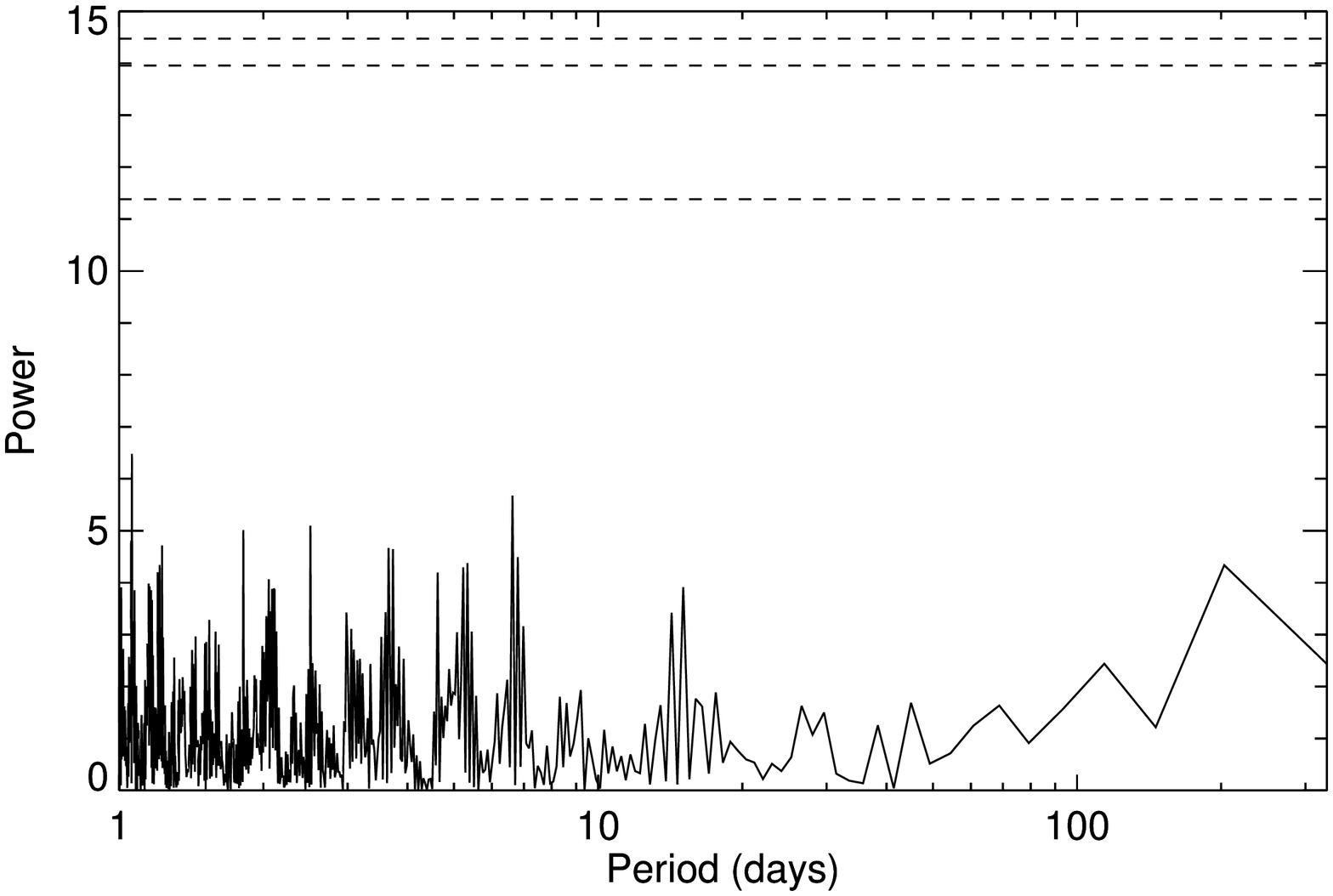}\\
    \end{tabular}
   \caption[]{Left: X-ray light curves in the 0.3--10 keV band for NGC~5408~X-1, split in 400 day windows. Only the light curves rebinned with one observation per day are shown here, for clarity. Right: corresponding Lomb-Scargle periodograms calculated from the unbinned light curves. Dashed, horizontal lines denote the false alarm probability (fap) corresponding to confidence levels of 0.99, 0.999, and 0.999936 (or significance of 2.6, 3.3, and $4 \sigma$), based on white noise simulations. The main peak in the top periodogram is at $\sim 115$~days, as originally reported by S09.} 
   \label{lc_periodograms_all}
\end{figure*}

\begin{figure*}
\centering
   \begin{tabular}{cc}
    \includegraphics[width=5cm,angle=270]{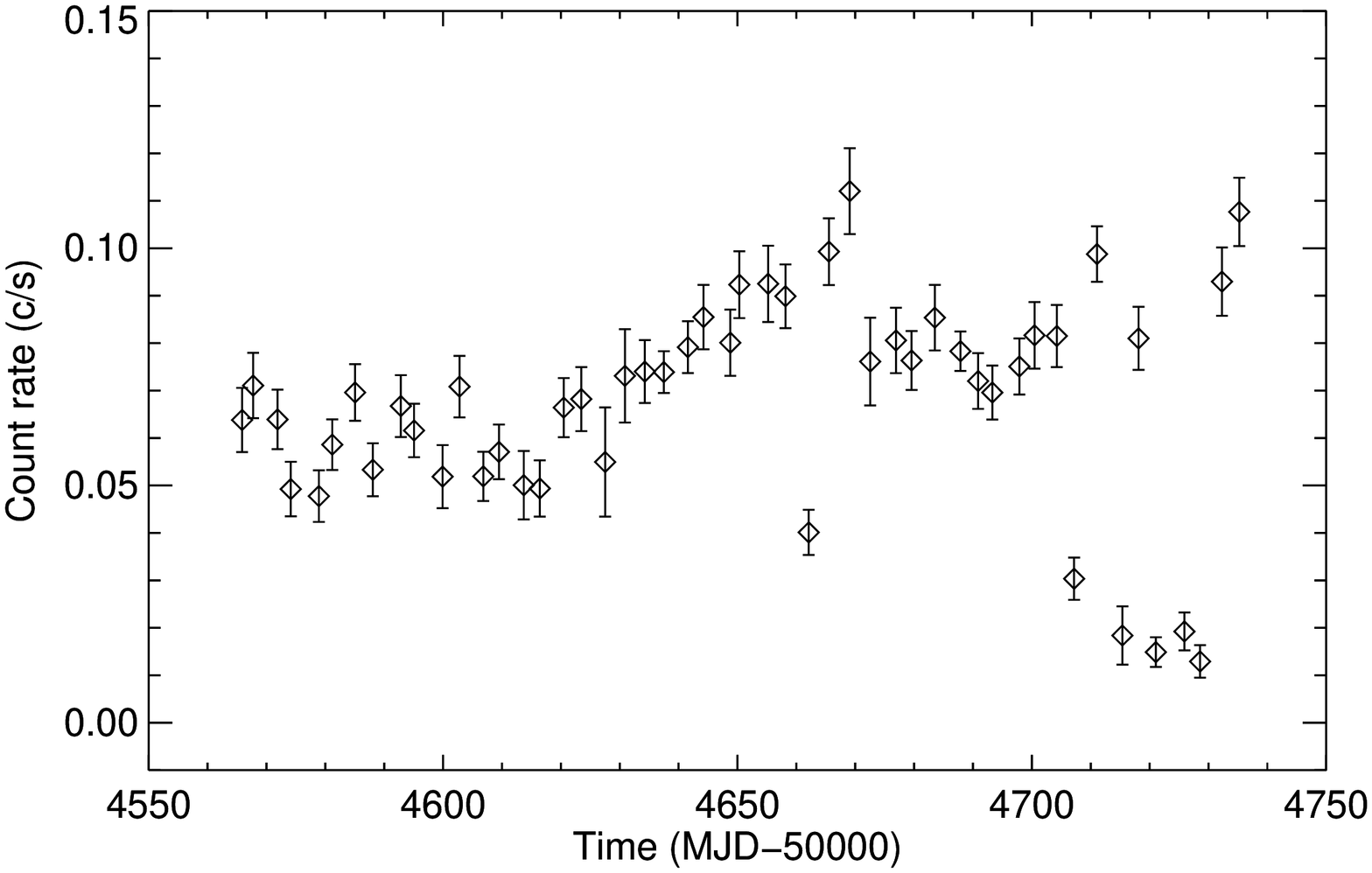}
   &\includegraphics[width=5cm,angle=270]{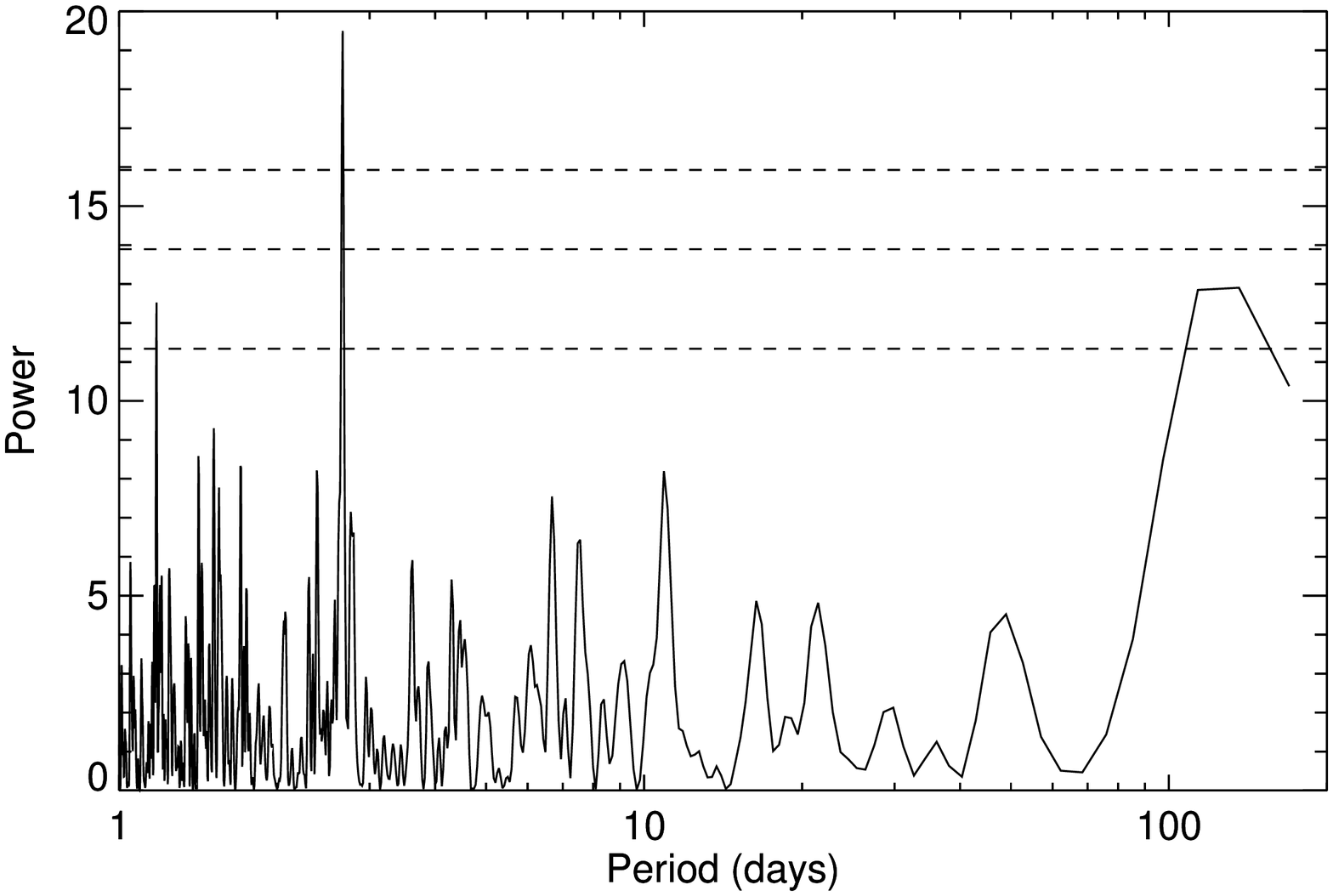}\\
    \includegraphics[width=5cm,angle=270]{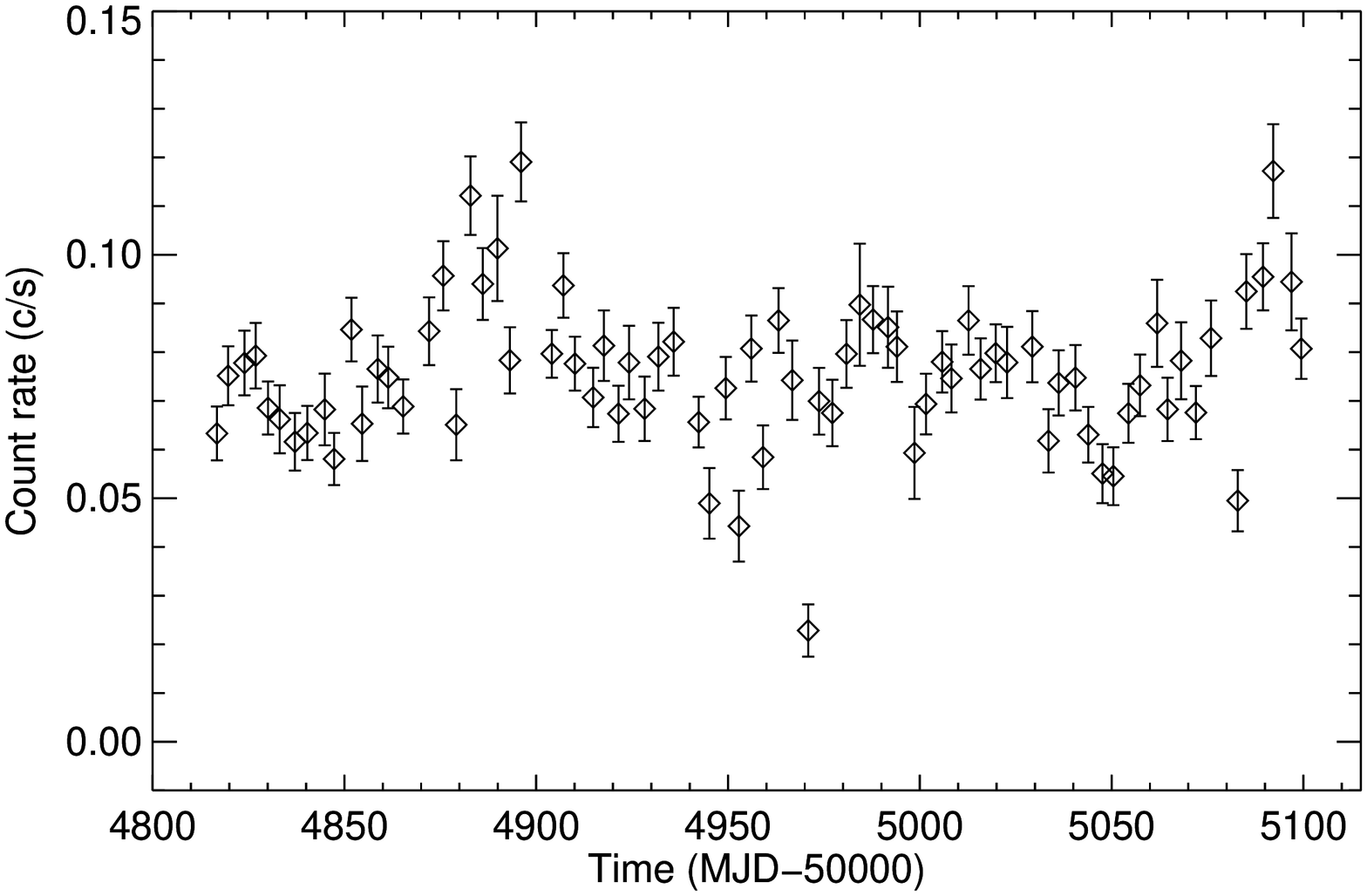}
   &\includegraphics[width=5cm,angle=270]{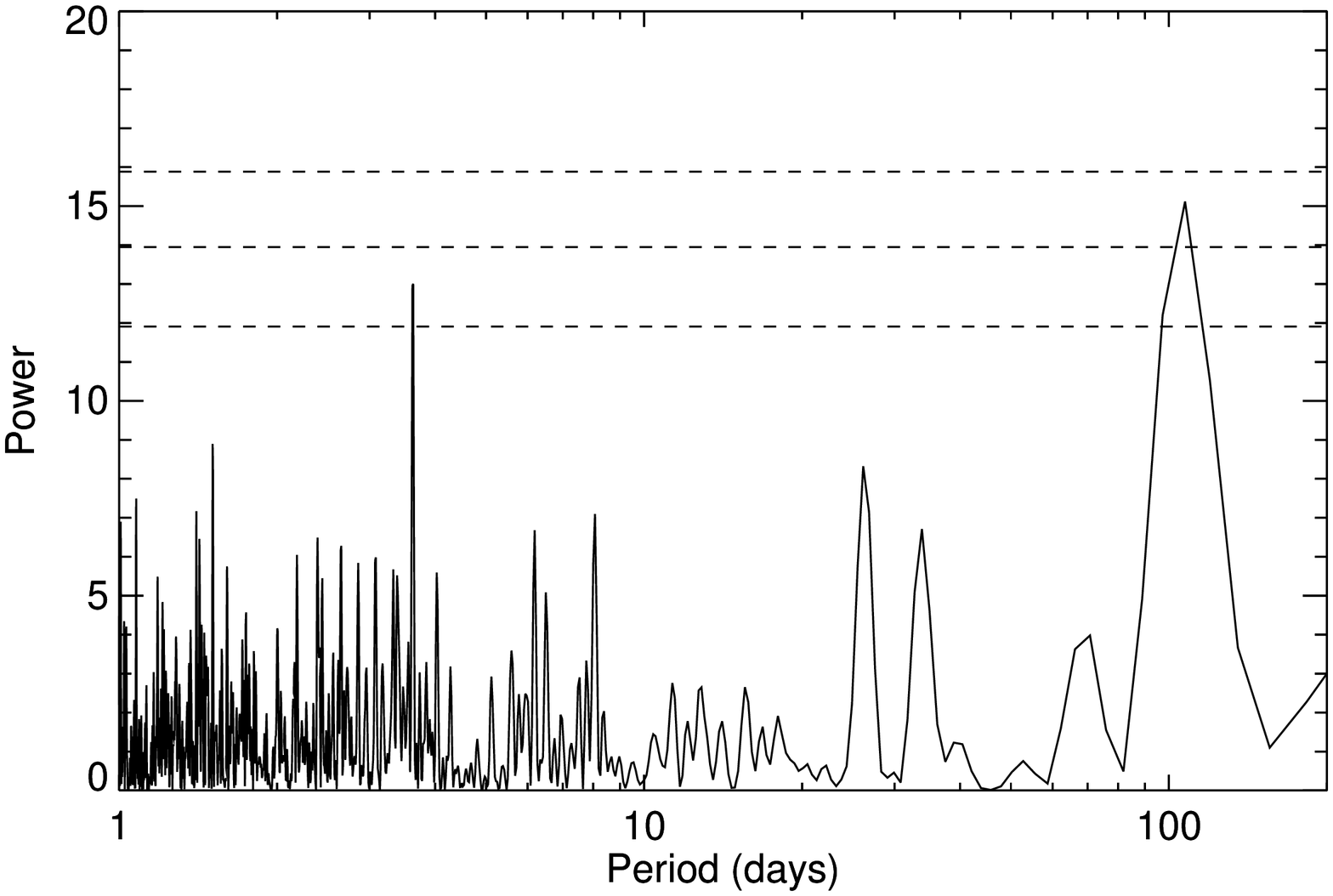}\\
    \includegraphics[width=5cm,angle=270]{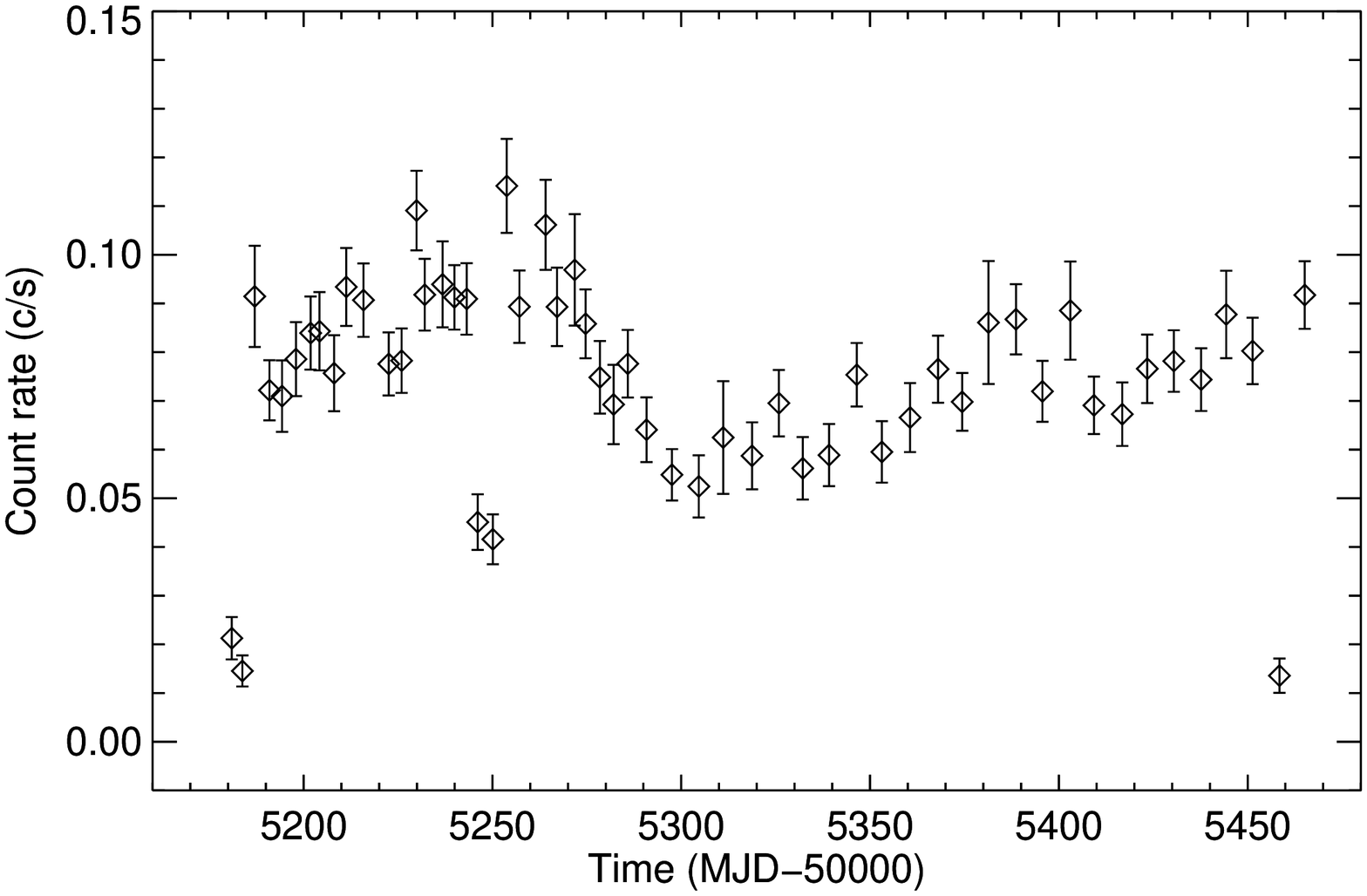}
   &\includegraphics[width=5cm,angle=270]{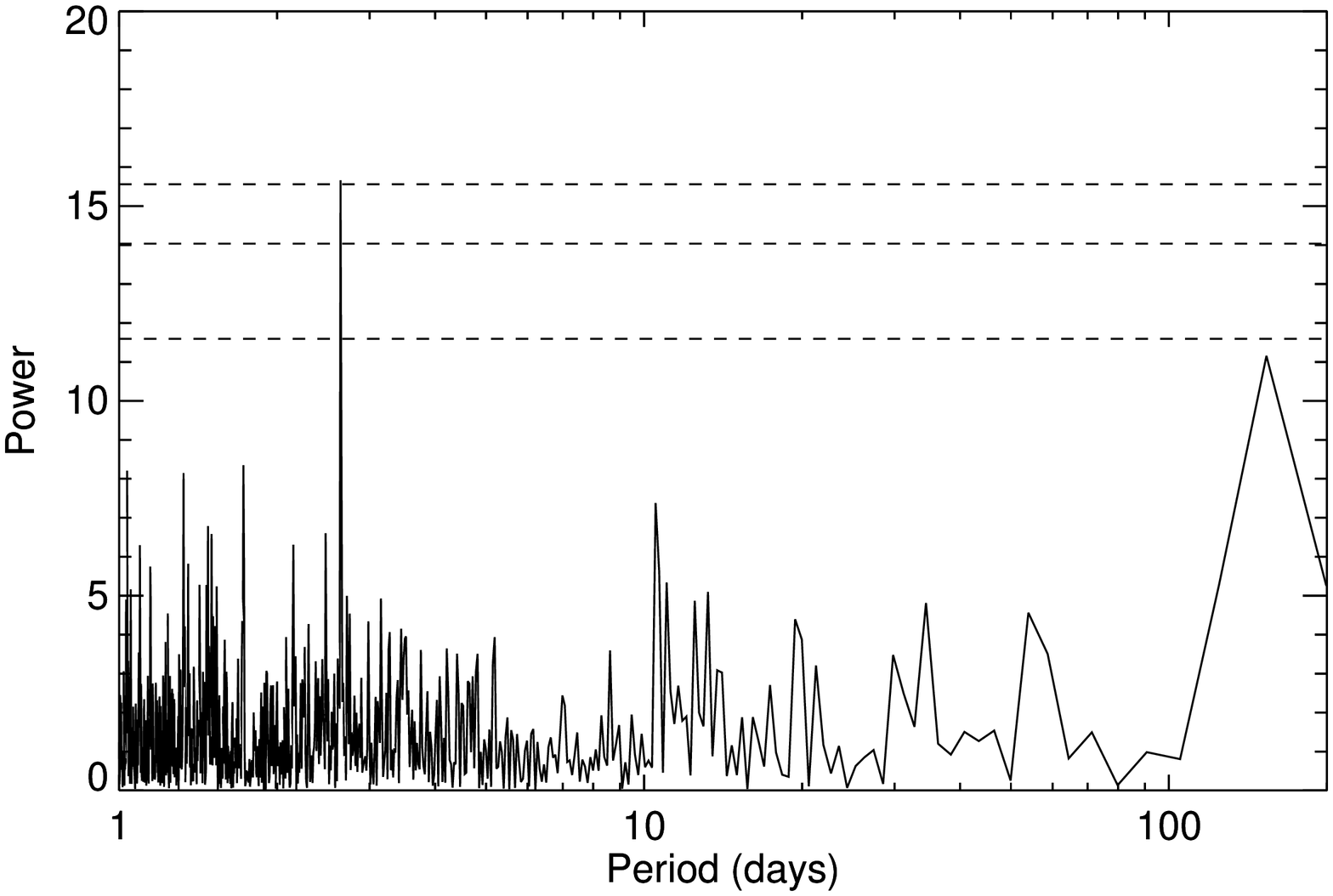}\\
    \end{tabular}
   \caption[]{Left: X-ray light curves in the 0.3--10 keV band for NGC~5408~X-1, split in contiguous windows (i.e. windows with no large gaps). Only the light curves rebinned with one observation per day are shown here, for clarity. Right: corresponding Lomb-Scargle periodograms calculated from the unbinned light curves. Dashed, horizontal lines denote the false alarm probability (fap) corresponding to confidence levels of 0.99, 0.999, and 0.999936 (or significance of 2.6, 3.3, and $4 \sigma$), based on white noise simulations. The two main peaks in the top periodogram are at $\sim 2.7$ and $\sim 136$~days.} 
   \label{lc_periodograms_contig}
\end{figure*}

\begin{figure*}
\centering
   \begin{tabular}{cc}
    \includegraphics[width=5cm,angle=270]{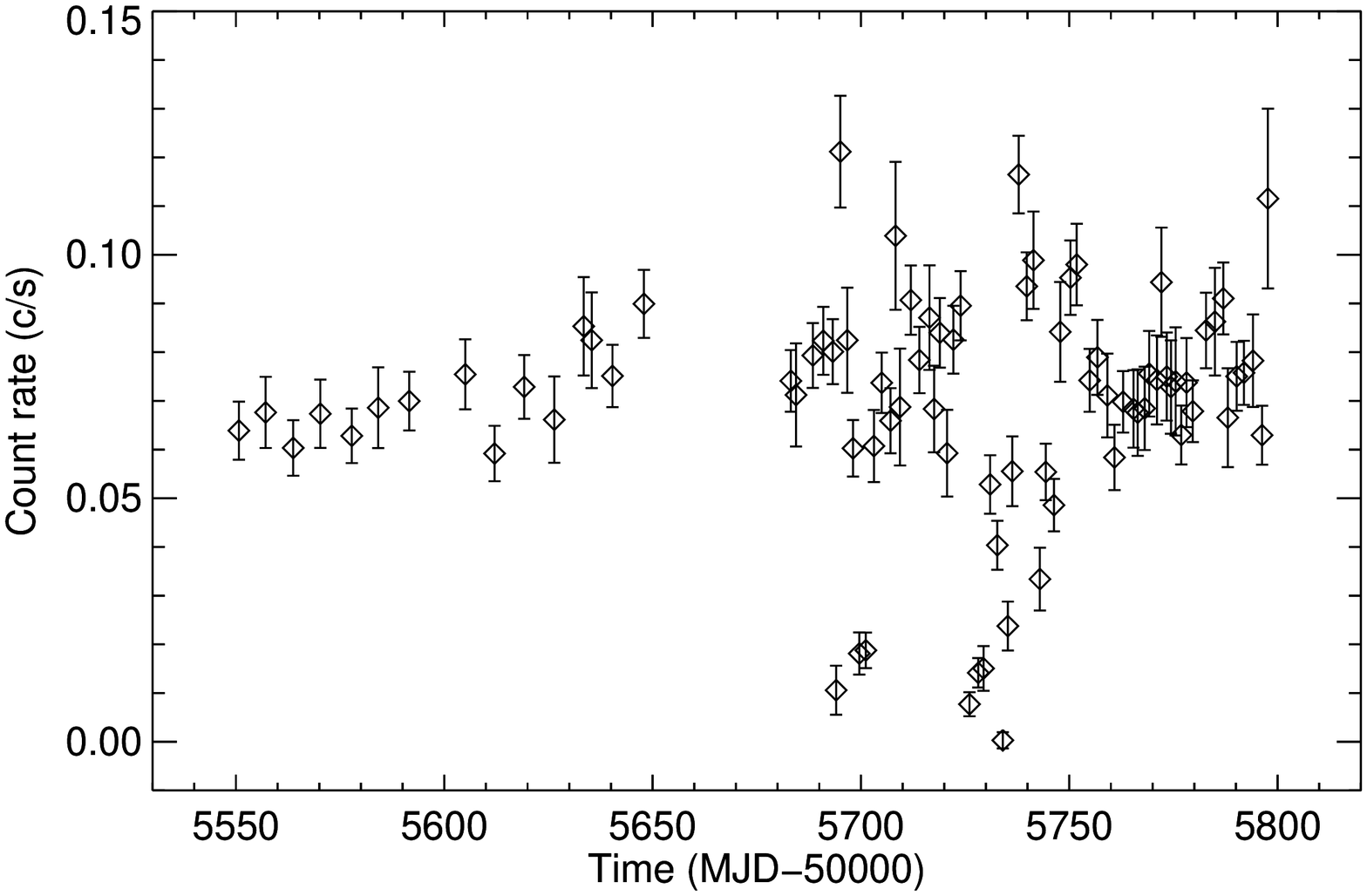}
   &\includegraphics[width=5cm,angle=270]{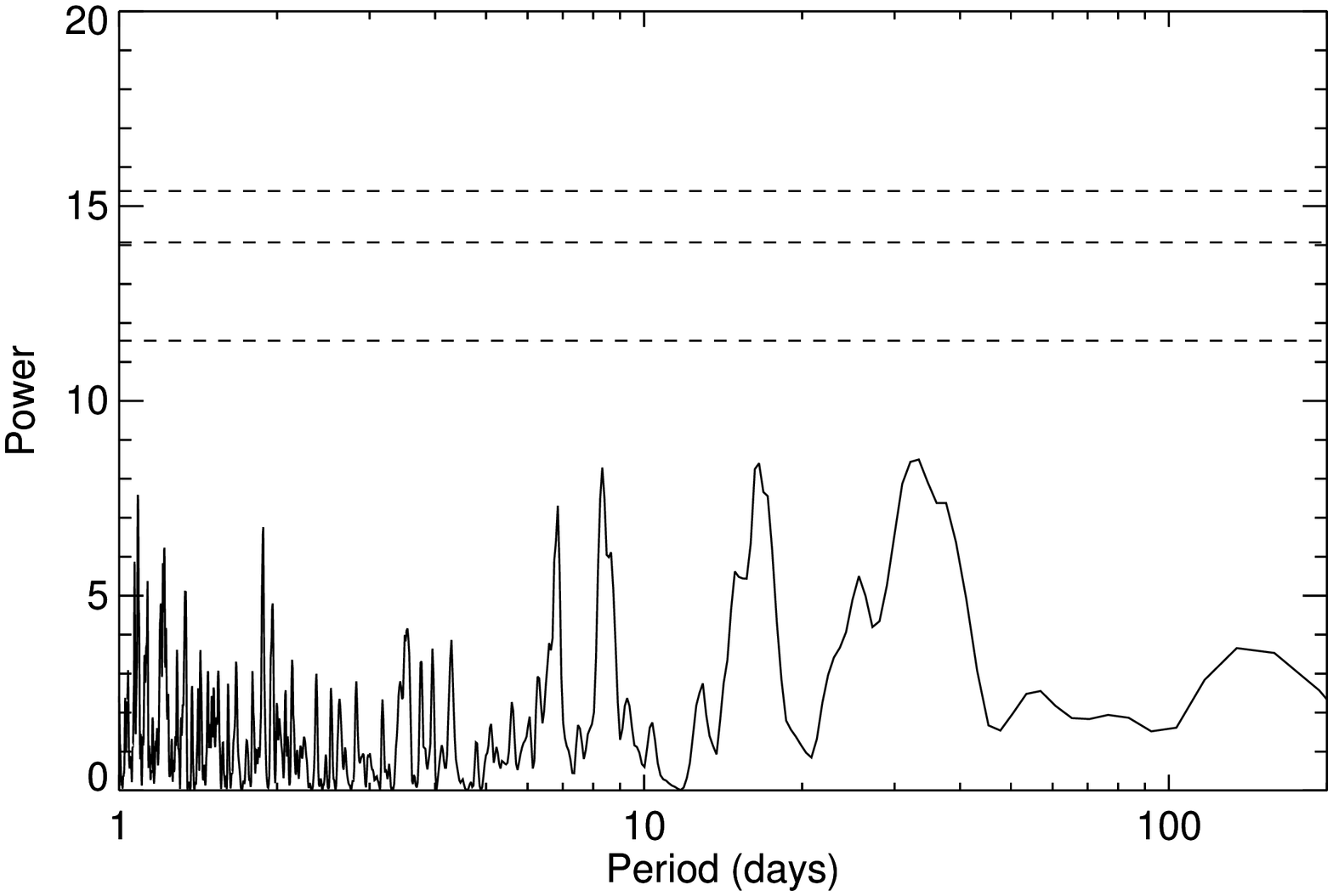}\\
    \includegraphics[width=5cm,angle=270]{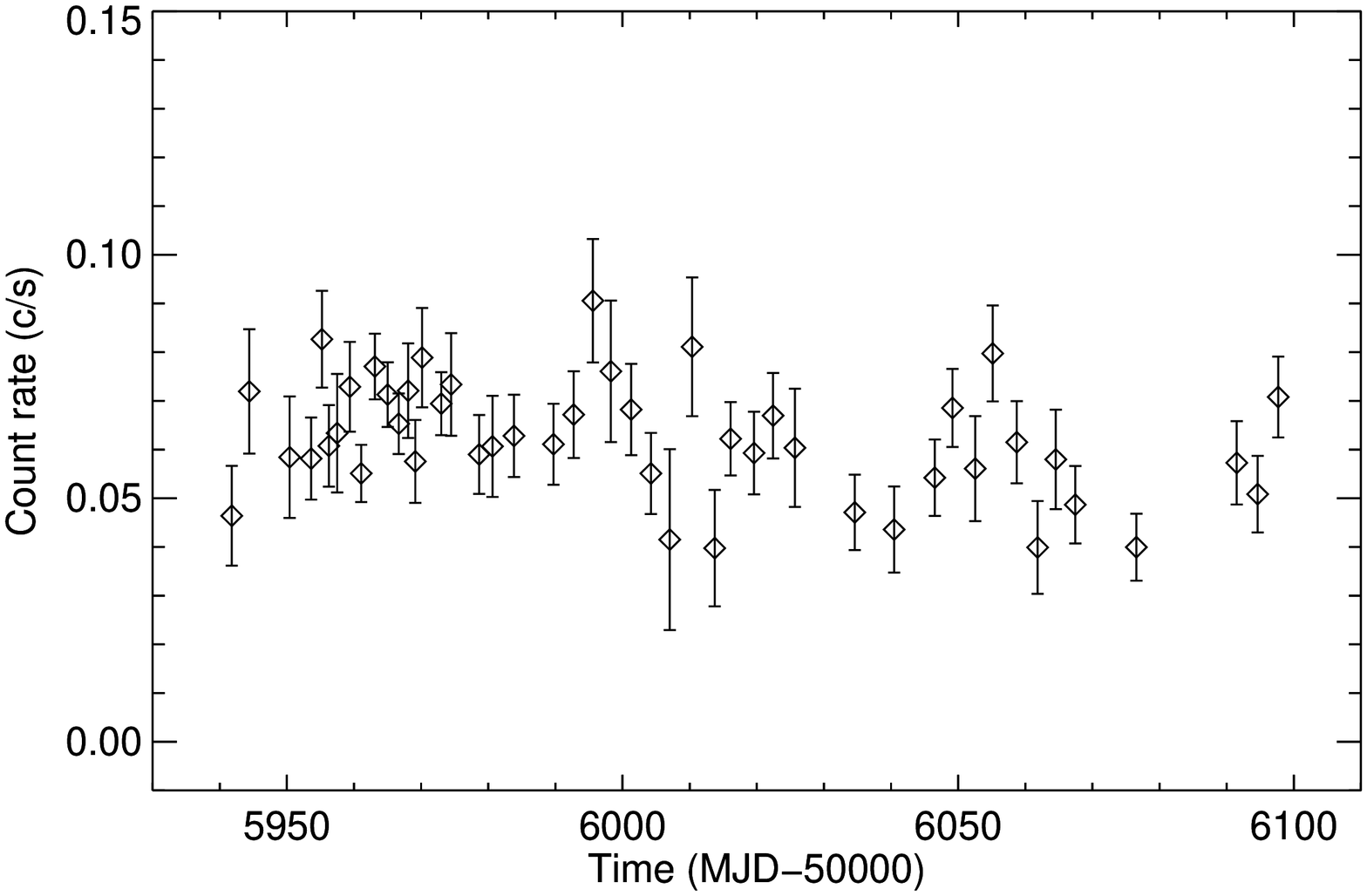}
   &\includegraphics[width=5cm,angle=270]{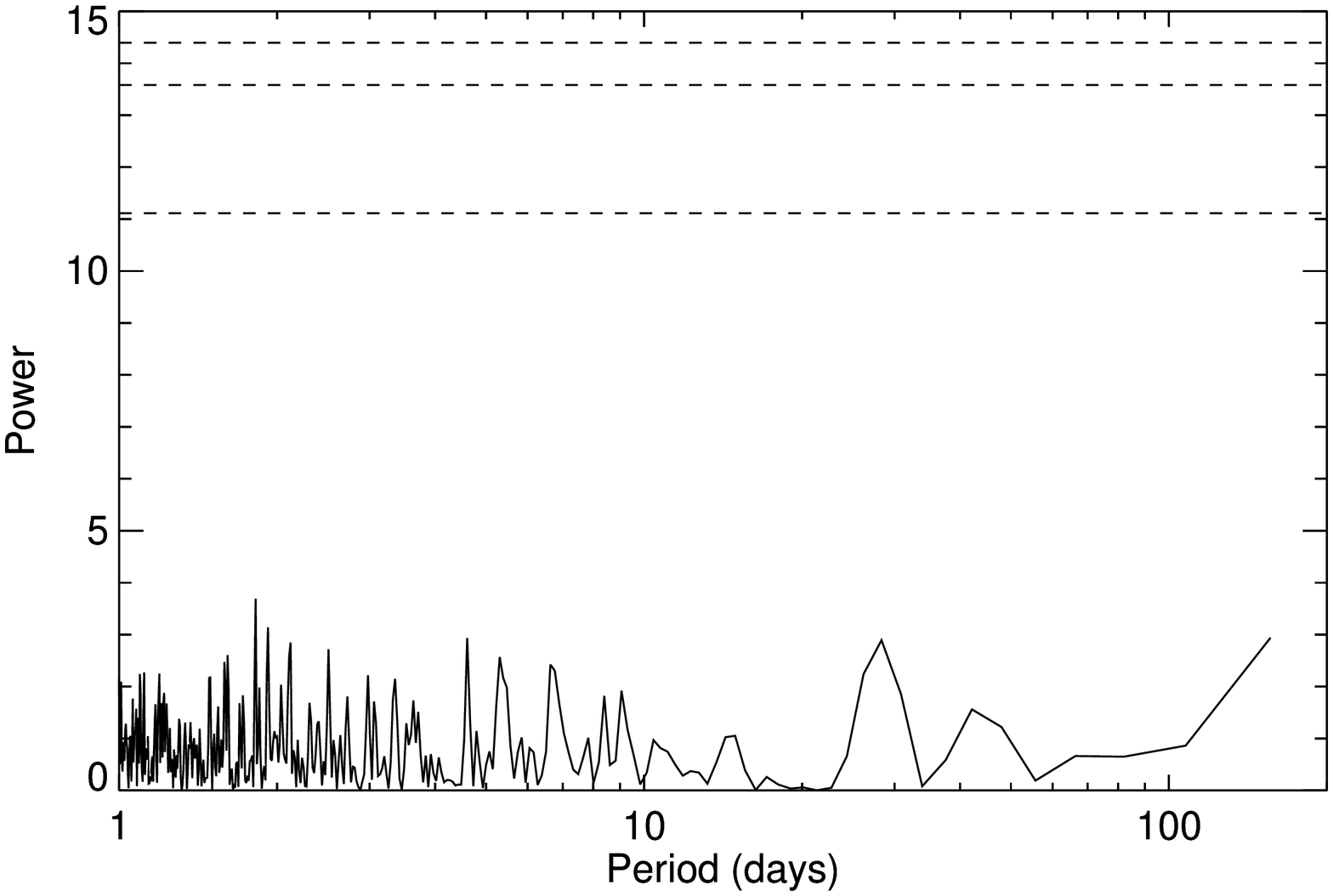}\\
    \end{tabular}
   \contcaption{} 
\end{figure*}

To understand better the behaviour of the periodicities in the long-term light curve, we calculated periodograms of different subsets of the light curve. There are evident gaps in the light curve (mostly due to observability constraints), but we decided to look at windows of 400 days similar to the original window in which the 115~day periodicity was discovered (S09). As can be seen on Figure~\ref{lc_periodograms_all}, the first 400~days indeed reveal only two clear peaks in the periodogram above the
$4 \sigma$ significance level of the white noise, with the highest one being consistent with the periodicity seen by S09. Subsequent periodograms show that the power in these peaks diminishes and becomes negligible after the second 400-day window. Instead, we see more power at shorter periods but in any case, there is no peak above the 0.99 significance level of the white noise for the three latest subsets. 
To understand the effect of the gaps on the power spectrum, we also recalculated periodograms using only intervals of contiguous observations (Figure~\ref{lc_periodograms_contig}). The result is roughly similar to the one above, albeit suffering from some of the windows being of short duration. We note that the 2.6~day periodicity behaviour is quite unstable, being the strongest peak in the first and third plots, albeit being insignificant in all other periodograms (Figure~\ref{lc_periodograms_contig}). We also note that the 3.6~day periodicity in the second periodogram may be related to the mean cadence of observations (see Table~\ref{tab_xrtdata}). However it is clear that there are no significant periodicities after MJD~$\sim 55500$.

In addition, and as described in more details in appendix~A, we simulated a fake data set containing a sinusoid with a period of 115.5~days and submitted this fake data set to the same treatment as discussed above. The 115~day period, if stable, should have been seen with high significance, at least up to $\sim \mathrm{MJD~55800}$. This confirms the Lomb-Scargle analysis of the observed data set that the temporal behaviour of N5408X1 is unlike that of a stable and persistent 115~day periodicity.

Finally, we also considered other algorithms to search for periodicities that are less biased towards sinusoidal variations, such as the phase dispersion minimisation (PDM) that was generalised by \citet{Stellingwerf78}. The strongest periods found using this method are the same as with the LS periodogram, with the addition of more aliases at multiple of the $\sim$~2.6, 112 and 190 day peaks seen in Figure~\ref{lc_periodogram_main}. However, and in agreement with the LS analysis, the PDM analysis does not indicate any significant and persistent periodicities in the data set.

\subsection{Dipping behaviour}

An interesting feature in the light curve already noted in \citet{Kong10} and \citet{Grise12} is the presence of several data points with a low count rate ($\la 0.03\ \mathrm{count \;s^{-1}}$; Figure~\ref{hid}). The average sampling of the light curve does not allow firm constraints on the duration and repeatability of such episodes, but it seems that there is a pattern of repeated low count rates. Using the midpoint of each of the five observed dipping intervals (Table~\ref{dips_details}), we find that this behaviour repeats on average every $250.5 \pm 27.8$~days (Figure~\ref{lc_periodogram_main}). Of specific interest, there are at least 2 episodes that last 20--50~days (see Figure~\ref{lc_dips}). Most notably, our interval with daily monitoring (May-August 2011 ; MJDs 55682--55797) reveals (Figure~\ref{lc_dips}, bottom panel) that there is a first interval of $\sim 7$~days in May (MJDs 55694--55701) where the count rate is highly variable on timescales of days and even hours, varying between count rates of $0.10\ \mathrm{count \; s^{-1}}$ and count rates of a few times $\sim 0.001\ \mathrm{count \; s^{-1}}$ (i.e, a few counts in a usual $\sim 1000$~s exposure). This is followed 24 days later (in June 2011, MJDs 55725--55735) by another $\sim 10$~day interval of highly variable count rate. In this second episode however, the count rate does not rise above $0.03\ \mathrm{count \; s^{-1}}$ for four consecutive days. It then varies between extreme count rates ($\sim 0.001$--$0.11\ \mathrm{count \; s^{-1}}$) for another few days before increasing up to average count rates. There is another, shorter interval of low count rates 9 days later that only lasts one day and a half (around MJD 55745). However, this daily monitoring continued until the end of August 2011 (MJD 55797) and did not reveal additional epochs of such variable activity (Figure~\ref{lc_dips}). The upper limit on the lowest count rate seen means that {\it the source was observed with a count rate that is clearly a factor of 35 or more below the average count rate of the source} ($\sim 0.07\ \mathrm{count \; s^{-1}}$).

Given the apparent average recurrence time of $\sim 250$~days of the dipping behaviour, we proposed and were awarded another series of observations from January 15 to June 19, 2012 (MJDs 55941--56097, see Figure~\ref{lc_periodogram_main}) as we expected a new series of dips. However, and as can be seen in Figure~\ref{lc_periodogram_main}, no observations with a count rate below $\sim 0.04\ \mathrm{count \; s^{-1}}$ were seen in all 2012 observations. This probably means that this is not a (strictly) periodic event, and/or that it can display variable lengths and amplitudes between episodes. We note that the third dipping interval (Figure~\ref{lc_periodogram_main}) also fell short of the expected recurrence time, being in advance by more than 10 days, although the gap between MJDs 55099 and 55180 does not allow us to comment on the duration of this dipping interval. We also note two possibly related dips, at MJD 55246 and 55250 that appear at $\sim 0.04\ \mathrm{c/s}$ in Figure~\ref{lc_periodogram_main} due to the binning, but where two consecutive snapshots show the count rate going from 0.01--0.02 to 0.06-0.10 c/s. However, the sampling was nearly as good during this interval (MJDs 55180--55250) as in the first dipping interval, so those may be unrelated to the apparently recurrent intervals of dips.

\begin{table*}
\centering
\caption[]{Observed X-ray dips in NGC~5408~X-1}
\label{dips_details}
\begin{threeparttable}[t]
\centering
\small
\begin{tabular}{lllclll}
\hline
\hline
Year & \# of 			  & \# of dips & \# of deep dips & Dipping interval & Midpoint & $\Delta T$ between \\
     & Snapshots (Obs.)\tnote{a}        &       ($< 0.03\ \mathrm{c/s}$)   &	($< 0.01\ \mathrm{c/s}$) & duration (days)        & (MJD)   &	midpoints (days)	\\
\hline
2008 &  151 (55) 	          & 12 (7)\tnote{b}	   & 2	& 21.5		 & 54717.9 $\pm$ 15.2  &	--	 \\
2009 &  205 (77) 	          & 7 (5)\tnote{b}	   & 1	& --		 & 54971.0	       & 253.1		\\
2010 &  126 (52) 	          & 6 (4)\tnote{b}	   & 2	& 2.9 (69.2?)		 & 55182.4 $\pm$ 2.0 (55215.5 $\pm$ 48.9?)   & 211.4 (244.5?)	\\
2010 &  	 	          & 		   & 	& --		 & 55458.4 	       & 276.0 (242.9?)		\\
2011 &  172 (117)	          & 24 (16)\tnote{b}	   & 8	& 51.9		 & 55720.0 $\pm$ 36.7  & 239.9		\\
2012 &  69  (53) 	          & 0	& 0	   & --			 & --	 	       &  $<221.8$ or none \\
\hline
\hline
Total &  723  (354) 	          & 49 (32)\tnote{b}	   & 13	   & --			 & --	 	       &  --			\\
\hline
\end{tabular}
\begin{tablenotes}
\footnotesize
\item[a] A {\it Swift} observation usually consists of several snapshots observed within several hours.
\item[b] In brackets, number of dips belonging to different observations (as opposed to different snapshots).
\end{tablenotes}
\normalsize
\end{threeparttable}
\end{table*}

\subsection{Spectral behaviour}

\begin{figure*}
\centering
   \begin{tabular}{cc}
   \includegraphics[width=6cm,angle=270]{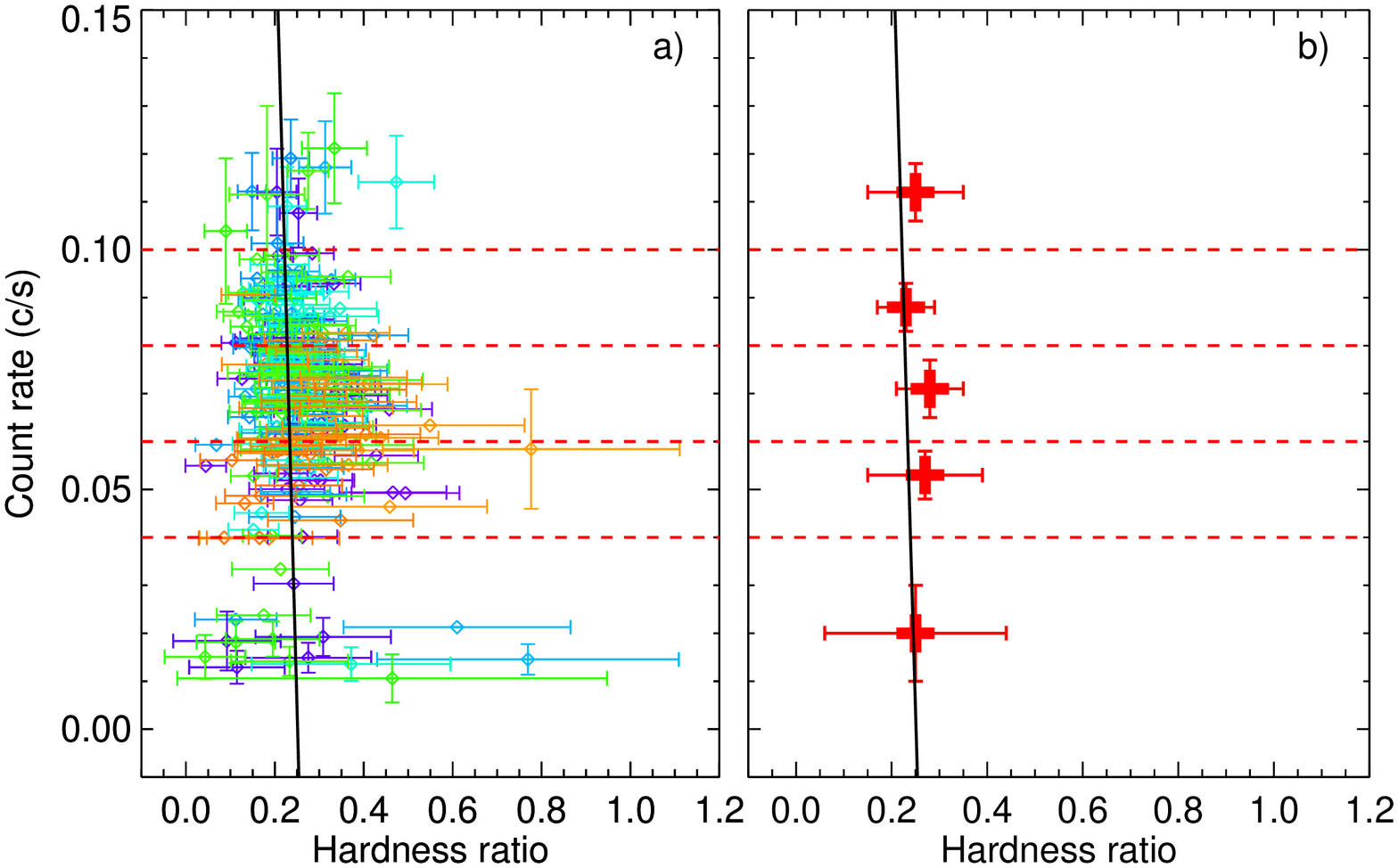} & \includegraphics[width=6cm,angle=270]{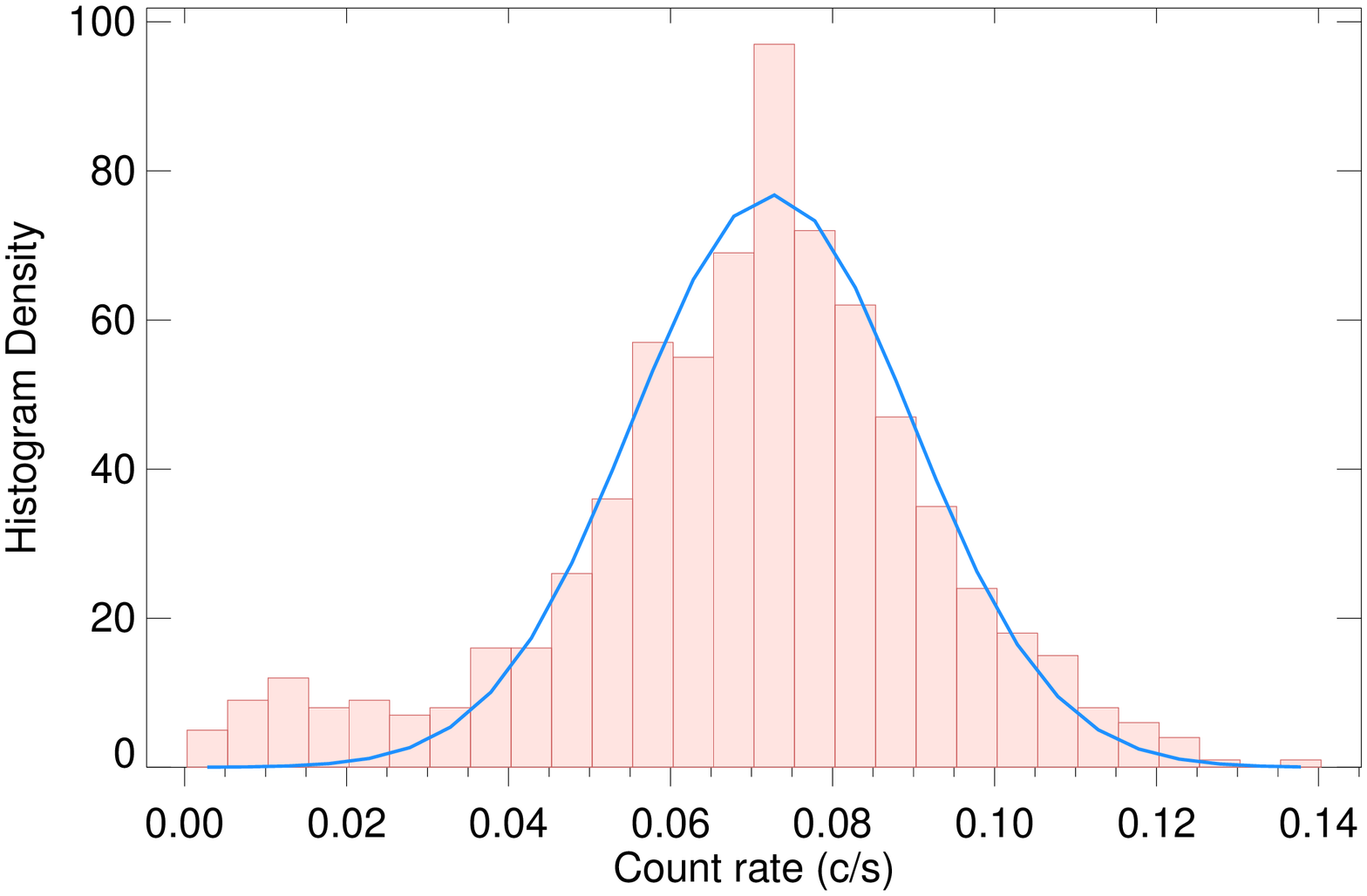} \\
   \end{tabular}
   \caption[]{Left: Hardness/intensity diagram for NGC~5408~X-1. The hardness ratio is defined as the ratio between the net count rate in the 1.5--10 keV band versus the 0.3--1.5 keV band. a) We used the light curves binned per day to calculate the corresponding hardness ratios. The color scheme represents the civil year the data were taken, as in Figure~\ref{lc_periodogram_main}. The red dashed horizontal lines show the count rate ranges used to co-add the spectra, and b) the red thick crosses represent the mean HR in each interval. The black solid line represents the best fit to all data points that is very close to a constant hardness. Note that in a) error bars on the count rates are only shown for a few data points for clarity, but were used in the fitting process. Right: Count rate histogram (from the unbinned data set). There is a net excess of data points at count rates $<0.03$~c/s - the X-ray dips - that form a separate population from the rest of the histogram, as can be already seen in the HID.} 
   \label{hid}
\end{figure*}

\begin{figure*}
\centering
   \begin{tabular}{cc}
   \includegraphics[width=8cm,angle=270]{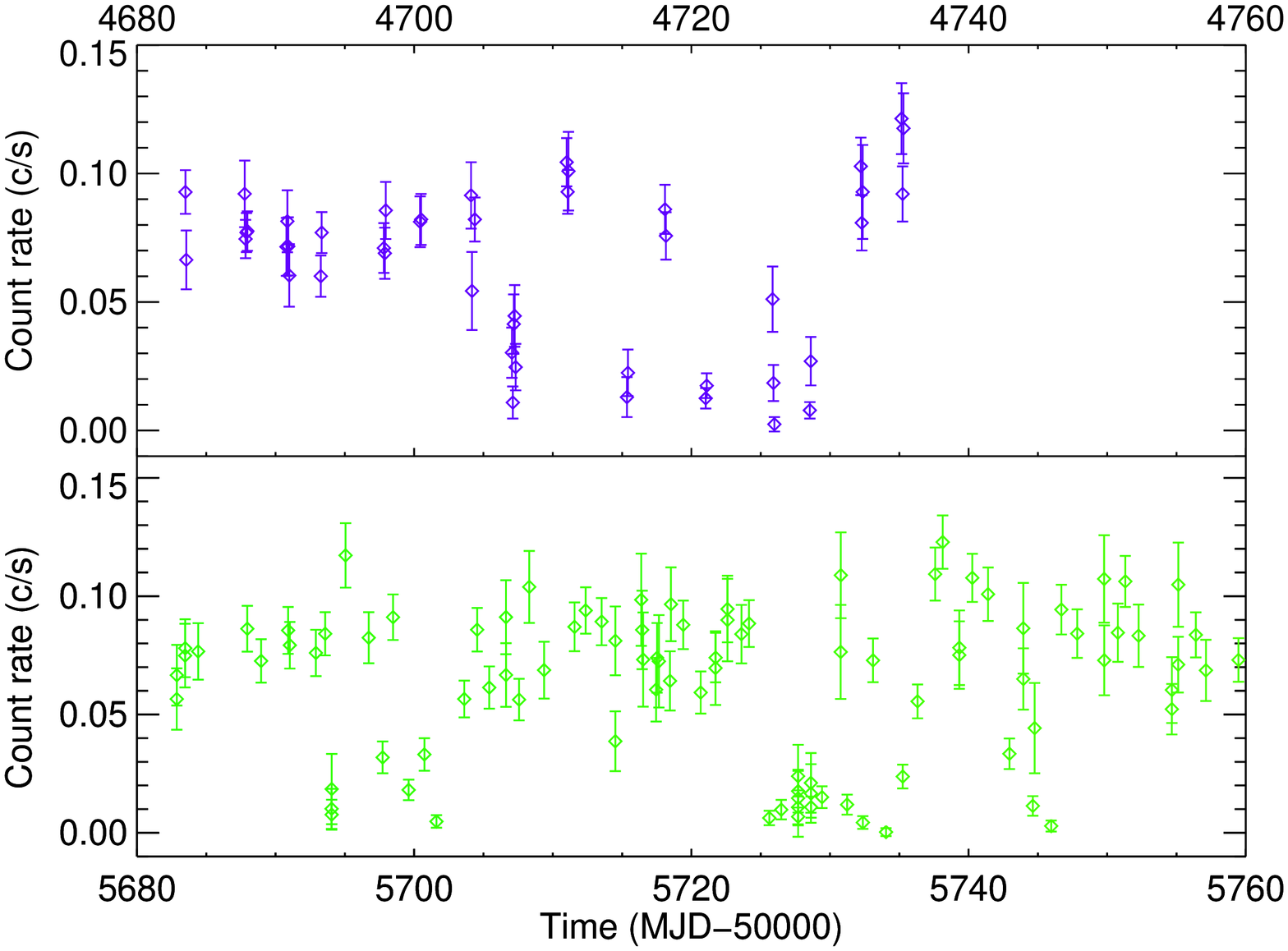} & \includegraphics[width=8cm,angle=270]{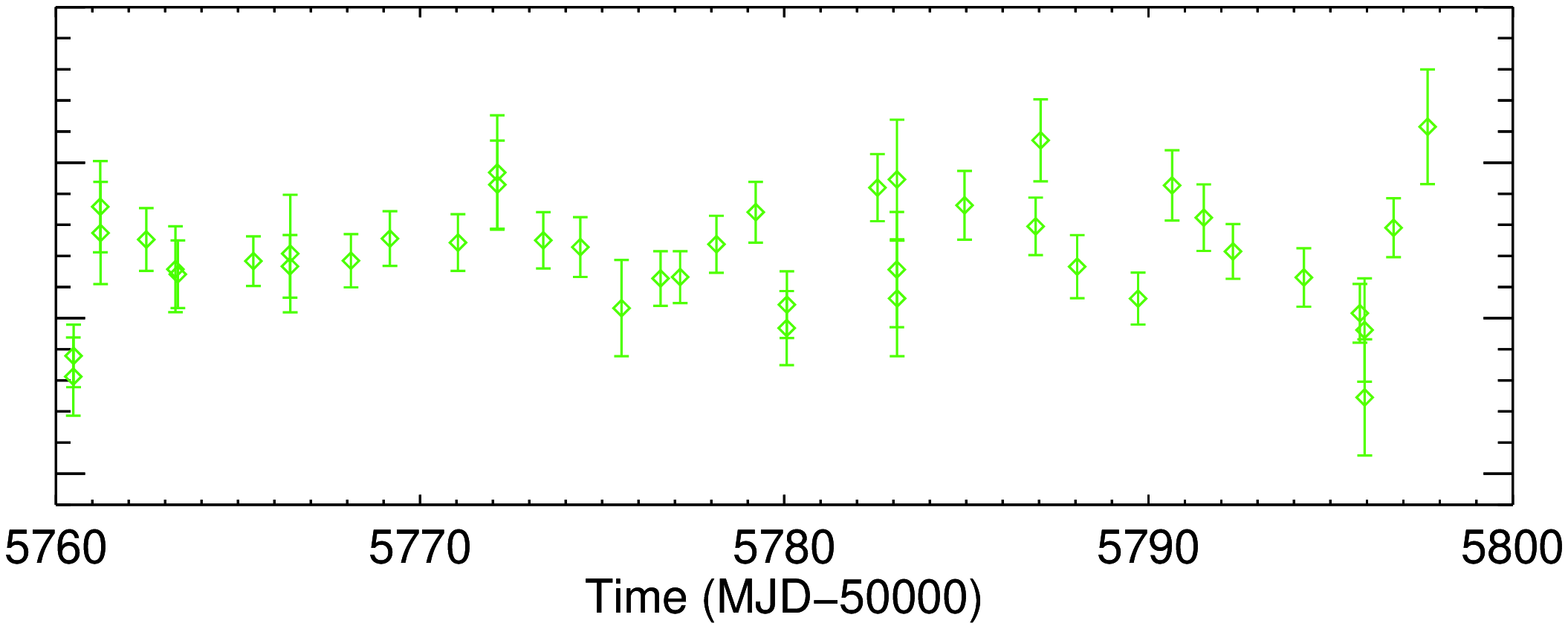} \\
   \end{tabular}
   \caption[]{X-ray light curves in the 0.3--10 keV band for NGC~5408~X-1, showing two main intervals where dips have been observed (in 2008 and 2011, resp. top and bottom left panels). Of specific interest is the lower left panel that shows the daily monitoring from May and June 2011, and which revealed two (plus one very short) interval of dips. On the right panel, it can be seen that the 2011 monitoring continued until the end of August without revealing new intervals of dips. The light curves are not rebinned, and show all measurements coming from snapshots longer than 100~s, hence the presence of multiple data points in one day.} 
   \label{lc_dips}
\end{figure*}

The variation of HR vs. time is shown in Figure~\ref{lc_periodogram_main}. The mean HR is $0.26 \pm 0.09$, and it is quite obvious that no significant spectral changes are occurring on timescales of months and years in N5408X1, as already seen in \citet{Kaaret09}.
The hardness/intensity diagram (HID) is presented in Figure~\ref{hid} (left panel). The mean count rate is $0.07 \pm 0.02\ \mathrm{count \; s^{-1}}$ and it can be seen that the X-ray dips clearly form a separate population from the persistent level (Figure~\ref{hid}).
The mean HR in each count rate strip (Figure~\ref{hid}, left panel) is consistent with a constant HR (covering HRs of 0.23--0.28) within the errors.
We performed a $\chi^2$ test, testing for the null hypothesis of constant hardness. This gives a $\chi^2 \sim 518$ for 305 degrees of freedom. If we let the HR varying as a free parameter, the best fit\footnote{We used the MPFIT package for IDL to estimate the best fit (\citealt{Markwardt09,More78}, and see http://purl.com/net/mpfit)} (shown on Figure~\ref{hid}) gives a $\chi^2 \sim 422$ for 304 degrees of freedom with a slope of $-3.4 \pm 2.4$. However, the best fit is very similar to a constant HR, ranging from 0.27 to 0.31 at respectively count rates of 0.14 and 0.01 c/s. To be more qualitative, we simulated X-ray spectra (based on an absorbed power-law plus multicolor disc blackbody, see below and Table~\ref{tab_fp}) and calculated the expected HR for different power-law indices (fixing the disc blackbody component to average values). The result is that the mean HR ($0.26 \pm 0.09$) corresponds to power-law indices $\sim 2.3$--$3.0$, therefore still consistent with a soft spectrum. Overall, there is no indication of significant spectral variations in N5408X1.

Given the large number of observations, we decided to explore possible subtle spectral changes by co-adding spectra based on count rate ranges. This is somewhat arbitrary but given the lack of significant spectral variation based on the HR, we should not introduce any significant bias by doing so. 
Five spectra were then created, with count rates ranges indicated in Table~\ref{tab_fp} and shown also on the HID (Figure~\ref{hid}). Two of the resulting spectra are of good quality (number of counts higher than 10000), two are of moderate quality ($\sim 4000$~counts) and the last one which spans the lowest count rates only has 698 counts (see Table~\ref{tab_fp} for details).
All five spectra were first fitted using empirical models usually used to describe ULXs spectra, such as an absorbed power-law and a power-law plus a multicolor disc blackbody (DISKBB,
\citealt{Mitsuda84}) combination to compare with previous results in the literature. Extinction was modeled using the Tuebingen-Boulder ISM absorption model (TBABS, \citealt{Wilms00}). We used a fixed component related to the Galactic extinction with $n_H = 0.057 \times 10^{22}\ \mathrm{cm^{-2}}$ \citep{Dickey90}, and let a second extinction component varying freely. 
Results for all spectra (Table~\ref{tab_fp}) appear consistent with previous observations (e.g. \citealt{Kajava09,Grise12}) and can be modeled with a soft power-law ($\Gamma \sim 2.4$--$2.6$) plus a cool blackbody component ($kT_{\mathrm{in}} \sim 0.2\ \mathrm{keV}$). Replacing the power-law by a more physically motivated Comptonization component such as {\small COMPTT} \citep{Titarchuk94}, we recover previously published results (Table~\ref{tab_fp} ; and see \citealt{Gladstone09,Middleton11,Grise12}) which can be interpreted with the presence of a cool, $kT \sim 1\ \mathrm{keV}$, optically thick ($\tau \sim 10$) corona. However, the constancy of the spectral parameters within the error bars do not allow further investigation.

\section[]{Discussion}
\subsection[]{Orbital period?}
Based on a $\sim 485$~day temporal baseline, S09 suggested that a detected 115~day periodicity was the orbital period of the system. However, \citet{Foster10} challenged this conclusion primarily based on a comparison with the Galactic source SS~433, and concluded that the periodicity was more likely to be a super-orbital period reflecting possibly the precession of the inner-disc/jet. \citet{Cseh11} also suggested that the 115~day periodicity was not compatible with the mass function constraints obtained with optical spectroscopy.
In this study, with a baseline of $\sim 1532$~days (three times the one used by S09), the significance of the $\sim 115$~day periodicity has not increased (Figure~\ref{lc_periodogram_main}), despite no significant change in the overall X-ray behaviour. There is a dominant, $\sim 2.6$~day periodicity when considering the whole data set, as well as another competing periodicity at $\sim 187$~days, with about the same power as the 115~day periodicity. Using different subsets of the light curve, it is quite clear that the periodicity at $\sim 115$~days disappears after only a few cycles, and that more power goes into shorter periodicities of a few tens of days. {\it Overall, there is no dominant periodicity that could be interpreted as a stable, orbital periodicity.}

This type of behaviour is not unlike that seen in many X-ray binaries (e.g. \citealt{Kotze12}). Super-orbital periods of tens to hundreds of days that evolve with time are, if not ubiquitous, largely present in these sources and are still not completely understood. One of the most cited phenomenon for such features is the precession of a tilted accretion disc (e.g. \citealt{Katz73}) even though the physical mechanism triggering the instability in the disc is still subject to debate. For instance, it has been shown that illumination by the central source creates an instability radiation-driven warping \citep{Ogilvie01} that could explain the 35 day super-orbital variability of the archetypal Her~X-1 (as first suggested by \citealt{Petterson75}). However, other mechanisms such as a tidal-force-induced precession of the accretion disc seems to also be a viable explanation (\citealt{Inoue12} ; see \citealt{Caproni06}, \citealt{Kotze12} and references therein for other possibilities).
It is not entirely clear what causes the (in)-stability of these super-orbital periods. In the radiation-driven warping model \citep{Ogilvie01}, there is a rather clear dependency on the mass ratio and separation of the binary system. Long-term X-ray observations from a sample of LMXBs/HMXBs \citep{Kotze12} seem to generally agree with the predictions of \citet{Ogilvie01} with HMXBs being more likely to produce stable precessing warped disc than LMXBs due mainly to their accretion discs being larger and therefore more likely to become unstable. However, \citet{Kotze12} also noted that accretion rate variations may also produce unstable periodic signals related to the evolutionary time-scales of these variations.
In that regard, perturbations by a third-body could be a good driver of super-orbital X-ray variation by modifying the accretion rate in a quasi-periodic manner, as hypothesized by \citet{Chou01} and confirmed by \citet{Zdziarski07} in the accreting neutron star 1820-303. Because the accretion rate depends quite sensitively on the degree of Roche lobe overflow \citep{Warner95}, a minor perturbation can drive major X-ray variations.

Another key point is that super-orbital periodicities of Galactic X-ray binaries are rather long compared to their orbital periods (a factor of $>3$ and more usually $>10$, see Table 1 \& 2 from \citealt{Kotze12}). This suggests that the orbital period of N5408X1 is likely to be $<40$~days (if we take the $115$~day super-orbital periodicity), or even much shorter ($<1$~day) if the 2.6~day periodicity also represents one of these unstable periodicities.	

\subsection[]{Nature of the dips}
A clear feature in the light curve of N5408X1 is the presence of (rare) periods of low count rates (see Figure~\ref{lc_periodogram_main}) that display a quasi-periodic behaviour and appear every $\sim 250$~days on average for at least four cycles. This behaviour is reminiscent of dipping X-ray binary systems. In LMXBs seen at high inclinations, it is believed that X-rays from the central object are being obscured by structures located in the disc, most likely from the stream of matter coming from the companion star and impacting the accretion disc \citep{White82}. In that case, the dipping behaviour usually repeats with the orbital period of the system (see e.g. \citealt{DiazTrigo06}). In HMXBs, eclipses of the compact object are more common than dips because of the size of the companion star. However, when observed, dips are usually related to clumps from the companion star wind intercepting X-rays from the compact object on our line of sight. This is what seems to happen in Cyg X-1 (e.g. \citealt{Balucinska-Church00}), although it could also be due to the interaction of this wind with the edge of the accretion disc \citep{Poutanen08}.

\begin{figure*}
\centering
   \includegraphics[width=10cm,angle=270]{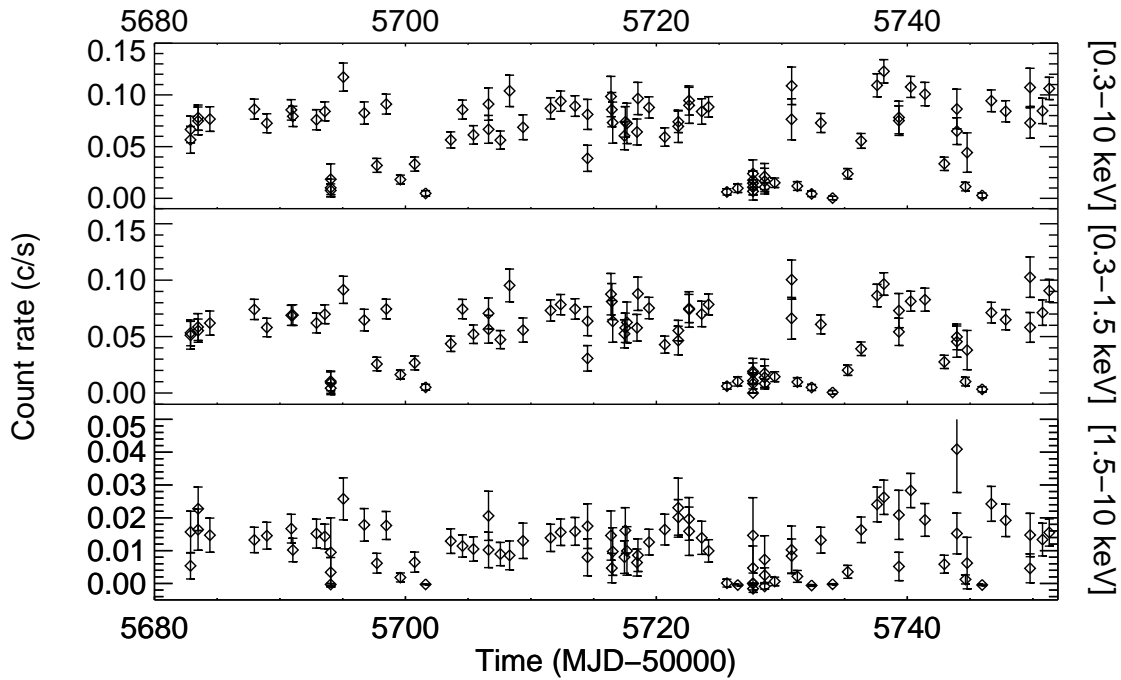}
   \caption[]{X-ray light curves in the 0.3--10 keV (top), 0.3--1.5 keV (middle) and 1.5--10 KeV (bottom) bands for NGC~5408~X-1, showing the main period of dips, as observed in 2011. The light curves are not rebinned, and show all measurements coming from snapshots longer than 100~s, hence the presence of multiple data points in one day.} 
   \label{lc_dips_energy}
\end{figure*}

The best coverage of these dipping episodes in N5408X1 comes from our daily monitoring that occurred in May and June 2011. Both episodes show fast variability, in the limit of the sampling of the {\it Swift} observations, with a minimum delay between low ($\la 0.01\ \mathrm{count \; s^{-1}}$) and `normal' ($> 0.04\ \mathrm{count \; s^{-1}}$) count rates of $\sim 11$~hours and several episodes of variability within a timescale of $\sim 17$--$36$~hours. We do not have strong constraints on the duration of individual dips, as count rates integrated over a snapshot may hide significant variability over timescales of minutes or even seconds (e.g. see \citealt{DiazTrigo09}, and most notably their Figure~2). However, there is some evidence that some dips (i.e., the deepest) last for at least 100~s to 15 minutes (i.e., an entire snapshot). A good example is the snapshot observed at MJD 55734, which shows that the upper limit on the integrated count rate over 960~s is $0.002\ \mathrm{count \; s^{-1}}$. This means that the intensity in the dip diminished by $\ga 97\%$ compared to the out-of-dip intensity. Also to be noted, the second dipping episode in 2011 shows a rather constant low count rate for 4 days (MJDs 55725--55729) in a row that could argue for a partial eclipse. However, we are lacking the continuous sampling needed to constrain this.

The high depth of some of the dips and their apparent fast variability are properties similar to the absorption dips seen in classical LMXB dippers (e.g. \citealt{Kuulkers13}). Therefore, it could be argued that the quite striking repeatability pattern of the dips in the long-term light curve (Figure~\ref{lc_periodogram_main}) is orbital in nature, and reflects the thickened outer region of the accretion disc passing in our line of sight every $\sim 250$~days.
If these dips are due to absorption, another diagnostic would be a change in hardness ratio since these dips should be energy-dependent (see e.g., \citealt{Boirin05} for a possible model). Unfortunately, the number of counts in the deep dips (count rate $\la 0.01\ \mathrm{count \; s^{-1}}$, or a few counts per integrated snapshot) does not allow a clear answer (Figure~\ref{hid}, left panel), mostly because the number of counts in the hard band ($1.5$--$10$~keV) is very close or equal to zero and does not allow to estimate the HR. Our averaged spectrum spanning count rates of $0.0$--$0.04\ \mathrm{count \; s^{-1}}$ (see section \ref{spectral_evol}) is also inconclusive in that regard. However, one can note that in the dips the soft and hard components seem to drop in similar ways (Figure~\ref{lc_dips_energy}) which would argue for energy-independent dips, although this needs to be confirmed with deeper observations.
The absence of dips during the latest observing campaign is also puzzling (see Figure~\ref{lc_periodogram_main}). We argue here that if the dips were a truly periodic event, we should have seen at least part of them in early 2012 because only $\sim 222$~days had passed since the midpoint of the last dipping periods (or 196/211/243 days if we consider separately all dipping (sub-)intervals seen in 2011). There is some apparent jitter in the first four dipping episodes ($\pm 28$~days) which is probably related to the time interval where dips can develop (the latest, detailed period where dips were observed indicates an interval of $\sim 52$~days). However, even when taking this into account, there are two expected dipping intervals that do not fit in this picture. It can be seen that the third epoch of dips was in advance of the mean recurrence time of the dipping behaviour, by $\ga 10$~days (Figure~\ref{lc_periodogram_main}). Perhaps the same behaviour happened in 2012, where it would mean that the dips were in advance of more than $\sim 3 $~days. We note that we can confirm the lack of dips in 2012 given our very good sampling (almost daily) for most of the expected dipping interval. Then, a more probable interpretation is that the dips display variable lengths and amplitudes between episodes, or even disappear at certain times.

The fact that dips do not repeat every ``cycle'' is not completely unusual in X-ray dippers. In the GBHB MAXIJ1659-152, absorption dips only appear during part of the outburst \citep{Kuulkers13}, as it has been seen in other GBHBs (4U~1630-47 and GRO~J1655-40, see e.g. \citealt{Kuulkers98, Kuulkers00}). However, in those cases, it seems likely that the transient behaviour of these sources may change the structure of their disc, which will likely play a role in the (dis)appearance of the dips. Here, on the contrary, N5408X1 has been seen to stay persistently (i.e. outside of the dips) ultraluminous on timescales of years without any significant change in its X-ray behaviour. Therefore, it is unclear what could cause the disappearance of an entire dipping interval in this source. Perhaps a warped, tilted and precessing disc may explain both the variability of the X-ray periodicities and the disappearance of the dips, as it was proposed in XB~1254-690 \citep{DiazTrigo09}.

Dips are quite rare among HMXBs. However, in the persistent HMXB Cyg~X-1, two different types of dips (type A and B, \citealt{Feng02}) have been discovered, being respectively energy-dependent and independent. If type A dips are probably related to the wind of the supergiant companion, type B dips might be due to partial covering of an extended X-ray emission by an opaque ``screen'' \citep{Feng02}. However, again, confirmation on the energy (in)dependence of the dips is needed for N5408X1 before being able to favour a single model.

In any case, it seems that long super-orbital periodicities as well as the presence of dips in the long-term X-ray light curve of N5408X1 is more reminiscent of high-inclination LMXBs rather than HMXBs (see e.g., \citealt{Foulkes10}). This is likely related to the similarity of the Roche lobe overflow accretion mechanism needed to power the black hole at ultraluminous luminosities. A deep X-ray observation during a dipping interval would likely be able to settle this issue.

We can also compare the long-term behaviour of N5408X1 studied in this paper, with the short-term behaviour studied in previous {\it XMM-Newton} observations.
It is, for instance, worth noting that \citet{Soria04} observed X-ray flares in light curves of N5408X1, based on {\it XMM-Newton} observations of a few ks. They found that the flaring, with changes in flux by a factor of $\sim 2$ over $\sim 100$~s, was due to the hard band, which can be seen to drop to a value consistent with zero at some times, while the soft band has always a persistent component \citep{Soria04}. The same behaviour was also seen in the study of \citet{Middleton11} using longer, $\sim 100$~ks {\it XMM-Newton} observations. They suggest that the strong (short-term) variability is due to obscuration of the Comptonized emission when viewed through the turbulent accretion disc wind, while the soft emission from the photosphere of the wind is not variable on short time-scales due to its large emission radius. Another ULX in M82, X37.8+54, also showed soft dips on timescales of several kiloseconds \citep{Jin10}.
In our case, it is quite clear that during the dips, the soft and hard components drop in similar ways (Figure~\ref{lc_dips_energy}). This probably argues for a different mechanism between the short-term and the long-term variability seen in the light curves. In the latter, there is the need for a structure large enough in our light of sight capable to block a significant fraction of the total X-ray flux coming from the system.

Finally, an interesting source to compare with is a bright X-ray source in NGC~55 \citep{Stobbart04}. Although in the lowest range of ultraluminous luminosities ($\sim 2 \times 10^{39}\ \mathrm{erg \; s^{-1}}$), it is possibly a good comparison to what we may be able to see in deeper X-ray observations for the source of our study. NGC~55 ULX displays $\sim 100$--$300$~s long dips. Thanks to the sensitivity of {\it XMM-Newton}, \citet{Stobbart04} were able to see that the relative depth of the dips increase with energy; in terms of the spectral modelling, the multicolor disc blackbody (the hard component, in their model) is more strongly obscured than the power-law component (the soft component, in their model). This leads to a few different scenarios that they were unable to constrain due to low statistics of the spectra in the dips. However, the duration of individual dips in N5408X1 in an interval such as observed in 2011 seem to be several times longer than in NGC~55 and therefore would allow to study the spectral behaviour of the source during the dips with more details.

To conclude on the dipping behaviour, a more detailed study of the dips, temporal as well as spectral, may help to understand the correct physical interpretation of the spectra of ULXs. In cool disc models (e.g.\ \citealt{Kaaret03}), the soft component must be more compact than the corona producing the hard component. On the other hand, the soft component of ULXs may be produced by strong outflows due to super-critical accretion (e.g. \citealt{Poutanen07, Middleton11}) in which case the opposite is expected since the hard component would be due to disc emission inside the spherization radius, i.e. would be on a much smaller scale. Depending on the nature and location of the structure blocking X-rays, we may be able to rule out one or the other model.

\subsection[]{Spectral evolution}
\label{spectral_evol}
As already suggested by the HID (Figure~\ref{hid}), the model-fitting of our five spectra at different count rates confirms the rather stable spectral parameters of N5408X1 vs. count rate and show that spectral parameters from these averaged spectra are completely consistent with spectra from pointed observations (e.g. see \citealt{Middleton11}). Obviously, one important question would be whether or not the X-ray spectrum of the source displays some distinct features (such as an increase in photoelectric absorption, or a change in disc temperature) at very low count rates, during the dips. Our fainter spectrum, spanning count rates between 0 and $0.04\ \mathrm{count \; s^{-1}}$ with 698 counts is not deep enough to answer to this question. In addition, the number of counts is clearly biased towards the highest count rates ($\sim 0.03$--$0.04\ \mathrm{count \; s^{-1}}$), so that no spectral information can be extracted from the very sharp dips ($\la 0.01\ \mathrm{count \; s^{-1}}$) where we would expect the largest spectral differences. The fit of this spectrum does not allow us to constrain the external column density (the $n_H$ value is pegged at 0 value). If we fix the intrinsic column density to N5408X1 to a value typical of the spectra at other count rates, spectral parameters are similar within the errors (Table~\ref{tab_fp}). Constraining the X-ray spectrum during the dips would be extremely useful to constrain the cause of the dips, but necessitate a larger X-ray telescope such as {\it XMM-Newton}. It would be interesting to study if the dips have really no dependence on energy, as it seems based on the light curves at different energies (Figure~\ref{lc_dips_energy}). 

As suggested by \citet{Grise12}, it would also be extremely interesting to get UV/optical observations during a period of X-ray dips, to test whether or not irradiation in the accretion disc contributes to the optical luminosity. It would also be a good test to study whether or not the companion star sees the same amount of X-rays from the compact object, which would give some constraints on the size of the structure blocking X-rays and constraints on the geometry of the system.

\begin{table*}
\centering
\caption[]{Spectral Fit Parameters}
\begin{threeparttable}[t]
\centering
\tiny
\begin{tabular}{l*{11}{c}r}
\hline
No.\tnote{a}  & CR range & $n_\mathrm{H}$\tnote{b}  & $\Gamma$\tnote{c} & $\Gamma_{\mathrm{norm}}$\tnote{d} & $T_{\mathrm{in}}$\tnote{e} & Disc$_{\mathrm{norm}}$\tnote{f} & & Flux\tnote{g} & $L_X$\tnote{h} & $f_{X_{\mathrm{MCD}}}$\tnote{i} & $\chi^2$/DoF\tnote{j}\\
     	& $\mathrm{count \; s^{-1}}$	& ($10^{22}\ \mathrm{cm^{-2}}$) &   & $10^{-4}$  &   &  &  & ($\times 10^{-12}$ & ($\times 10^{40}\ \mathrm{erg\ s^{-1}}$) & &\\
& & & & & & & & $\mathrm{erg\ s^{-1}\ cm^{-2}}$) & & & \\
\hline
\multicolumn{11}{c}{Absorbed power-law + multicolor disc blackbody ({\texttt tbabs*tbabs*(po+diskbb)})} \\
\hline
1	 & 0.00--0.04 	&0.0$^{+0.05}_{-0.00}$	     & 2.01$^{+0.48}_{-0.55}$ 	& 0.6$^{+0.4}_{-0.3}$     	  & 0.22 $^{+0.03}_{-0.04}$ & 10$^{+22}_{-6}$ & & 0.5$^{+0.0}_{-0.3}$ & 0.18$^{+0.01}_{-0.01}$  & 0.47$^{+0.11}_{-0.11}$  & 34.2 (26) \\
1\tnote{0}	 & 0.00--0.04 	&0.05	     		     & 2.17$^{+0.50}_{-0.50}$ 	& 0.8$^{+0.4}_{-0.4}$     	  & 0.18 $^{+0.03}_{-0.03}$ & 29$^{+34}_{-16}$ & & 0.5$^{+0.05}_{-0.1}$ & 0.22$^{+0.02}_{-0.02}$  & 0.49$^{+0.11}_{-0.11}$  & 37.1 (27) \\
2	 & 0.04--0.06 	&0.07$^{+0.04}_{-0.03}$	     & 2.46$^{+0.19}_{-0.19}$ 	& 3.0$^{+0.6}_{-0.6}$ 	  & 0.18 $^{+0.03}_{-0.02}$ & 74$^{+114}_{-41}$ &  &   1.39$^{+0.04}_{-0.37}$ & 0.67$^{+0.02}_{-0.02}$  & 0.40$^{+0.05}_{-0.05}$  & 115.1 (130) \\
3	 & 0.06--0.08 	&0.06$^{+0.02}_{-0.02}$	     & 2.55$^{+0.10}_{-0.10}$ 	& 4.3$^{+0.4}_{-0.4}$	  & 0.18 $^{+0.02}_{-0.01}$ & 82$^{+55}_{-31}$  &  &  1.86$^{+0.02}_{-0.18}$ & 0.87$^{+0.01}_{-0.01}$  & 0.34$^{+0.03}_{-0.03}$  & 309.6 (223) \\
4	 & 0.08--0.10 	&0.07$^{+0.02}_{-0.02}$	     & 2.61$^{+0.11}_{-0.12}$ 	& 5.4$^{+0.6}_{-0.6}$	  & 0.17 $^{+0.02}_{-0.01}$ & 185$^{+152}_{-79}$ &  &  2.29$^{+0.04}_{-0.27}$ & 1.17$^{+0.02}_{-0.02}$  & 0.39$^{+0.03}_{-0.03}$  & 192.3 (193) \\
5	 & 0.10-0.14 	&0.05$^{+0.03}_{-0.03}$	     & 2.58$^{+0.19}_{-0.19}$ 	& 5.9$^{+1.3}_{-1.2}$	  & 0.20 $^{+0.03}_{-0.03}$ & 73$^{+82}_{-36}$   & &   2.74$^{+0.06}_{-0.45}$ & 1.22$^{+0.03}_{-0.03}$  & 0.36$^{+0.05}_{-0.05}$  & 115.1 (133) \\
\hline
\hline
No.\tnote{a}  & CR range & $n_\mathrm{H}$\tnote{b}  & kT$_e$\tnote{k} & Compt$_{\mathrm{norm}}$\tnote{l} & $T_{\mathrm{in}}$\tnote{e} & Disc$_{\mathrm{norm}}$\tnote{f} & $\tau$\tnote{m} & Flux\tnote{g} & $L_X$\tnote{h} & $f_{X_{\mathrm{MCD}}}$\tnote{i} & $\chi^2$/DoF\tnote{j}\\
     	& $\mathrm{count \; s^{-1}}$	& ($10^{22}\ \mathrm{cm^{-2}}$) & (keV)  & $10^{-4}$  &   &  &   & ($\times 10^{-12}$ & ($\times 10^{40}\ \mathrm{erg\ s^{-1}}$) & &\\
& & & & & & & & $\mathrm{erg\ s^{-1}\ cm^{-2}}$) & & & \\
\hline
\multicolumn{11}{c}{Absorbed Comptonization component + multicolor disc blackbody ({\texttt tbabs*tbabs*(comptt+diskbb)})} \\
\hline
1	& 0.00--0.04 & \multicolumn{10}{c}{no possible constraints} \\
2	& 0.04--0.06 & 0.04$^{+0.04}_{-0.03}$	     & 1.04$^{+0.69}_{-0.25}$ 	& 3.8$^{+1.6}_{-1.6}$       & 0.20 $^{+0.04}_{-0.03}$ & 64$^{+89}_{-36}$ & 11.0$^{+15.0}_{-4.7}$ & 1.34$^{+0.01}_{-0.80}$ &0.55$^{+0.02}_{-0.02}$ & 0.64$^{+0.04}_{-0.04}$  & 107.3 (129) \\
3	& 0.06--0.08 & 0.02$^{+0.02}_{-0.02}$	     & 1.07$^{+0.31}_{-0.18}$ 	& 5.8$^{+1.2}_{-1.1}$       & 0.19 $^{+0.02}_{-0.02}$ & 90$^{+53}_{-33}$ & 9.9$^{+3.6}_{-2.4}$  & 1.81$^{+0.01}_{-0.34}$ & 0.71$^{+0.01}_{-0.01}$ & 0.60$^{+0.02}_{-0.02}$  & 282.6 (222) \\
4	& 0.08--0.10 & 0.04$^{+0.02}_{-0.02}$	     & 1.29$^{+1.71}_{-0.39}$ 	& 8.1$^{+2.2}_{-3.9}$ 	    & 0.17 $^{+0.02}_{-0.02}$ & 230$^{+171}_{-95}$ & 7.5$^{+3.5}_{-3.6}$  & 2.24$^{+0.01}_{-0.59}$     & 0.98$^{+0.02}_{-0.02}$ & 0.60$^{+0.03}_{-0.03}$  & 182.8 (192) \\
5	& 0.10--0.14 & \multicolumn{10}{c}{no possible constraints} \\\\
\end{tabular}
\label{tab_fp}
\begin{tablenotes}
\footnotesize
\item[0] The external absorption column was fixed in this case to $n_H = 0.05 \times 10^{22}\ \mathrm{cm^{-2}}$
\item
\item[a] Spectrum index - Total number of counts in each spectrum is 698 (1) ; 3918 (2) ; 13330 (3) ; 9991 (4) ; 4324 (5)
\item[b] External absorption column (internal = $0.057 \times 10^{22}\ \mathrm{cm^{-2}}$)
\item[c] Power-law photon index
\item[d] Power-law normalization
\item[e] Inner-disc temperature
\item[f] Disc normalization
\item[g] Absorbed flux (0.3--10 keV) 
\item[h] Unabsorbed luminosity (0.3--10 keV) for $D=4.8\ \mathrm{Mpc}$
\item[i] Fraction of the total unabsorbed flux (0.3--10 keV) in the disc component
\item[j] $\chi^2$ and degrees of freedom
\item[k] Electron temperature in keV
\item[l] Comptonization normalization
\item[m] Plasma optical depth
\item
\item All errors are at the 90 per cent confidence level.
\end{tablenotes}
\end{threeparttable}
\end{table*}

\section[]{Conclusion}

We performed a study using the longest X-ray monitoring of a ULX available to date. Thanks to more than four years of {\it Swift} observations, it becomes clear that the 115~day periodicity seen in the first 500~days (S09) is not stable and disappears only after a few cycles. Therefore, it cannot be considered to be the orbital period of the system, but is probably related to a super-orbital phenomenon as suggested by \citet{Foster10}.
The last 600~days of the monitoring do not reveal any significant periodicities in the periodograms and suggest that long periodicities are chaotic, or that there is no strict periodicity in this ULX.

Perhaps a more interesting feature in the light curve is the presence of repeated intervals of dips. Apparently, the dips form during relatively short times (sub-periods of $\sim 1$~week within longer periods of $\sim 50$~days), are highly variable (amplitude of $\ga 35$ in count rate) on timescales of hours and days, and possibly repeat every $\sim 250$~days, albeit they are not strictly periodic. This is different from the short-term variability that has been seen in previous studies, because the count rate in the soft (0.3-1.5~keV) and hard (1.5-10~keV) bands during the dips seem to drop in a similar way, suggesting the need for a structure large enough to block both soft and hard X-rays.

It appears that monitoring ULXs with good sampling (a daily monitoring is optimal) is crucial to finding behaviour that departs from their usually persistent X-ray luminosity and spectral parameters measured in discrete, pointed observations, and {\it Swift} is the ideal telescope in that regard. These could be subsequently studied with larger X-ray telescopes and at other wavelengths.
For N5408X1, an interesting follow-up would be dedicated deep X-ray observations during a dip. This may help us understand the correct physical interpretation of the spectra of ULXs. Along with simultaneous optical observations, this should shed some light on the poorly known geometry and parameters of this system.

\section*{Acknowledgments}
The authors of this paper are extremely grateful to the {\it Swift} team and in particular to Neil Gehrels for awarding us a long and detailed monitoring of this source (in 2011 and 2012) which has led to this paper. FG and PK acknowledge support from NASA grant NNX10AF86G. FG also acknowledges partial support from the Spanish Ministry of Science and Innovation (MICINN) under grant AYA 2010-18080. We acknowledge the use of the Coyote Graphics Library (available at http://www.idlcoyote.com) for the making of the plots. FG thanks Laurence Boirin for a careful read of this paper. We thank the referee for a very thorough read of our paper and for numerous comments that helped improve the clarity of this paper.

\appendix
\section{Fake data set}

\begin{figure*}
\centering
   \begin{tabular}{cc}
    \includegraphics[width=5cm,angle=270]{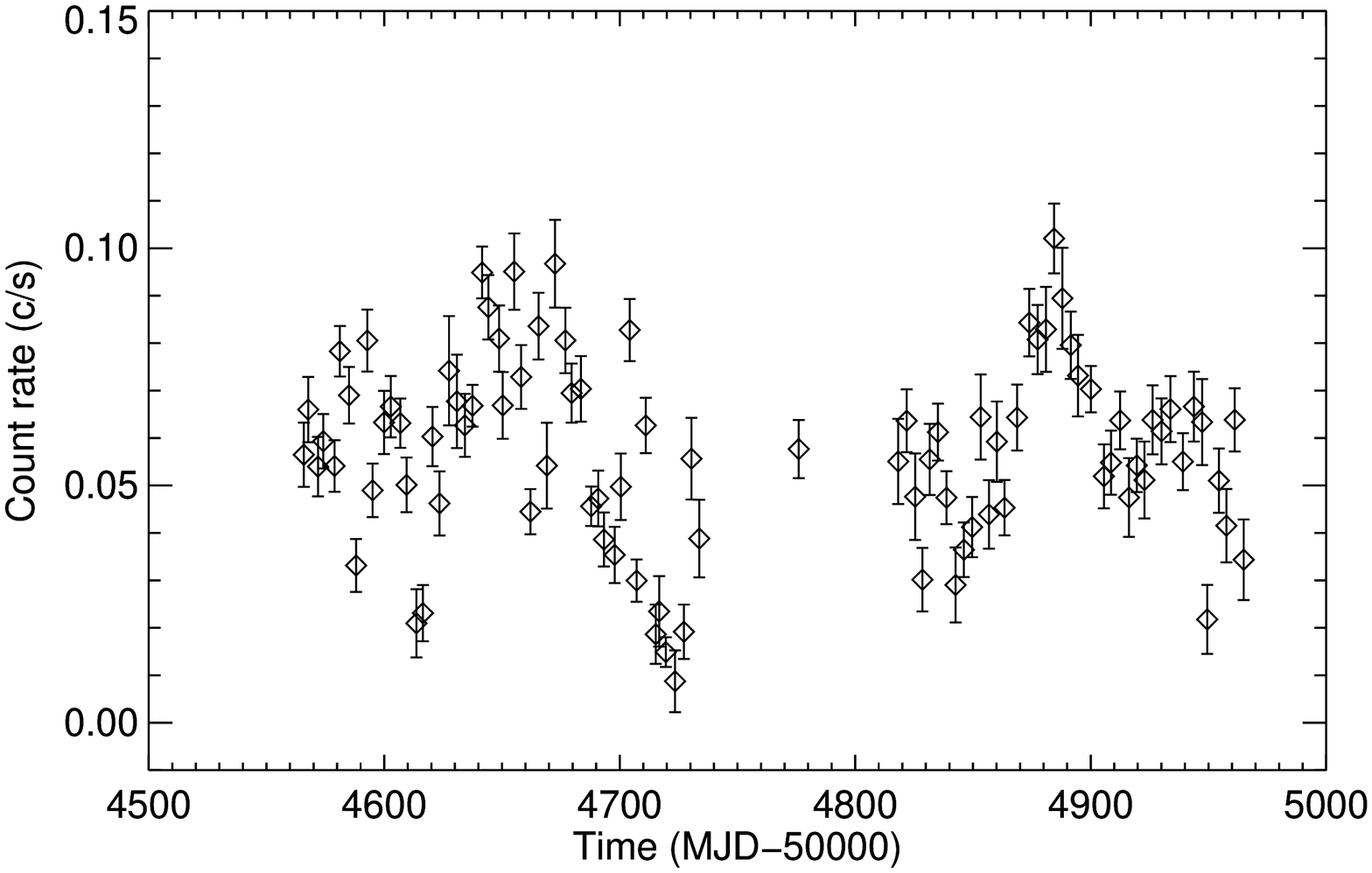}
   &\includegraphics[width=5cm,angle=270]{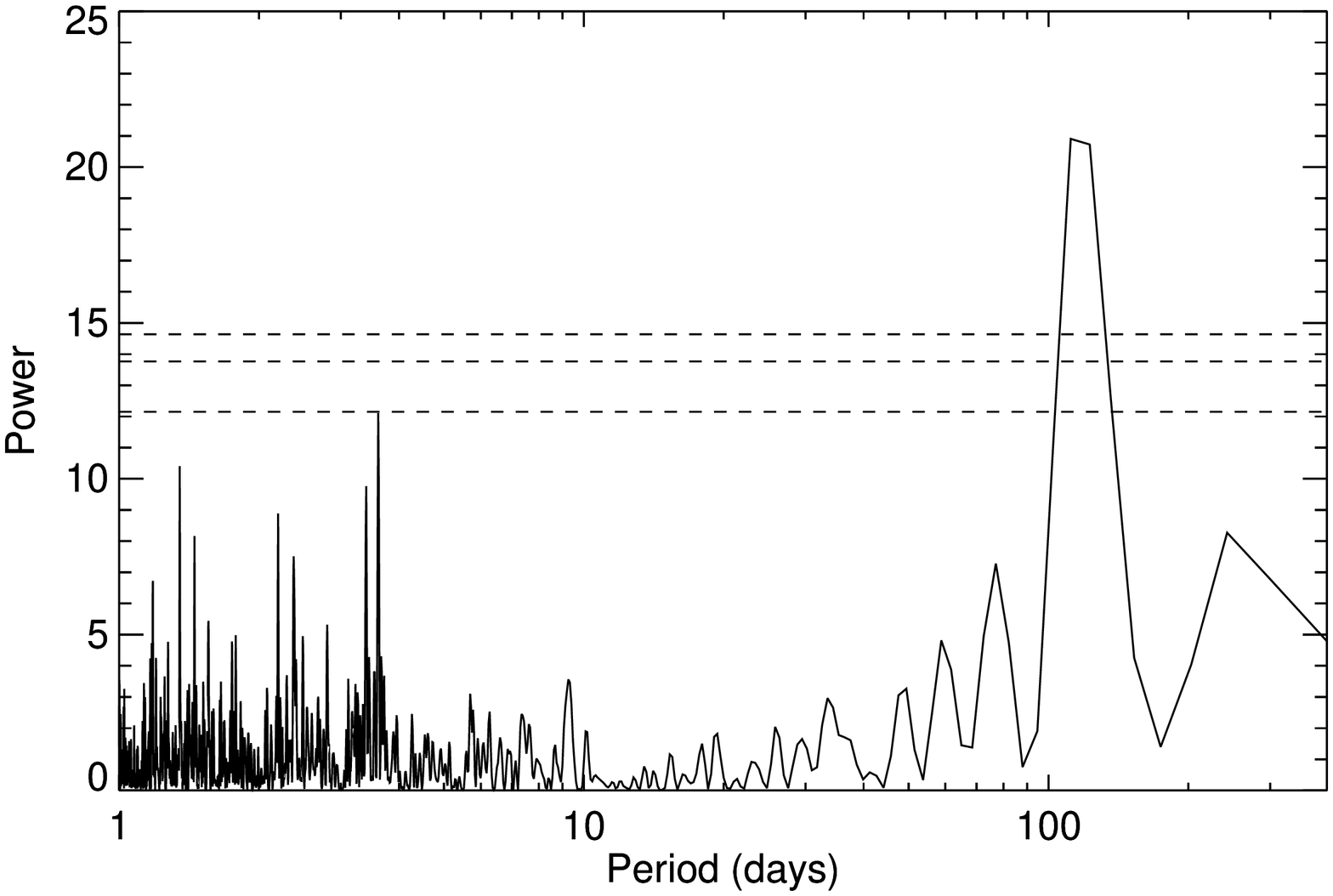}\\
    \includegraphics[width=5cm,angle=270]{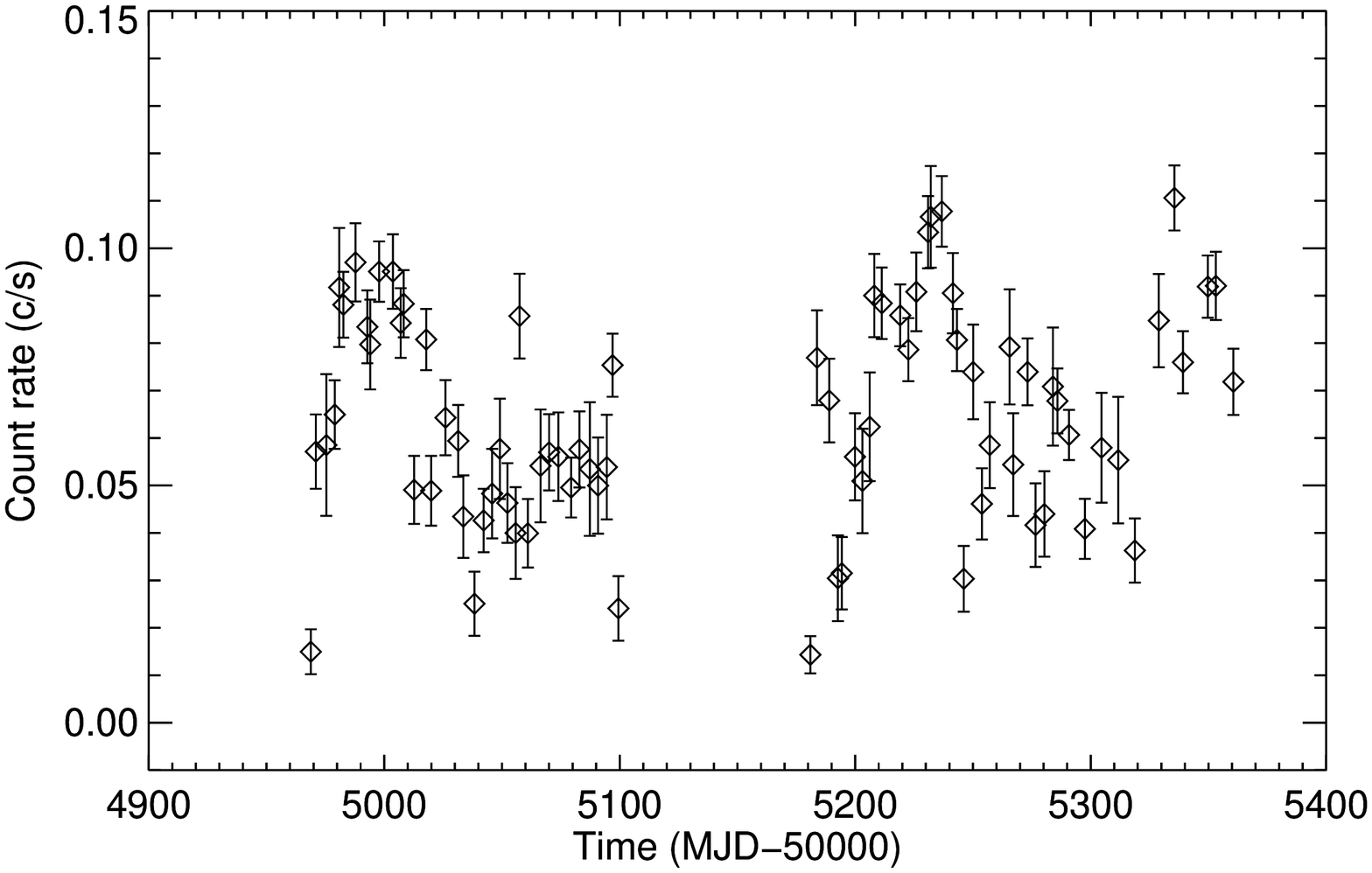}
   &\includegraphics[width=5cm,angle=270]{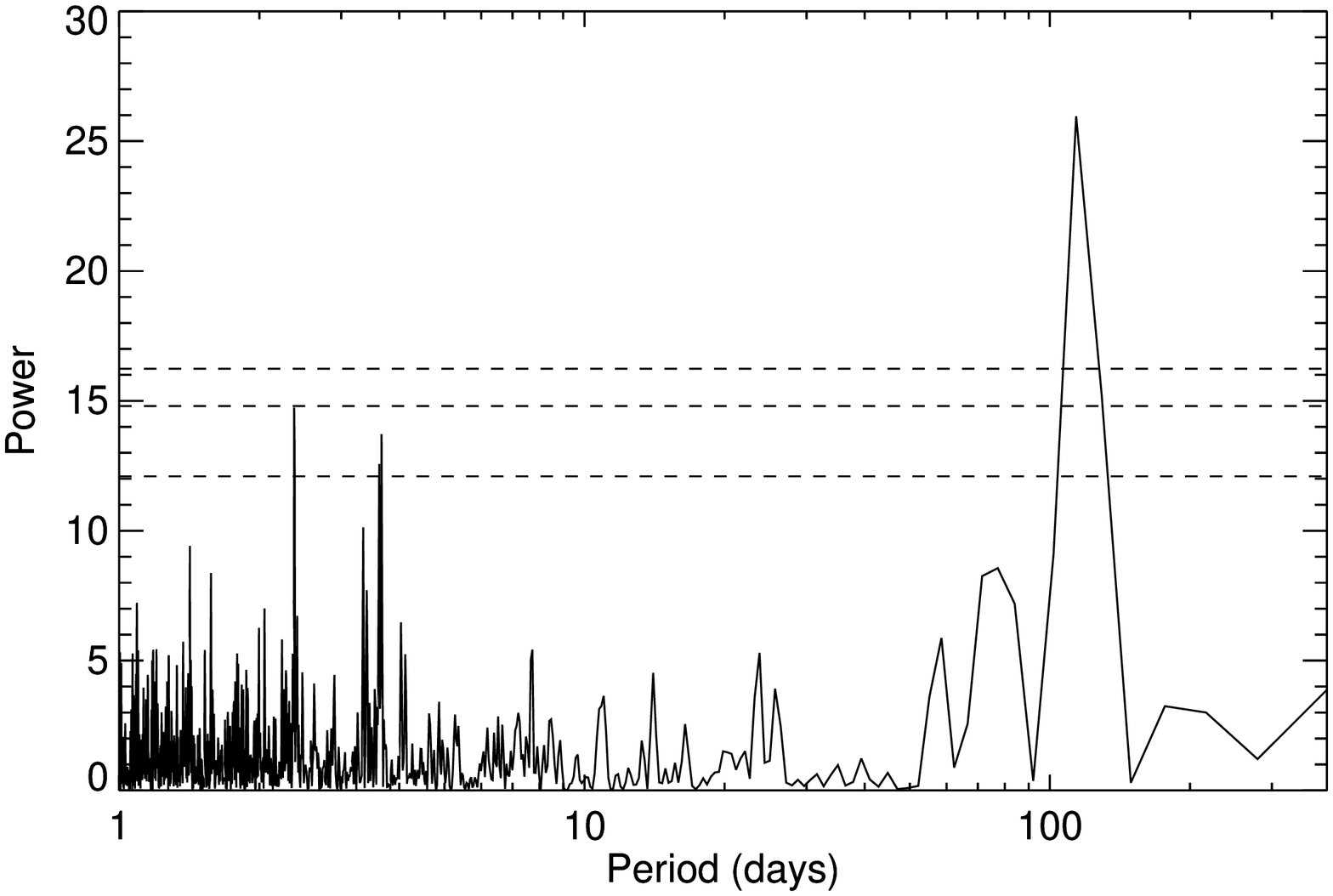}\\
    \includegraphics[width=5cm,angle=270]{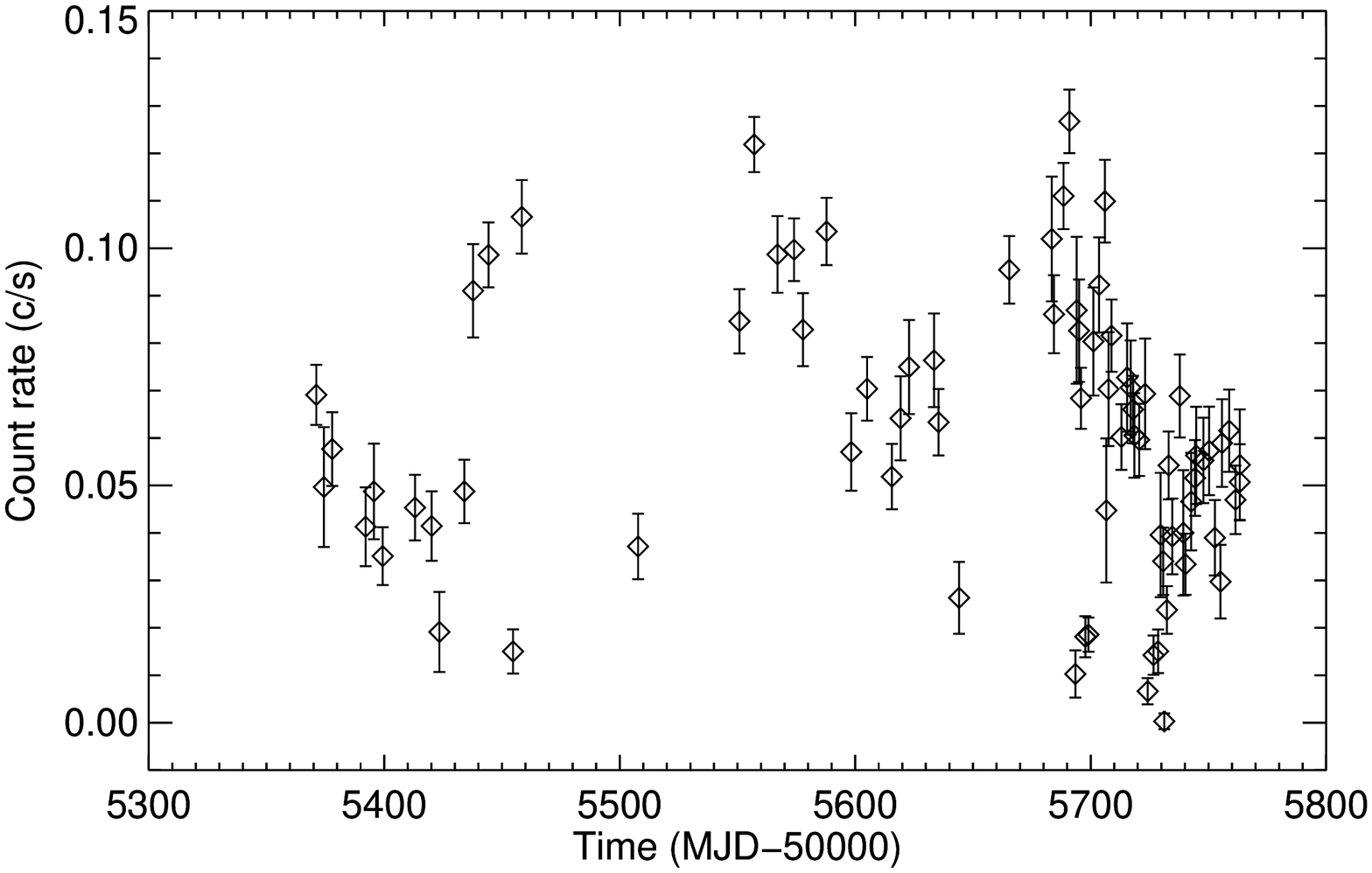}
   &\includegraphics[width=5cm,angle=270]{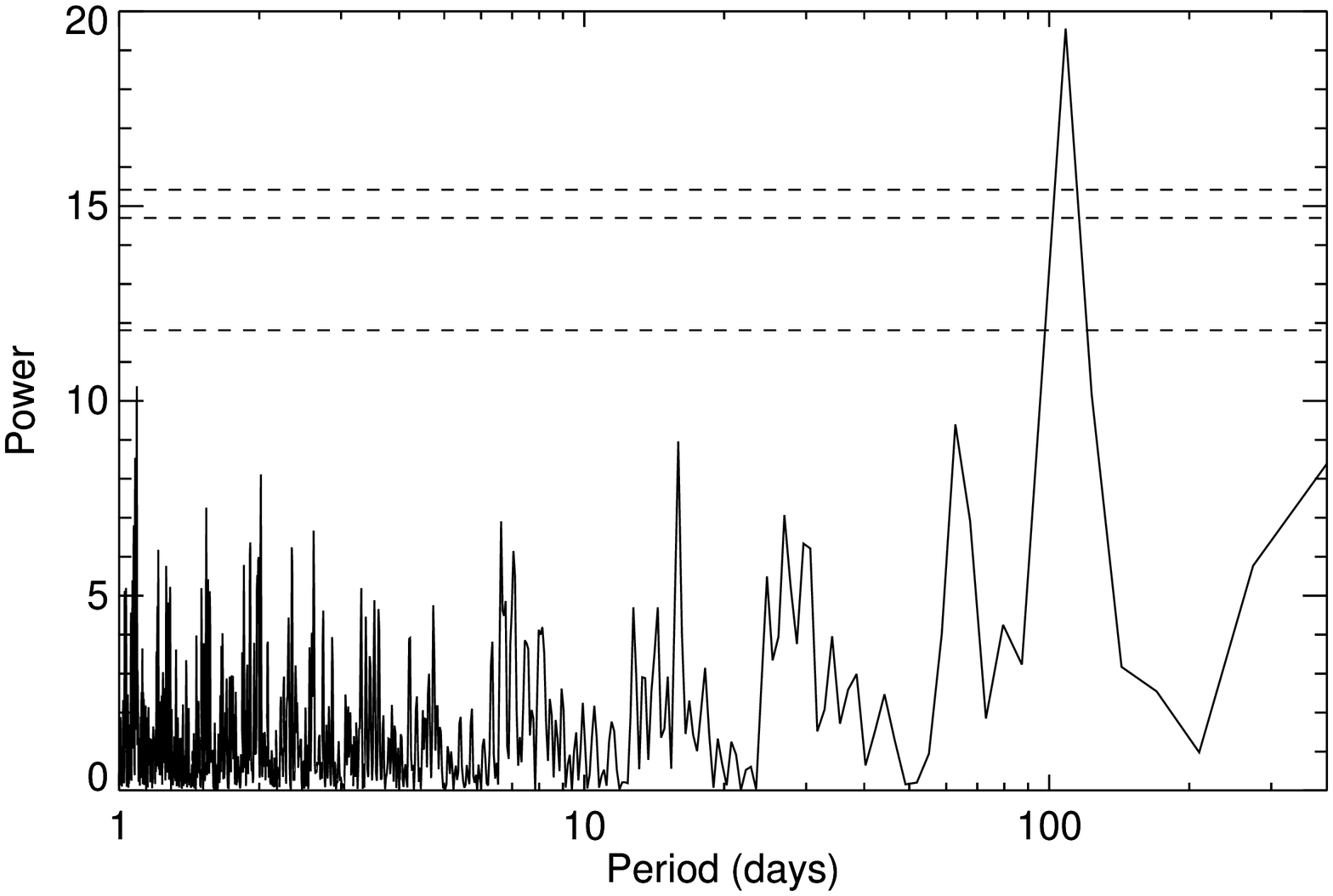}\\
    \includegraphics[width=5cm,angle=270]{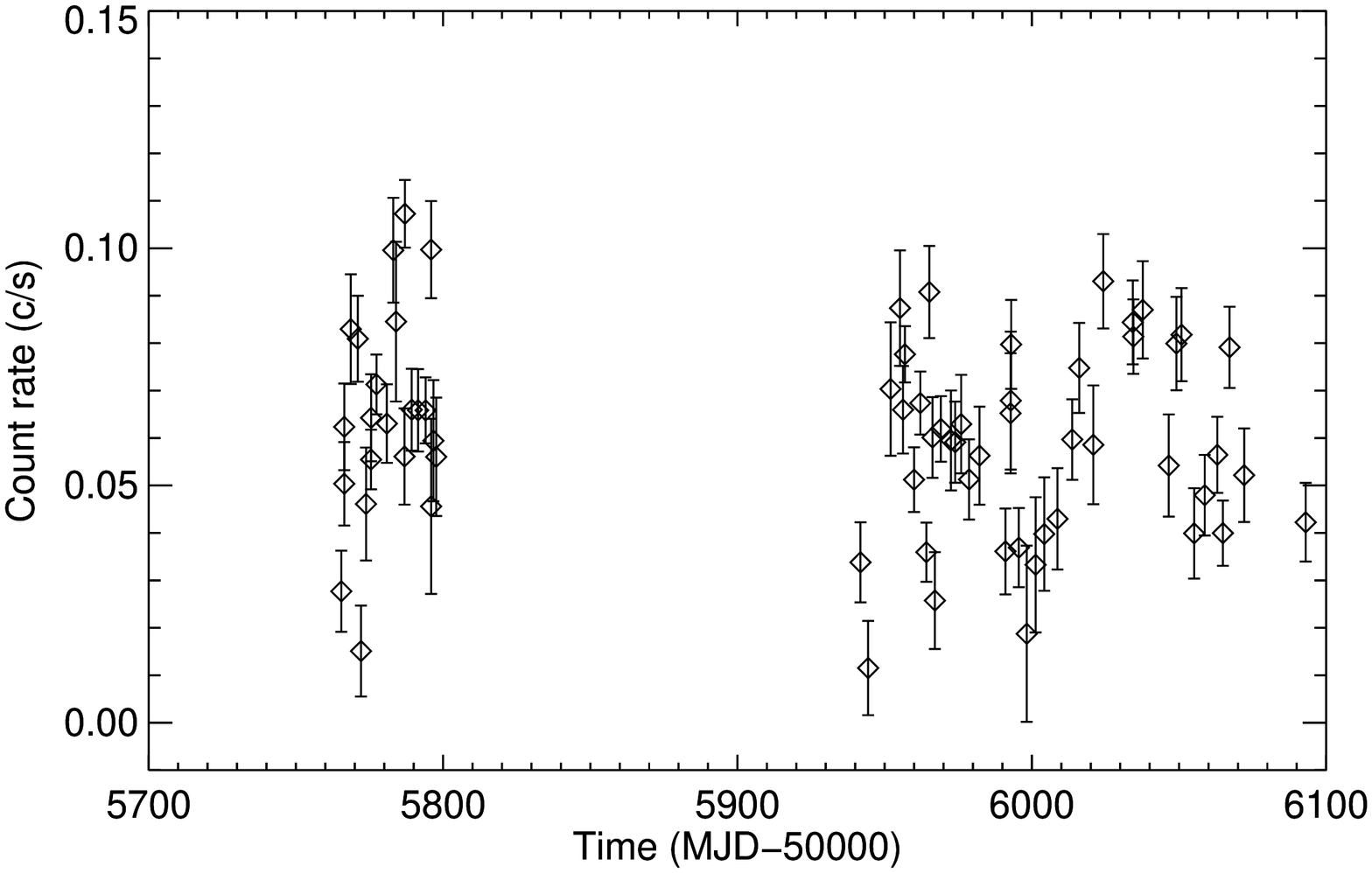}
   &\includegraphics[width=5cm,angle=270]{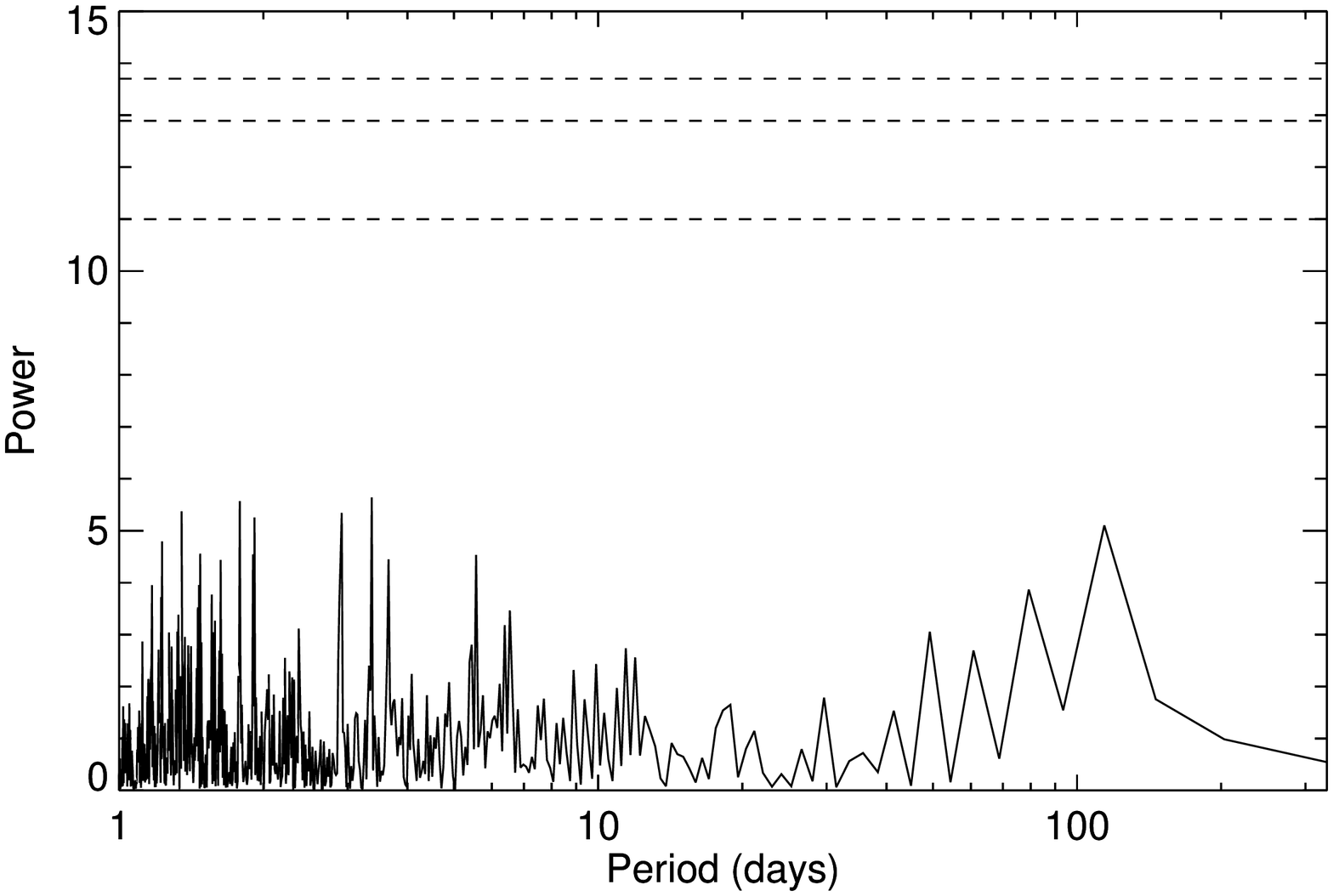}\\
    \end{tabular}
   \caption[]{Left: X-ray light curves in the 0.3--10 keV band for the fake data set, split in 400~day windows. Only the light curves rebinned with one observation per day are shown here, for clarity. Right: corresponding Lomb-Scargle periodograms calculated from the unbinned light curves. Dashed, horizontal lines denote the false alarm probability (fap) corresponding to confidence levels of 0.99, 0.999, and 0.999936 (or significance of 2.6, 3.3, and $4 \sigma$), based on white noise simulations. The main peak in the top periodogram is at $\sim 112$~days.} 
   \label{lc_periodograms_fake}
\end{figure*}

We simulated a fake data set containing a sinusoid with a period of 115.5~days  using the best fit to the folded light curve using the S09 interval only, including the variance of the original data set, and we sampled it at the 723 epochs of the (unbinned) observations. We also replaced intervals of dips (count rate $< 0.03$) with data coming from the observations, and we submitted this fake data set to a Lomb-Scargle analysis using 400~day windows, as well as contiguous windows (see Figures \ref{lc_periodograms_fake} \& \ref{lc_periodograms_fake_contig} for associated plots). The main conclusion is that the 115~day period, if stable, should be seen with high significance in all periodograms from all temporal sub-divisions, except for the very last part of the light curve that is simply too short and with a wide gap.

\begin{figure*}
\centering
   \begin{tabular}{cc}
    \includegraphics[width=5cm,angle=270]{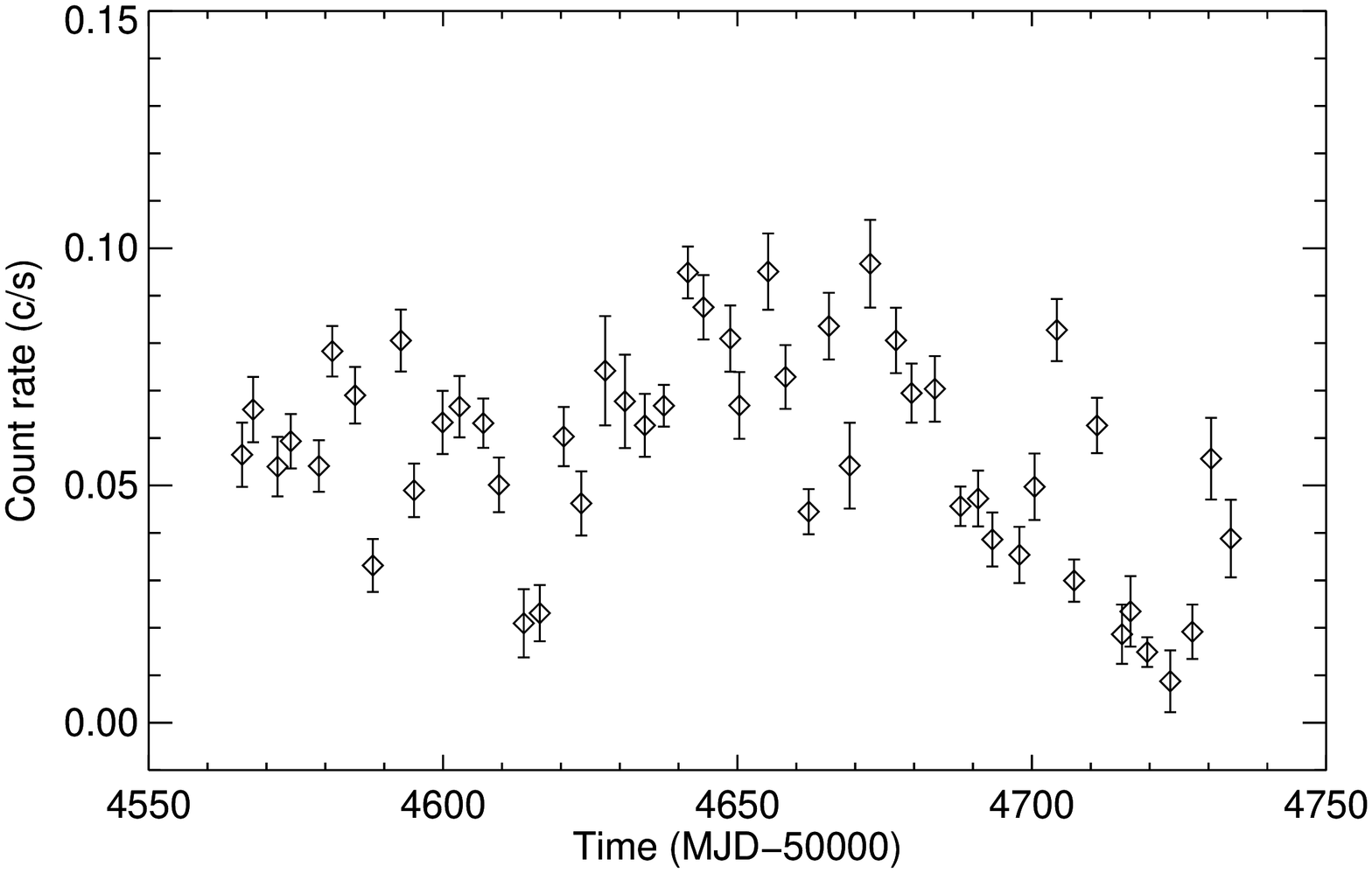}
   &\includegraphics[width=5cm,angle=270]{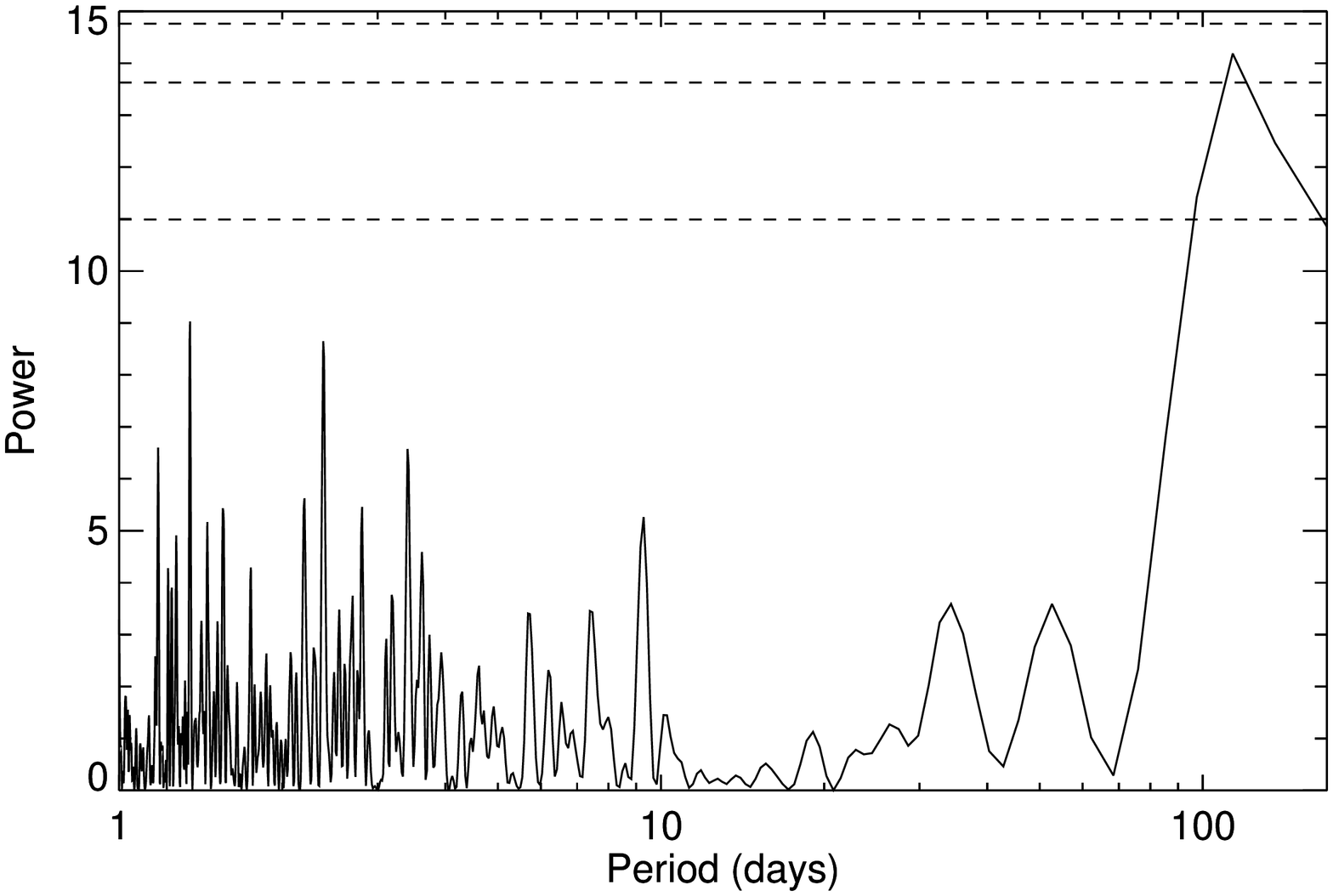}\\
    \includegraphics[width=5cm,angle=270]{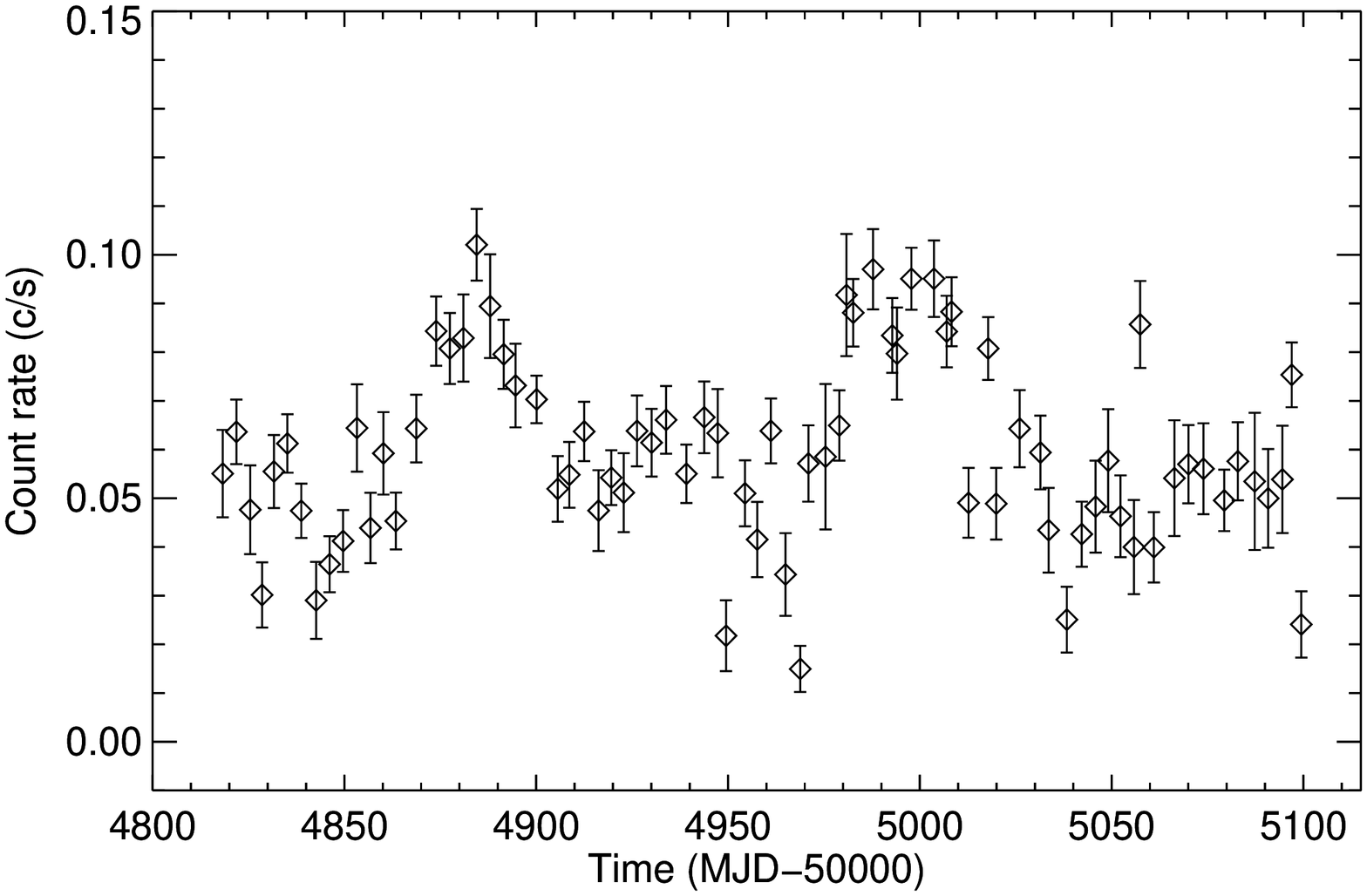}
   &\includegraphics[width=5cm,angle=270]{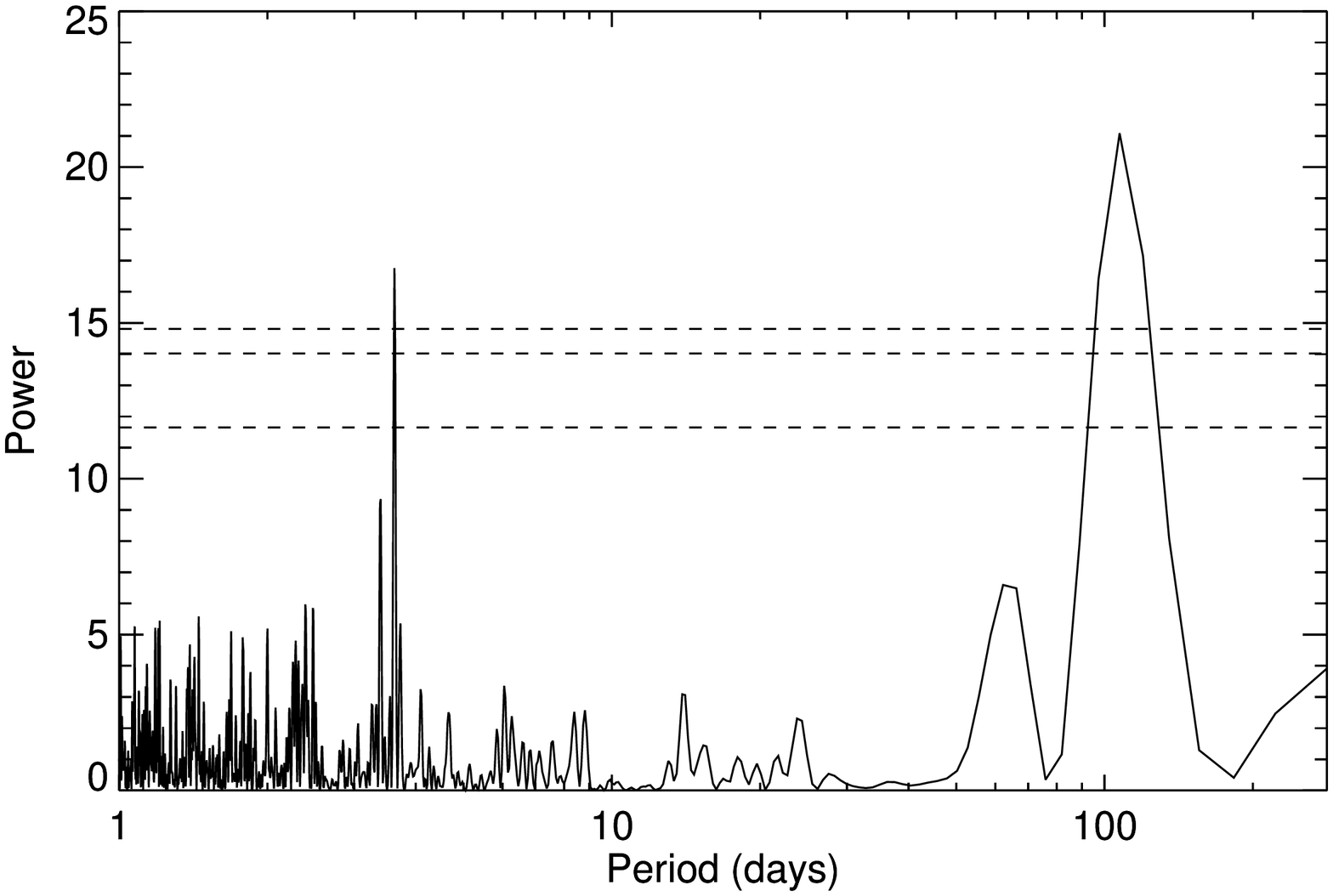}\\
    \includegraphics[width=5cm,angle=270]{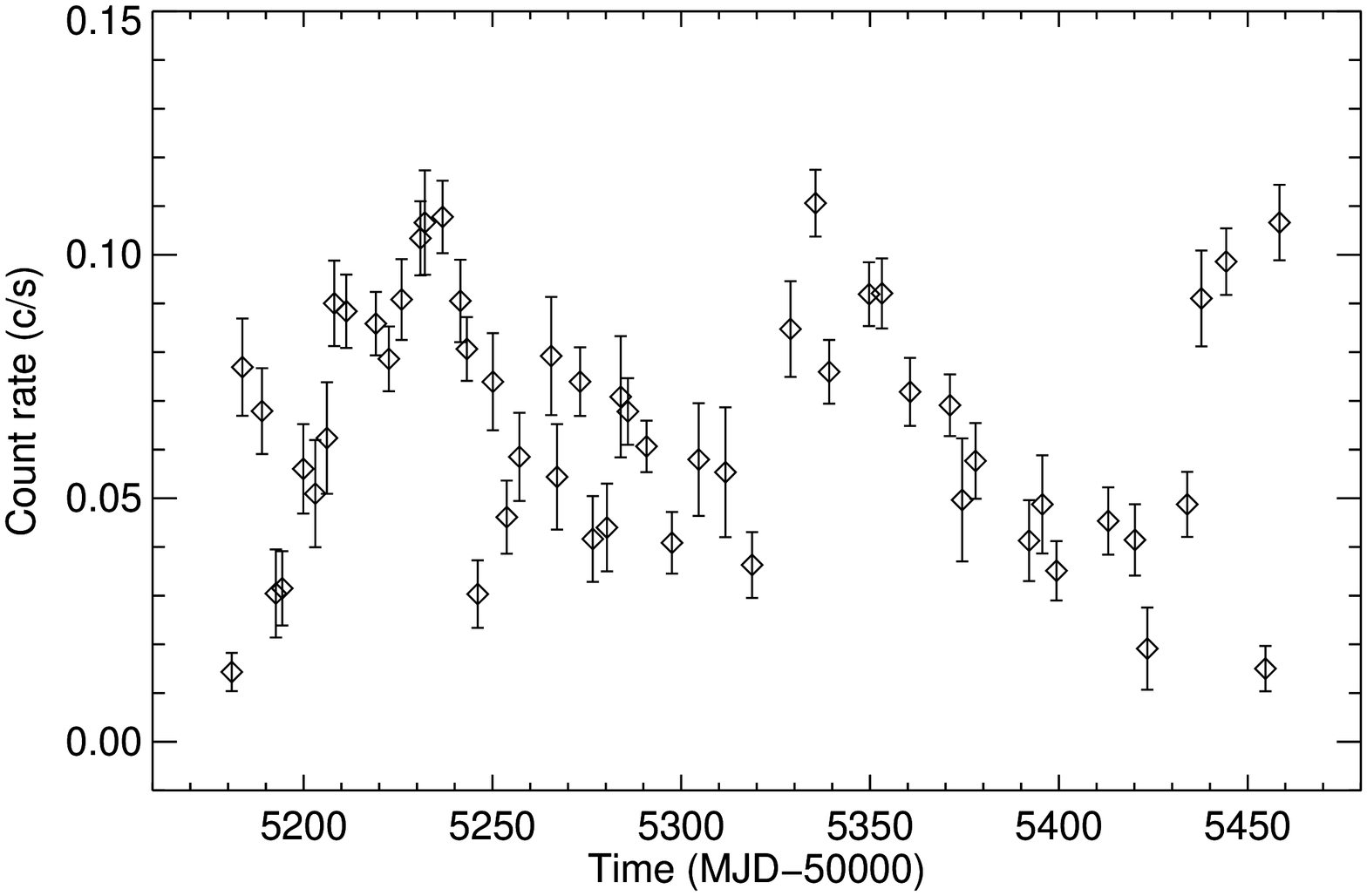}
   &\includegraphics[width=5cm,angle=270]{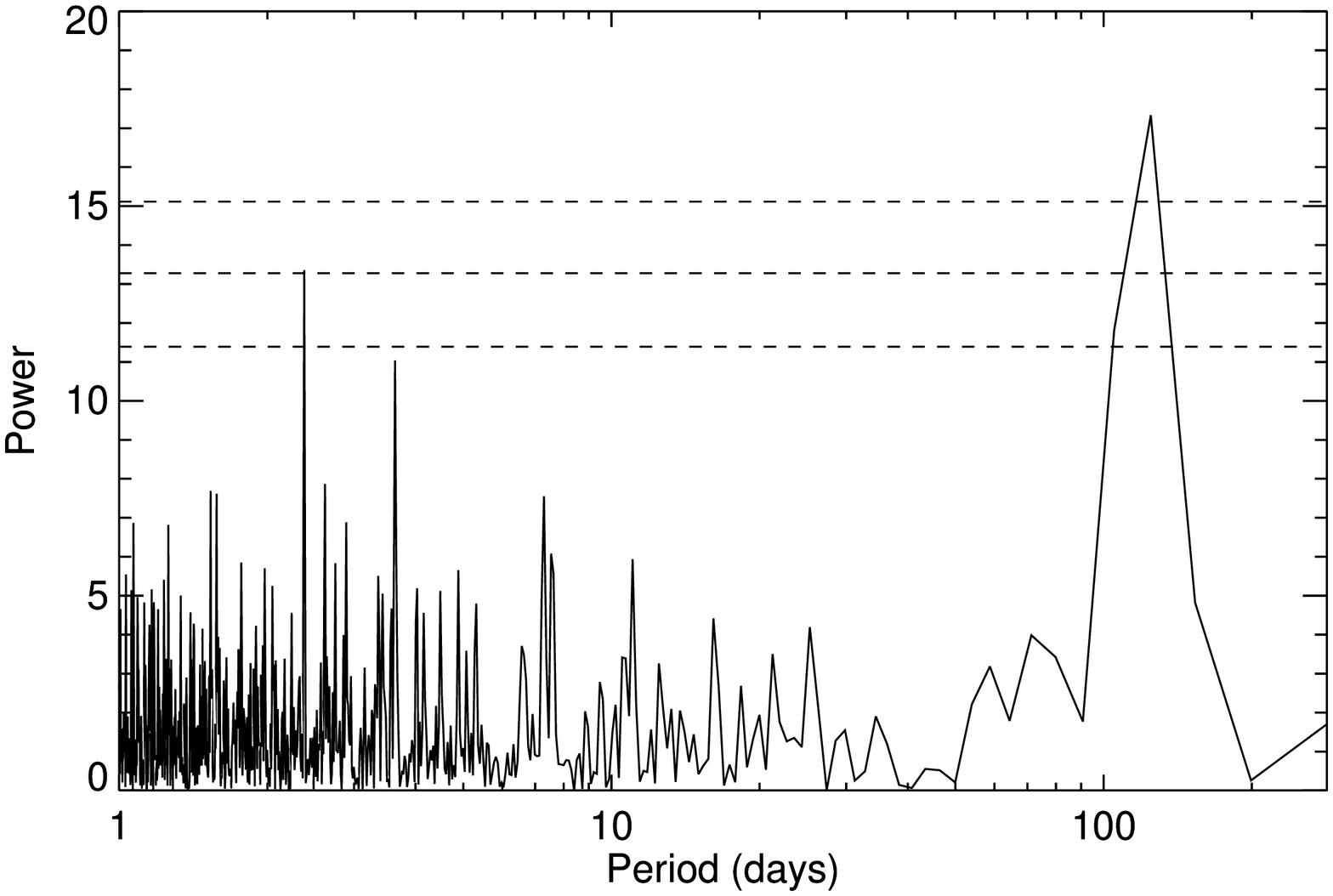}\\
    \end{tabular}
   \caption[]{Left: X-ray light curves in the 0.3--10 keV band for the fake data set, split in contiguous windows (i.e. windows with no large gaps). Only the light curves rebinned with one observation per day are shown here, for clarity. Right: corresponding Lomb-Scargle periodograms calculated from the unbinned light curves. Dashed, horizontal lines denote the false alarm probability (fap) corresponding to confidence levels of 0.99, 0.999, and 0.999936 (or significance of 2.6, 3.3, and $4 \sigma$), based on white noise simulations. The main peak in the top periodogram is at $\sim 114$~days.} 
   \label{lc_periodograms_fake_contig}
\end{figure*}

\begin{figure*}
\centering
   \begin{tabular}{cc}
    \includegraphics[width=5cm,angle=270]{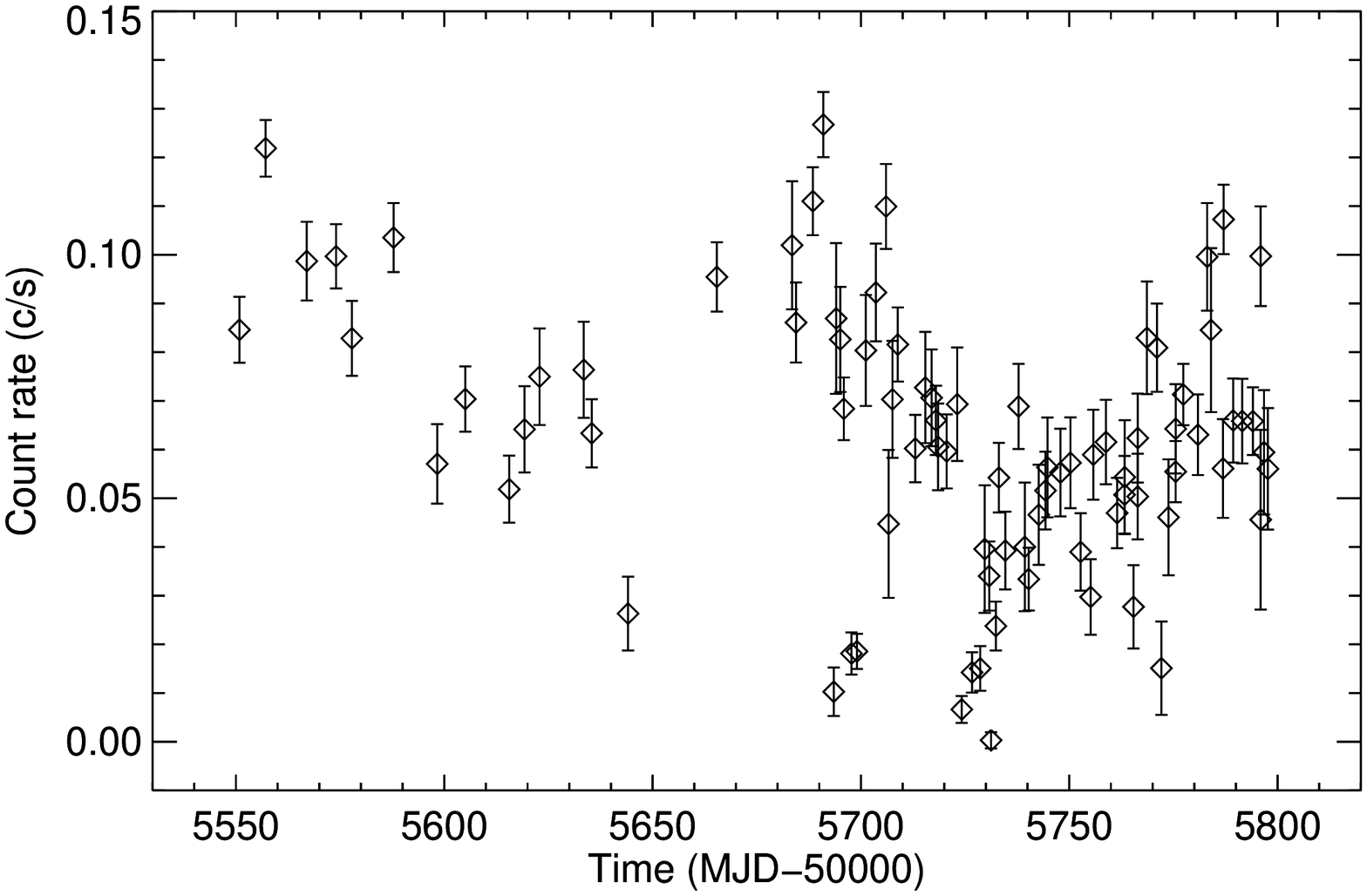}
   &\includegraphics[width=5cm,angle=270]{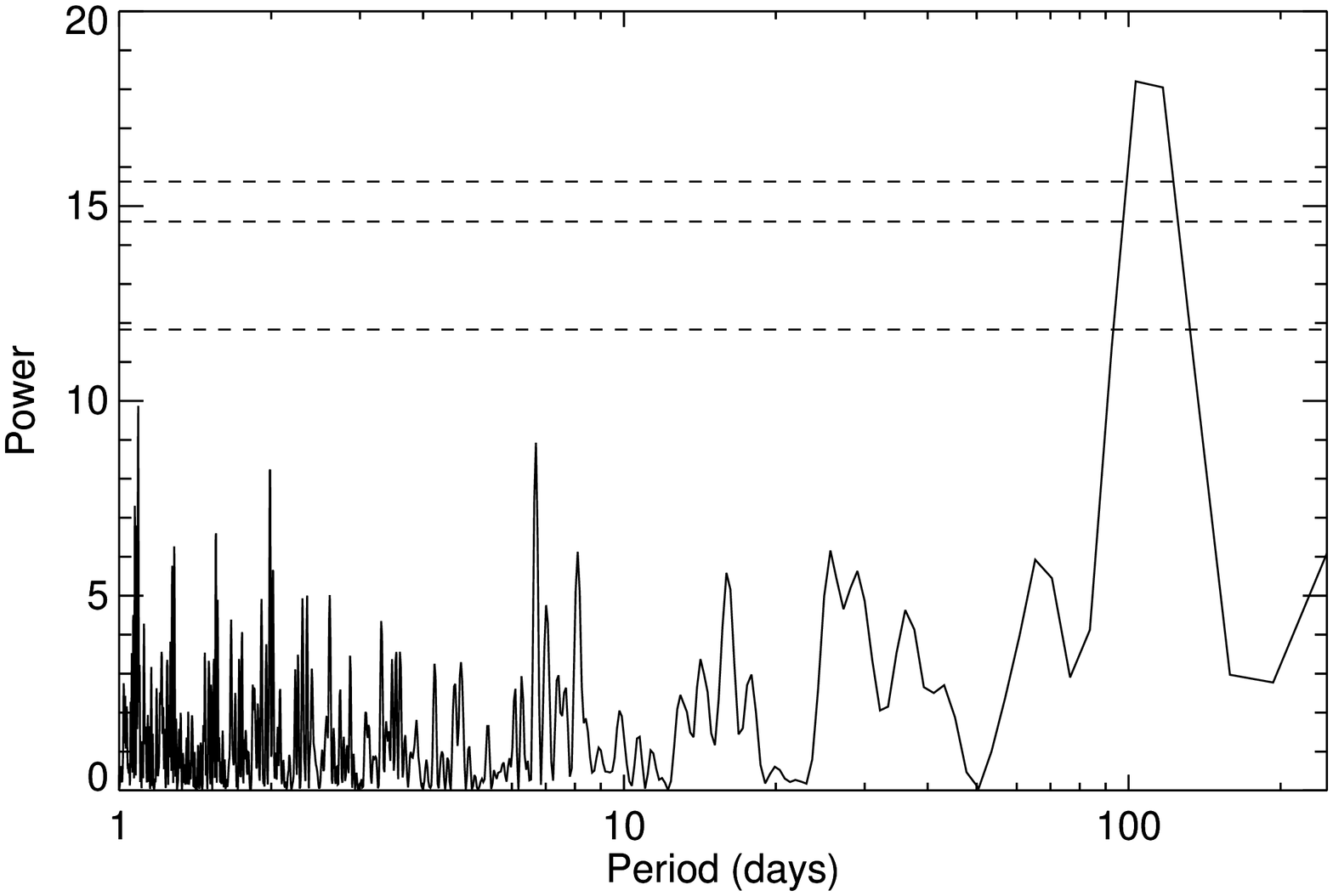}\\
    \includegraphics[width=5cm,angle=270]{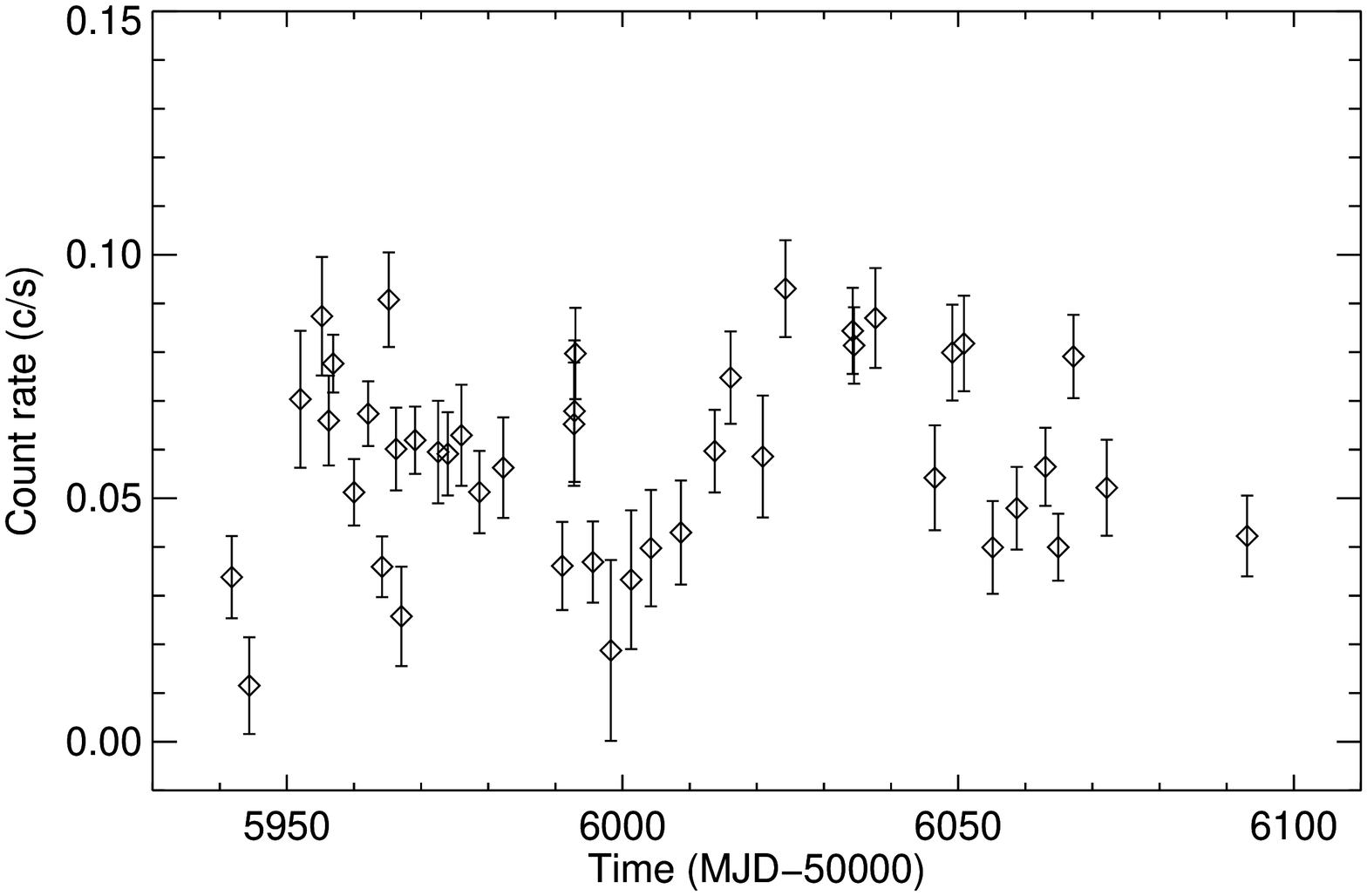}
   &\includegraphics[width=5cm,angle=270]{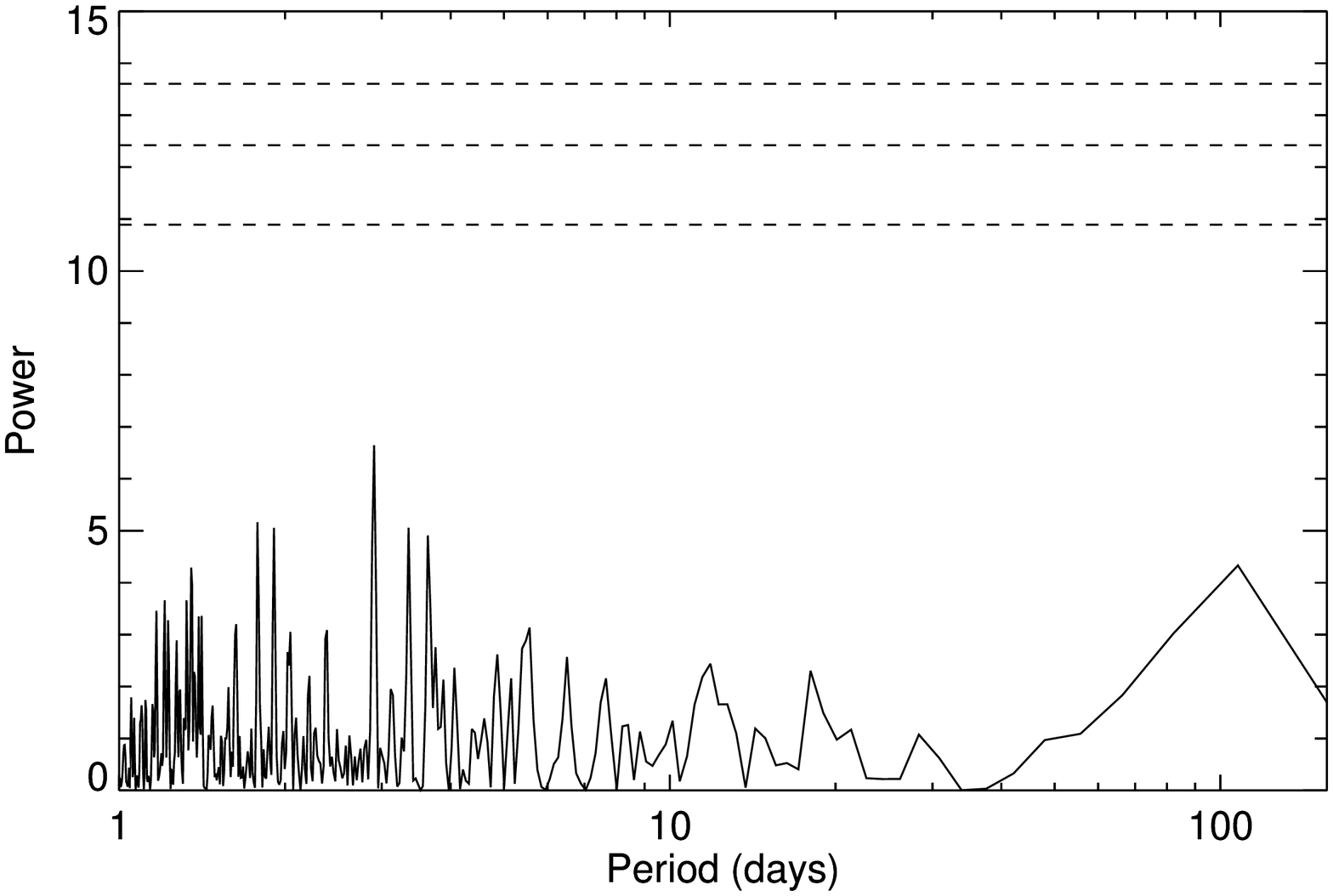}\\
    \end{tabular}
   \contcaption{} 
\end{figure*}

\end{document}